\documentclass[12pt]{article}
\usepackage{amssymb}
\usepackage{amsmath}
\usepackage{cprotect}
\usepackage{cite}
\usepackage{axodraw}
\usepackage{youngtab}
\usepackage{hyperref}
\input xy
\xyoption{all}
\usepackage{enumitem}

\newcommand{\bea}{\begin{eqnarray}}
\newcommand{\eea}{\end{eqnarray}}

\def\sqr#1#2{{\vcenter{\vbox{\hrule height.#2pt
            \hbox{\vrule width.#2pt height#1pt \kern#1pt
                  \vrule width.#2pt}\hrule height.#2pt}}}}

\def\sqra#1#2#3{{\vcenter{\vbox{\hrule height.#2pt
            \hbox{\vrule width.#2pt height#1pt \kern5pt 
#3
                  \vrule width.#2pt}\hrule height.#2pt}}}}

\numberwithin{equation}{section}

\numberwithin{table}{section}

\begin{document}

\begin{center}

{\large\bf Decomposition and the Gross-Taylor string theory}

\vspace*{0.2in}

Tony Pantev$^1$,  
Eric Sharpe$^2$

\begin{tabular}{cc}
{\begin{tabular}{l}
$^1$ Department of Mathematics\\
209 South 33rd Street\\
Philadelphia, PA  19104-6395, USA
\end{tabular}}
&
{\begin{tabular}{l}
$^2$ Department of Physics MC 0435\\
850 West Campus Drive\\
Virginia Tech\\
Blacksburg, VA  24061, USA \end{tabular}}
\end{tabular}

{\tt tpantev@math.upenn.edu},
{\tt ersharpe@vt.edu}

\end{center}

It was recently argued by Nguyen-Tanizaki-\"Unsal that two-dimensional
pure Yang-Mills theory is equivalent to (decomposes into) a disjoint
union of (invertible) quantum field theories, known as universes.
In this paper we compare this decomposition to the Gross-Taylor expansion
of two-dimensional pure $SU(N)$ Yang-Mills theory in the large $N$ limit as
the string field theory of a sigma model.  Specifically,
we study the Gross-Taylor expansion of  
individual
Nguyen-Tanizaki-\"Unsal universes.
These differ from the Gross-Taylor expansion of the full Yang-Mills theory
in two ways: a restriction to single instanton degrees, and some
additional contributions not present in the expansion of the full
Yang-Mills theory.
We propose to interpret the restriction to single instanton degrees
as implying a constraint,
namely that the
Gross-Taylor string has a global (higher-form) symmetry with Noether current
related to the worldsheet instanton number.
We compare two-dimensional pure Maxwell
theory as a prototype obeying such a constraint,
and also discuss in that case an analogue of the Witten effect
arising under two-dimensional theta angle rotation. 
We also propose a geometric interpretation of the additional terms,
in the special case of Yang-Mills theories on two-spheres.
In addition, also for the case of theories on two-spheres,
we propose a reinterpretation of the terms in the Gross-Taylor
expansion of the Nguyen-Tanizaki-\"Unsal universes, 
replacing sigma models on branched covers by counting disjoint unions of stacky
copies of the target Riemann surface,
that makes the Nguyen-Tanizaki-\"Unsal decomposition into invertible
field theories more nearly manifest.
As the Gross-Taylor string is a sigma model coupled
to worldsheet gravity, we also briefly outline the tangentially-related topic 
of decomposition in two-dimensional
theories coupled to gravity.

\begin{flushleft}
July 2023
\end{flushleft}

\newpage

\tableofcontents

\newpage

\section{Introduction}

The papers \cite{Gross:1992tu,Gross:1993hu,Gross:1993yt,Cordes:1994sd,Moore:1994dk,Cordes:1994fc}
proposed that pure two-dimensional Yang-Mills theory could be understood
as the string field theory of a string theory.  This was demonstrated by
first expanding the $SU(N)$ pure Yang-Mills partition function 
in the large $N$ limit as a formal
sum of correlation functions in two-dimensional untwisted
$S_n$ Dijkgraaf-Witten theories, summed
over all $n$, an expansion we will refer to as the
Gross-Taylor expansion.
Those correlation functions were then interpreted
combinatorially in the form of a sum over maps
\cite{Gross:1992tu,Gross:1993hu,Gross:1993yt},
and then later, using an interpretation of
the $\Omega$-points,
in terms of branched
covers \cite{Cordes:1994sd,Moore:1994dk,Cordes:1994fc,Ramgoolam:1994ba},
suggesting the interpretation as a string field theory of some sigma
model.  
Specific proposals have been made for the corresponding sigma model, 
at least at the level of proofs of
principle, 
see \cite{Cordes:1994sd,Moore:1994dk,Cordes:1994fc} for a sigma model
localizing on holomorphic maps, and
\cite{Horava:1993aq,Horava:1995ic} for a sigma model localizing
on harmonic maps.

In this paper we will reexamine these arguments in the context of
decomposition \cite{Hellerman:2006zs} of
two-dimensional pure Yang-Mills theories.
Decomposition is
the observation that $d$-dimensional quantum field theories with
global $(d-1)$-form symmetries are equivalent to (``decompose into'')
disjoint unions of other theories, known in this context as universes.
It was first described in \cite{Hellerman:2006zs} as part of an effort
to resolve questions concerning the consistency of
string propagation on stacks.
Decomposition has been checked in a wide variety of contexts
(including not only orbifolds and gauge theories but also 
e.g.~open string theory
and K theory \cite{Hellerman:2006zs}) 
via techniques including for example
mirror symmetry \cite{Pantev:2005zs,Hellerman:2006zs},
supersymmetric
localization \cite{Sharpe:2014tca}, 
and numerical/lattice computations \cite{Honda:2021ovk}, 
in not only two dimensional
theories but also in three (see 
e.g.~\cite{Pantev:2022kpl,Pantev:2022pbf,Perez-Lona:2023llv})
and four dimensions
(see e.g.~\cite{Tanizaki:2019rbk,Cherman:2020cvw}).
Its applications have included phases of gauged linear
sigma models (see e.g.~\cite{Caldararu:2010ljp,Addington:2012zv,Sharpe:2012ji,Hori:2011pd,Chen:2020iyo,Guo:2021aqj,Halverson:2013eua,Hori:2013gga,Hori:2016txh,Wong:2017cqs,Kapustka:2017jyt,Katz:2022lyl,Katz:2023zan,Lee:2023piu}),
predictions for Gromov-Witten invariants
(see e.g.~\cite{ajt1,ajt2,ajt3,t1,gt1,xt1}), 
IR limits of pure supersymmetric gauge theories and
elliptic genera \cite{Eager:2020rra},
adjoint QCD$_2$
\cite{Komargodski:2020mxz},
and anomalies in orbifolds \cite{Robbins:2021xce}.
(See also \cite{Meynet:2022bsg} for a recent relation to 
quivers with multiple components.)
See e.g.~\cite{Sharpe:2010zz,Sharpe:2010iv,Sharpe:2019yag,Sharpe:2022ene} 
for reviews.

In particular, decomposition has been applied to argue that
two-dimensional pure Yang-Mills theories
are equivalent to disjoint unions of invertible field theories
(meaning, trivial field theories with only a vacuum state)
\cite{Cherman:2020cvw,Nguyen:2021yld,Nguyen:2021naa},
with universes in one-to-one correspondence with irreducible representations
of the gauge group.  We will refer to those universes of the
decomposition of two-dimensional
pure Yang-Mills as Nguyen-Tanizaki-\"Unsal universes.

We begin the paper with a short review of decomposition in two-dimensional
pure Yang-Mills in section~\ref{sect:rev}.  
The rest of this paper is organized into three main sections:
\begin{enumerate}
\item First, in section~\ref{sect:revisit-gt},
we discuss the combinatorics of the Gross-Taylor expansion,
deriving expressions for the expansion of the individual Nguyen-Tanizaki-\"Unsal
universes.  

Viewed as a sum of two-dimensional untwisted Dijkgraaf-Witten theory
correlation functions, the result is extremely natural:
one restricts to a value of $n$ defined by the representation of
$SU(N)$ defining the universe, corresponding to the symmetric
group $S_n$, and then in correlation functions,
one inserts projectors onto a single $S_n$ Dijkgraaf-Witten universe,
using the known fact that two-dimensional Dijkgraaf-Witten theory also
decomposes.
In other words, a single Nguyen-Tanizaki-\"Unsal universe receives contributions
only from a single Dijkgraaf-Witten theory universe in the Gross-Taylor
expansion -- the two decompositions intertwine naturally.

Interpreted in terms of sigma models,
the resulting expressions for the separate universes are more subtle.
They
appear naively to describe (1) restrictions of the Gross-Taylor theory to maps
(worldsheet instantons)
of fixed degrees, plus (2) some additional contributions.
Both require further explanation, to which we turn
in the next section.
\item In section~\ref{sect:interp},
we 
suggest interpretations of the two points raised above.
We propose that the restrictions to maps of fixed degree
be interpreted physically
as a new constraint on the Gross-Taylor sigma model,
that it possess a higher-form
symmetry with Noether current coupling to the pullback of the K\"ahler
form (which integrates to worldsheet instanton number, the
covering map degree).
We discuss the prototypical example of two-dimensional pure Maxwell
theory, which has a one-form symmetry coupling to $U(1)$ bundle curvature.

In hindsight, existence of such a symmetry in the Gross-Taylor
string is expected from the yoga relating target-space
and worldsheet symmetries, as we discuss, though the coupling to
worldsheet instanton degree is not predicted by yoga alone. 

We also discuss 
the interpretation of the additional 
contributions mentioned above.
Such additional contributions are typical in a decomposition,
and cancel out when one sums over universes, as we review,
so their existence is not a surprise, but they do require interpretation.
In the special case of 
Yang-Mills theories on $S^2$,
we propose a geometric interpretation
involving
stacky worldsheets.
\item Finally, in section~\ref{sect:disjoint},
in the special case of Yang-Mills theories on $S^2$, 
we propose an alternative geometric interpretation of the
Gross-Taylor expansion of the Nguyen-Tanizaki-\"Unsal universes,
not in terms of sigma models on
branched covers of $\Sigma_T$, but instead in terms
of some sort of counting problem, 
counting stacky copies of $\Sigma_T$, in line with the interpretation as 
invertible field theories.

\end{enumerate}

In subsection~\ref{sect:interp:proto}, we also discuss an analogue of the
Witten effect in two-dimensional pure Maxwell theory, in which universes
of the decomposition are interchanged under theta angle rotation.

In appendix~\ref{app:identities} we include some relevant group algebra
identities, of use in the series expansions in section~\ref{sect:revisit-gt}.
In appendix~\ref{app:basics-stacks}, we collect some pertinent basics
of stacks, to assist in understanding stacky worldsheets.

Appendix~\ref{app:grav:decomp} discusses examples of decomposition in
two-dimensional theories coupled to worldsheet gravity.
(This is not necessarily our prediction for the Gross-Taylor string,
but is certainly a possibility, 
as the Gross-Taylor sigma model
is coupled to worldsheet gravity.)
Briefly, unlike a typical field-theoretic decomposition in which the
universes are completely decoupled, after coupling to gravity, the
universes have gravitational interactions (but no non-gravitational
interactions).

Appendix~\ref{app:ginsu} discusses possible alternative interpretations
of the restriction on maps in the Gross-Taylor expansion of the
Nguyen-Tanizaki-\"Unsal universes, as different quantum field theories
constructed to localize onto particular map degree sectors.
These alternatives are not completely satisfactory, for reasons discussed
there, so we do not advocate them, but we do include them for completeness.

\section{Short review of decomposition in two-dimensional
pure Yang-Mills}
\label{sect:rev}

Consider two-dimensional pure Yang-Mills with gauge group $G$ and
action
\begin{equation}
S \: = \: \frac{1}{g_{YM}^2} \int_{\Sigma}  {\rm Tr}\, F^2.
\end{equation}
On a Riemann surface $\Sigma$ of genus $g$ with $b$ boundaries,
along which are associated group elements $U_1, \cdots, U_b$
the partition function 
has the form\footnote{
For gauge groups with $U(1)$ factors, we shall discuss how the $\theta$
angle appears in the exact expression
in section~\ref{sect:interp:proto}.
} 
\cite[section 3.7]{Cordes:1994fc},
\cite{Migdal:1975zg,Drouffe:1978py,Lang:1981rj,Menotti:1981ry,Rusakov:1990rs,Witten:1991we,Witten:1992xu,Blau:1993hj,Aganagic:2004js},
\cite[section 2]{Vafa:2004qa},
\begin{equation}  \label{eq:pureym:z}
Z(U_1, \cdots, U_b) \: = \: 
\sum_R \left( \dim R \right)^{2-2g-b} \exp\left[- g_{YM}^2 A C_2(R)
 \right] \, \chi_R(U_1) \cdots \chi_R(U_b).
\end{equation}
where the sum is over irreducible representations $R$ of the gauge group $G$.
One can glue the Riemann surfaces along boundaries using
(see e.g.~\cite[equ'ns (3.17)-(3.19)]{Cordes:1994fc})
\begin{eqnarray}   \label{eq:orthog11}
\int dU \, \chi_R(U) \chi_S(U^{-1}) & = &
\delta_{RS},
\\
\int dU \, \chi_R(VU) \chi_S(U^{-1} W) & = &
\delta_{RS} \frac{ \chi_R(VW) }{ \dim R},
\\
\int dU \, \chi_R(U V U^{-1} W) & = &
\frac{ \chi_R(V) \chi_R(W) }{\dim R}.
\end{eqnarray}

Recently, it was observed in \cite{Cherman:2020cvw} (for abelian cases) and
\cite{Nguyen:2021yld,Nguyen:2021naa} (for nonabelian cases)
that pure two-dimensional gauge theories are equivalent to disjoint unions of
physical theories; in other words, they ``decompose,'' in the sense of
\cite{Hellerman:2006zs,Sharpe:2022ene}.  Specifically, the
component physical theories (``universes'') are in one-to-one
correspondence with irreducible representations $R$, which is naturally
reflected in the form of the partition function expressions above.
For example, the universe associated to irreducible representation $R$ has
partition function
\begin{equation}
\left(\dim R \right)^{2-2g}  \exp\left[- g_{YM}^2 A C_2(R)
 \right] 
\end{equation}
on a closed Riemann surface $\Sigma$ of genus $g$,
so that summing over universes reproduces the
Yang-Mills partition function~(\ref{eq:pureym:z})..
To be clear, decomposition in this context is the statement that
the basis of irreducible representations diagonalizes correlation functions;
every correlation function (e.g.~(\ref{eq:pureym:z}))
can be written in terms of a sum of
contributions from the constitutent universes, indexed by irreducible
representations.  This implies, but is very much stronger than, the statement
that the partition function can be written as a certain sum.

Decomposition is also visible at the level of Hilbert spaces.
The Hilbert space of two-dimensional pure Yang-Mills consists of the
class functions on $G$, meaning functions 
which are invariant under conjugation.
Such functions can be expanded in a Fourier series in characters $\chi$ of
$G$, as \cite[chapter II]{hermann}
\begin{equation}
f(g) \: = \: \sum_R c_R \chi_R(g),
\end{equation}
where the $c_R$ are constants determined by $G$,
and the sum is over irreducible representations of $G$.
Certainly decomposition into universes indexed by irreducible
representations is consistent with the structure of the Fourier series
above  More to the point, the universe associated to irreducible
representation $R$ has a one-dimensional Hilbert space, generated\footnote{
For example, the QFT admits a set of orthogonal projectors,
which in bra-ket notation have the form $| R \rangle \langle R |$ for
$R$ an irreducible representation.
}
by the character $\chi_R$.  A quantum field theory with a 
one-dimensional Hilbert space is known as an invertible field theory,
so we see that the universes of the decomposition are
each examples of
invertible field theories (as defined by that property).

More generally, unitary two-dimensional topological field theories
(with semisimple local operator algebras) decompose into
a disjoint union of invertible field theories
\cite{Durhuus:1993cq,Moore:2006dw,Huang:2021zvu,Komargodski:2020mxz}.
The partition functions of such theories are determined by an Euler
number counterterm and an area counterterm, so that on a worldsheet $\Sigma$
the partition function is just the exponential of the integral of those
counterterms, and has the form
\begin{equation}
Z(\Sigma) \: = \: (\phi_1)^{\chi(\Sigma)} 
\exp\left(- \phi_2 \mbox{Area} \right),
\end{equation}
universally, for some constants $\phi_{1,2}$.
We can see this structure in the partition function in pure
two-dimensional Yang-Mills, where the contribution from any fixed
universe / irreducible representation $R$ is
\begin{equation}
(\dim R)^{\chi(\Sigma)} \exp\left( - C_2(R) \mbox{Area} \right).
\end{equation}
In the language above, for any irreducible representation $R$,
$\phi_1 = \dim R$ and $\phi_2 = C_2(R)$.

In passing, there exist generalizations of two-dimensional pure Yang-Mills
which are also area-preserving-diffeomorphism invariant and
exactly soluble, see \cite{Witten:1992xu,Ganor:1994bq,Sugawara:1996id}.
Their classical actions are of the form
\cite[equ'n (5)]{Ganor:1994bq}
\begin{equation}
\int_{\Sigma} {\rm tr}\, \left( i \phi F \: - \Phi(\phi) \right),
\end{equation}
where $\phi$ is an auxiliary Lie-algebra-valued scalar, and $\Phi$ a function
of the form \cite[equ'n (7)]{Ganor:1994bq}
\begin{equation}
\Phi(\phi) \: = \: \sum_{\{k_i \}} a_{\{k_i\}} \prod_i 
{\rm tr}\, \left( \phi^i \right)^{k_i}.
\end{equation}
This reduces to pure Yang-Mills in the special case that
$\Phi(\phi) \propto {\rm tr}\, \phi^2$
(see e.g.~\cite[section 2.1]{Blau:1993hj}).
Their partition functions are of the form
\cite[equ'n (10)]{Ganor:1994bq}
\begin{equation}
Z \: = \: \sum_R \left( \dim R \right)^{\chi(\Sigma)}
\exp\left( - A \Lambda(R) \right),
\end{equation}
for
\begin{equation}
\Lambda(R) \: = \: \sum_{\{k_i\}} a_{\{k_i\}} 
C_{\{k_1-1 + k_2-2 + k_3-3 + \cdots\}}(R).
\end{equation}
At least naively, the arguments of \cite{Nguyen:2021yld,Nguyen:2021naa}
appear to apply, hence it is natural to conjecture
\cite{yuyapriv}
that these theories decompose, 
into universes indexed by irreducible representations $R$.

\section{Revisiting the Gross-Taylor argument}
\label{sect:revisit-gt}

In this section, we review the Gross-Taylor asymptotic series expansion
of the partition function of both pure two-dimensional $SU(N)$ Yang-Mills theory,
as well as partition functions of Nguyen-Tanizaki-\"Unsal universes therein,
in the large $N$ limit,
on a genus $p$ Riemann surface $\Sigma_T$.  We also review the interpretation
of the terms in that series expansion
in terms of two-dimensional Dijkgraaf-Witten theory,
and in terms of a sigma model (the `Gross-Taylor sigma model')
mapping a branched cover $\Sigma_W$ to 
$\Sigma_T$.

Interpreted as an expansion in Dijkgraaf-Witten theories, 
the series expansion of
the Nguyen-Tanizaki-\"Unsal universes is extremely natural, as we explain.
However, we encounter two puzzles when we interpret the expension of the
Nguyen-Tanizaki-\"Unsal universes in terms of maps from a branched cover:
\begin{itemize}
\item the expansion of a fixed universe involves maps of a fixed degree,
(so that summing over universes recovers maps of all degrees,)
and
\item there are additional\footnote{
Additional contributions to the series expansions of
individual universes which cancel out
when universes are summed over, as these do,
are common in decomposition, as we
review in detail in section~\ref{sect:interp:stack}.
The puzzle here is the
interpretation of the extra contributions, not their existence per se.
} contributions in the large $N$ limit which
cannot be ascribed to maps from smooth branched covering surfaces
$\Sigma_W$.
\end{itemize}
We will propose resolutions of those puzzles
in section~\ref{sect:interp}.

Also, in order to help clarify the interpretation, we keep track of the
leading area ($A$) dependence in the large $N$ limit.
(The careful reader will recall that in the full Yang-Mills partition 
function, summed over all universes, there is a phase transition
as a function of area \cite{Gross:1980he,Wadia:1980cp,Brezin:1980rk,Gross:1990md,Douglas:1993iia}.)

Similar expansions have also been studied for gauge groups
$SO(N)$ and $Sp(N)$, 
see e.g.~\cite{Naculich:1993ve,Naculich:1993uu,Naculich:1994kd,Ramgoolam:1993hh,Crescimanno:1996hx}.
We have not worked through them carefully, but we expect that
their analyses should be similar to what we present for $SU(N)$ theories.

\subsection{Series expansion of partition functions}

In this section we will review the pertinent large $N$
asymptotic series expansion
of the partition functions of both the full two-dimensional
pure $SU(N)$ Yang-Mills theory, as well as those of the Nguyen-Tanizaki-\"Unsal
universes.  Along the way, we will 
develop some identities
that will be used in both.

First, it will be useful to rescale certain normalizations and
write the partition function~(\ref{eq:pureym:z})
of two-dimensional pure $SU(N)$ Yang-Mills
in the form
\cite{Migdal:1975zg,Rusakov:1990rs,Gross:1992tu},
\cite[equ'n (3.20)]{Cordes:1994fc}
\cite[equ'n (2.51)]{Witten:1991we}
\begin{equation}  \label{eq:2d-ym-sun}
Z \: = \: \sum_R (\dim R)^{2-2p} \exp\left( - g_{YM}^2 \frac{A}{2N} C_2(R)
 \right),
\end{equation}
on a closed genus-$p$ worldsheet $\Sigma_T$ of area $A$.

The references \cite{Gross:1992tu,Gross:1993hu,Gross:1993yt,Cordes:1994sd,Moore:1994dk,Cordes:1994fc,Ramgoolam:1994ba} rewrite the
two-dimensional pure $SU(N)$ Yang-Mills partition function in the large $N$
limit in a form that
looks like a string field theory, namely as a sum over other
worldsheets $\Sigma_W$, suggesting the existence of a
two-dimensional sigma model of maps $\Sigma_W \rightarrow \Sigma_T$,
for which an expression was given in \cite{Cordes:1994sd,Moore:1994dk,Cordes:1994fc}.
In this expansion, $1/N$ plays the role of string dilaton.

As part of that, to get the correct\footnote{
See also \cite{Kimura:2008gs} which discussed a reformulation of the non-chiral
expansion in terms of purely holomorphic maps plus some line defects.
} $1/N$ asymptotics, it was
argued in \cite{Gross:1993hu,Gross:1993yt,Cordes:1994sd,Moore:1994dk,Cordes:1994fc,Ramgoolam:1994ba} 
that one should
replace the sum over irreducible representations by a sum over
`coupled' representations.  For the moment, to try to make the analysis more
clear, we will set aside the use of coupled representations, and formally
derive an expansion using just the naive sum.  
We will rederive the expansion including coupled
representations shortly.

Setting aside coupled representations for the moment,
the first step is to use Schur-Weyl duality to show
that for $SU(N)$ representations
\cite[equ'n (6.5)]{Cordes:1994fc}
\begin{eqnarray}
\dim R(Y) 
& = & \frac{1}{n!} \sum_{\sigma \in S_n}
\chi_{r(Y)}(\sigma) \, N^{\sum_i k_i(\sigma)},
\\
& = & \frac{N^n}{n!} \chi_{r(Y)}(\Omega_n),
\end{eqnarray}
where
$Y$ is the Young tableau associated with representation $R$ of
$SU(N)$, $n$ is the number of boxes in $Y$,
$r(Y)$ is the representation of the symmetric group $S_n$ defined
by the same Young tableau $Y$, and
\cite[equ'n (6.5)]{Cordes:1994fc}
\begin{equation}  \label{eq:omega-defn}
\Omega_n \: = \: \sum_{\sigma \in S_n} N^{K_{\sigma} - n} \sigma,
\end{equation}
for $K_{\sigma}$ the total number of cycles in the cycle decomposition
of $\sigma$,
\begin{equation}
K_{\sigma} \: = \: \sum_i k_i(\sigma).
\end{equation}
As an aside, later it may be helpful to note that\footnote{
From \cite[section 6.1.2]{Cordes:1994fc}, $\Omega_n$ is invertible
for $N > n$, which we will always assume.
}
\cite[equ'n (5.2)]{Cordes:1994sd}
\begin{equation}
\Omega_n^{-1} \: = \: 1 \: + \:
\sum_{k=1}^{\infty} {\sum_{v_1,\cdots,v_k \in S_n}}^{\!\!\!\prime}
\left( \frac{1}{N} \right)^{\sum_j (n - K_{v_j})} \left( v_1 \cdots v_k
\right) (-)^k, 
\end{equation}
where the primed sum means that the $v_i \neq 1$.

More generally \cite[equ'n (6.6)]{Cordes:1994fc},
\begin{equation}
( \dim R(Y) )^m \: = \:
\left( \frac{ N^n \dim r(Y) }{ |S_n| } \right)^m
\frac{ \chi_{r(Y)}( \Omega_n^m ) }{ \dim r(Y) }.
\label{eq:power-of-R}
\end{equation}
In a moment, we will also need the identity
\begin{eqnarray}
\sum_{s,t \in G} \chi_r(s t s^{-1} t^{-1}) 
& = &
\frac{|G|}{\dim r} \sum_{s \in G} \chi_r(s) \chi_r(s^{-1}),
\\
& = &
\left( \frac{ |G| }{\dim r} \right)^2 \chi_r(1)
\: = \:
\left( \frac{ |G| }{\dim r} \right)^2  \dim r,
\label{eq:orthog-imp1}
\end{eqnarray}
which follows
from the orthogonality relations~(\ref{eq:orthog1}), (\ref{eq:orthog2}),
(\ref{eq:char-master2}).

With this in mind, we can now expand the $R$ contribution to the
pure $SU(N)$ Yang-Mills partition function~(\ref{eq:2d-ym-sun}),
which we will write for any positive integer exponent $m$:
\begin{eqnarray}
\lefteqn{
(\dim R(Y))^{m}
} \nonumber \\
& = &
\left( \frac{ N^n \dim r(Y) }{ |S_n| } \right)^{m}
\frac{ \chi_{r(Y)}( (\Omega_n)^{m} ) }{ \dim r(Y) }
\mbox{ using~(\ref{eq:power-of-R})},
\\
& = &
N^{n m} \left( \frac{ \dim r(Y) }{n!} \right)^{m + 2p}
\left[ \prod_{i=1}^p \sum_{s_i,t_i \in S_n} \left(
\frac{ \chi_{r(Y)}( s_i t_i s_i^{-1} t_i^{-1} ) }{ \dim r(Y) } \right)
\right]
\frac{ \chi_{r(Y)}( (\Omega_n)^{m} ) }{ \dim r(Y) },
\nonumber
\\ & & \mbox{   using (\ref{eq:orthog-imp1})}, \nonumber
\\
& = &
N^{n m} \left( \frac{ \dim r(Y) }{n!} \right)^{m+2p}
\left[
\sum_{s_1, t_1  \cdots \in S_n}
\prod_{i=1}^p \frac{ \chi_{r(Y)}( s_i t_i s_i^{-1} t_i^{-1} ) }{ \dim r(Y) } 
\right]
\frac{ \chi_{r(Y)}( (\Omega_n)^{m} ) }{ \dim r(Y) },
\nonumber
\\
& = &
N^{n m} \left( \frac{ \dim r(Y) }{n!} \right)^{m+2p}
\sum_{s_1, t_1  \cdots \in S_n}
\frac{ \chi_r\left( (\Omega_n)^{m} \prod_{i=1}^p s_i t_i s_i^{-1}
t_i^{-1} \right) }{\dim r(Y) }
\end{eqnarray}
using~(\ref{eq:char-central-product}) and
the fact that $\Omega_n$
is central.

Next, we include the finite area corrections.

For $SU(N)$ representations \cite[equ'n (6.11)]{Cordes:1994fc},
\begin{equation}
C_2(R(Y)) \: = \: n N \: + \: 2 \frac{ \chi_{r(Y)}(T_2) }{\dim r(Y) }
\: - \: \frac{n^2}{N},
\end{equation}
so we can write
\begin{eqnarray}
\lefteqn{
(\dim R(Y))^{2-2p} \exp\left( - g_{YM}^2 \frac{A}{2N} C_2(R(Y)) \right)
} \nonumber \\
& = & 
N^{n(2-2p)} \left( \frac{ \dim r(Y) }{n!} \right)^{2}
\sum_{s_1, t_1  \cdots \in S_n}
\frac{ \chi_r\left( (\Omega_n)^{2-2p} \prod_{i=1}^p s_i t_i s_i^{-1}
t_i^{-1} \right) }{\dim r(Y) }
\nonumber \\
& & \hspace*{1in} \cdot
\exp\left( - g_{YM}^2 \frac{A}{2} n \: + \: g_{YM}^2 A \frac{1}{N} \frac{  \chi_{r(Y)}(T_2) }{\dim r(Y) }
\: - \: g_{YM}^2 \frac{A}{2} \frac{ n^2 }{N^2} \right).
\end{eqnarray}

Let us take a moment to interpret the expression above.
The reader will note that,
to leading order in $1/N$, the second two terms in
the exponential can\footnote{
In other treatments, they are retained in a power series expansion.
For our purposes, it will suffice to only keep the leading order
area dependence.
} be dropped.  The area dependence is precisely what
one would expect from sigma model contributions if one replaces
$g_{YM}^2/2$ with $1/\alpha'_{GT}$, which we will do henceforward,
following e.g.~\cite[section 2]{Matsuo:2004nn}.
(As a consistency test, note that $\alpha'_{GT}$ has units of area,
so $A/\alpha'_{GT}$ is unitless, as needed for these expressions to be
consistent.) 

So, with that interpretation, we can write the partition function
of the universe associated to $R$ as
\begin{eqnarray}
\lefteqn{
(\dim R(Y))^{2-2p} \exp\left( - g_{YM}^2 \frac{A}{2N} C_2(R(Y)) \right)
} \nonumber \\
& = & 
N^{n(2-2p)} \left( \frac{ \dim r(Y) }{n!} \right)^{2}
\sum_{s_i, t_i  \cdots \in S_n}
\frac{ \chi_r\left( (\Omega_n)^{2-2p} \prod_{i=1}^p s_i t_i s_i^{-1}
t_i^{-1} \right) }{\dim r(Y) }
\nonumber \\
& & \hspace*{1in} \cdot
\exp\left( -  \frac{A}{\alpha'_{GT}} n \right) \: + \: {\rm subleading}.
\end{eqnarray}
Using equation~(\ref{eq:delta-vs-chi}), we can write this as
\begin{eqnarray}
\lefteqn{
(\dim R(Y))^{2-2p} \exp\left( - g_{YM}^2 \frac{A}{2N} C_2(R(Y)) \right)
} \nonumber \\
& = & 
N^{n(2-2p)} \left( \frac{ \dim r(Y) }{n!} \right)
\sum_{s_i, t_i  \cdots \in S_n}
\frac{ \delta\left( (\Omega_n)^{2-2p} \prod_{i=1}^p s_i t_i s_i^{-1}
t_i^{-1} P_r \right) }{\dim r(Y) }
\nonumber \\
& & \hspace*{1in} \cdot
\exp\left( -  \frac{A}{\alpha'_{GT}} n \right) \: + \: {\rm subleading},
\end{eqnarray} 
where $P_r$ is the projector defined in appendix~\ref{app:projs}.

In this language, the zero-area limit is an $\alpha'_{GT} \rightarrow \infty$ limit.
The result is reminiscent of results on high energy symmetries
\cite{Gross:1987kza,Gross:1987ar,Gross:1988ue} which argue that
in that limit,
sums over Riemann surfaces are dominated by particular surfaces (points on the
moduli space of curves).

More to the point, we see that the contributions to a single universe
differ from contributions to the Gross-Taylor expansion of the full
two-dimensional Yang-Mills theory in two ways:
\begin{enumerate}
\item First, $n$ is fixed, meaning contributions
are from maps of a fixed degree (equal to the number of boxes
in the Young tableau).  Restricting a sigma model to maps of
a single degree is equivalent to restricting a gauge theory to
instantons of a single degree, which ordinarily would not be consistent.
We will return to this in section~\ref{sect:interp}, where we will argue
that this implies a new constraint on the Gross-Taylor sigma model.
\item Second, the $\delta$ function constraint on the group combinatorics
now has an insertion of $P_r$.  We shall see later that this implies
that there exist additional contributions, not interpretable in terms
of smooth branched covers.  We will return to this also in
section~\ref{sect:interp}.
\end{enumerate}

The next key observation, detailed in \cite[section 2]{Gross:1993hu},
\cite[section 6.3]{Cordes:1994fc}, is that to get the correct
$1/N$ asymptotics, one should replace the sum over irreducible
representations $R$ by a sum over `coupled representations' $S \overline{T}$.
As explained in the references,
this resolves technical difficulties in the summation involving
contributions from representations $R$ such that
$C_2(R)/N$ is ${\cal O}(1)$ in the $1/N$ expansion, but whose contribution
is expanded in the $1/N$ expansion across infinitely many terms.
To that end, we will next repeat the analysis above using coupled
representations.  The results will have the same form as obtained above.

Such coupled representations are described in detail in
\cite[section 2]{Gross:1993hu},
\cite[section 6.3]{Cordes:1994fc}.
Briefly, following \cite[section 2]{Gross:1993hu},
if we let $L$ denote the length of the first row of the Young diagram
for $T$, then the Young diagram for $S \overline{T}$ is defined as follows:
\begin{enumerate}
\item start with an $N \times L$ rectangle,
\item subtract the Young diagram from $T$ from the bottom of the rectangle,
\item add the Young diagram for $S$ to the top right of the rectangle.
\end{enumerate}
More formally, if $c_i$, $\tilde{c}_i$
denote the height of the $i$th column of $S$, $T$, respectively, then
the height of the $i$th column of the coupled representation is
\cite[equ'n (2.6)]{Gross:1993hu}
\begin{equation}
\left\{ \begin{array}{cl}
N - \tilde{c}_{L+1-i} & i \leq L,
\\
c_{i-L} & i > L.
\end{array}
\right.
\end{equation}
Some pertinent properties of coupled representations include
\cite[equ'n (2.7)]{Gross:1993hu}
\begin{equation}
C_2(R \overline{S}) \: = \: C_2(R) + C_2(S) \: + \: 2 \frac{n_R n_S}{N},
\end{equation}
where $n_R$, $n_S$ are the numbers of boxes in the Young diagrams for
$R$, $S$, respectively, and
\cite[equ'n (2.8)]{Gross:1993hu}
\begin{equation}
\dim R \overline{S} \: = \: (\dim R) (\dim S) \left( 1 \: + \: 
{\cal O}(1/N^2) \right).
\end{equation}
Furthermore, as should be clear from the construction above,
in the large $N$ limit
a single coupled representation $R \overline{S}$ uniquely determines
both $R$ and $S$ separately.  As a result, so long as we are working at large
$N$, we can replace the original sum over irreducible representations
with a sum over $R$ and $S$ separately.

One can then extract an expression for $\dim R \overline{S}$ in the
same fashion as we did previously for $\dim R$.  As details of
e.g.~projectors will be important for our later analysis, we repeat
the key steps here.

From \cite[equ'n (6.17)]{Cordes:1994fc}, we can write
\begin{equation}
\dim R \overline{S} \: = \: \frac{N^{n_+ + n_-}}{n_+! \, n_-!}
\chi_{r \otimes s^*}\left( \Omega_{n_+ n_-} \right).
\end{equation}
where $n_+$ is the number of boxes in the Young diagram for $R$,
and $n_-$ is the number of boxes in the Young diagram for $S$,
and $r$, $s$ are representations of $S_{n_+}$, $S_{n_-}$ determined by
$R$, $S$.  More generally \cite{sanpriv},
\begin{eqnarray}
( \dim R \overline{S} )^m 
& = &
\left( \frac{ N^{n_+ + n_-} (\dim r) (\dim s) }{ n_+! \, n_-!} \right)^m
\frac{ \chi_{r \otimes s^*}\left( \Omega_{n_+,n_-}^m \right) }{
(\dim r)(\dim s) },
\end{eqnarray}
and then, using~(\ref{eq:orthog-imp1}),
\begin{eqnarray}
( \dim R \overline{S} )^m 
& = &
N^{(n_+ + n_-)m}
\left( \frac{\dim r}{n_+!} \right)^{m + 2 p}
\left( \frac{\dim s}{n_-!} \right)^{m + 2 p}
\nonumber \\
& & \cdot
\left[ \prod_{i=1}^p \sum_{s^+_i, t^+_i \in S_{n_+}}
\left( \frac{ \chi_r([s^+_i,t^+_i]) }{ \dim r } \right) \right]
\left[ 
 \prod_{i=1}^p \sum_{s^-_i, t^-_i \in S_{n_-}}
\left( \frac{ \chi_s([s^-_i,t^-_i]) }{ \dim s } \right) \right]
\nonumber \\
& & \cdot
\frac{ \chi_{r \otimes s^*}\left( \Omega_{n_+,n_-}^m \right)  }{
(\dim r)(\dim s) },
\end{eqnarray}
Then, since $\Omega_{n_+,n_-}$ is central in $S_{n_+} \otimes S_{n_-}$
\cite{sanpriv},
using~(\ref{eq:char-central-product}),
\begin{eqnarray}
( \dim R \overline{S} )^m 
& = &
N^{(n_+ + n_-)m}
\left( \frac{\dim r}{n_+!} \right)^{m + 2 p}
\left( \frac{\dim s}{n_-!} \right)^{m + 2 p}
\frac{1}{(\dim r)(\dim s)}
\nonumber \\
& & \cdot
\sum_{s^{\pm}_i, t^{\pm}_i \in S_{n_{\pm}}}
\chi_{r\otimes s^*}\left( (\Omega_{n_+,n_-})^m
\prod_{i=1}^p [s^+_i,t^+_i] \otimes [s^-_i,t^-_i] \right).
\end{eqnarray}
From its definition in \cite[equ'n (10.20)]{Cordes:1994sd},
\begin{equation}
\Omega_{n_+,n_-} \: = \: \Omega_{n_+} \otimes \Omega_{n_-} \: + \:
\mbox{subleading in }1/N,
\end{equation}
so, to leading order in $1/N$, we have
\begin{eqnarray}
( \dim R \overline{S} )^m 
& = &
N^{(n_+ + n_-)m}
\left( \frac{\dim r}{n_+!} \right)^{m + 2 p}
\left( \frac{\dim s}{n_-!} \right)^{m + 2 p}
\frac{1}{(\dim r)(\dim s)}
\nonumber \\
& & \cdot
\sum_{s^{\pm}_i, t^{\pm}_i \in S_{n_{\pm}}}
\chi_r\left( (\Omega_{n_+})^m \prod_{i=1}^p [s^+_i,t^+_i] \right)
\chi_{s^*}\left( (\Omega_{n_-})^m \prod_{i=1}^p [s^-_i,t^-_i] \right)
\nonumber \\
& & \: + \: \mbox{subleading}.
\end{eqnarray}

We are now ready to describe the series expansion of both
the partition function of a 
single Nguyen-Tanizaki-\"Unsal universe
on a Riemann surface
$\Sigma_T$ of genus $p$, as well as 
the pure Yang-Mills partition function.

We begin with the series expansion of a Nguyen-Tanizaki-\"Unsal universe.
We define
\begin{eqnarray}
Z^+_R(A,p,N) & = &
N^{n_+ (2-2p)}\left( \frac{\dim r}{n_+!} \right)^{2}
\sum_{s^+_i, t^+_i \in S_{n_+}}
\frac{
\chi_r\left( (\Omega_{n_+})^{2-2p} \prod_{i=1}^p [s^+_i,t^+_i] \right)
}{
\dim r }
\nonumber \\
& & \hspace*{2in} \cdot
\exp\left( -  \frac{A}{\alpha'_{GT}} n_+ \right),
\label{eq:chiral0}
\\
Z^-_S(A,p,N) & = &
N^{n_- (2-2p)} \left( \frac{\dim s}{n_-!} \right)^{2}
\sum_{s^-_i, t^-_i \in S_{n_-}}
\frac{
\chi_{s^*}\left( (\Omega_{n_-})^{2-2p} \prod_{i=1}^p [s^-_i,t^-_i] \right)
}{
\dim s }
\nonumber \\
& & \hspace*{2in} \cdot
\exp\left( -  \frac{A}{\alpha'_{GT}} n_- \right),
\end{eqnarray}
where $n_+$, $n_-$ are the number of boxes in the Young diagrams for
$R$, $S$, respectively,
then we see that the partition function of a Nguyen-Tanizaki-\"Unsal
universe associated to the coupled representation $R \overline{S}$ is
\begin{equation} \label{eq:ntu-part:rs}
( \dim R \overline{S} )^{2-2p} 
\exp\left( - g_{YM}^2 \frac{A}{2N} C_2(R \overline{S}) \right)
\: = \:
Z^+_R(A,p,N) \, Z^-_S(A,p,N) \: + \: \mbox{subleading}.
\end{equation}
(We are, again, using the fact that in the large $N$ limit,
a coupled representation $R \overline{S}$ uniquely determines both
$R$ and $S$ separately.  We are also, again, only keeping the leading
area dependence in the large $N$ limit, absorbing the rest into
subleading terms.)

The full nonchiral partition function of the zero-area limit of
pure Yang-Mills is then giving by summing over coupled representations:
\begin{eqnarray}
Z_{\rm YM}(\Sigma_T) & = &
\sum_{R\overline{S}} (\dim R\overline{S})^{2-2p}
\exp\left( - g_{YM}^2 \frac{A}{2N} C_2(R \overline{S}) \right),
\\
& = & \left( 
\sum_R Z^+_R(A,p,N) \right)
\left( \sum_S Z^-_S(A,p,N) \right)
\nonumber \\
& & \: + \: \mbox{subleading}.
\label{eq:fullym:exp}
\end{eqnarray}
(The reader may compare e.g.~\cite[equ'n (10.18)]{Cordes:1994sd}).
We have implicitly used the fact that in the large $N$ limit,
a coupled representation $R \overline{S}$ uniquely determines $R$ and
$S$ separately.

In passing, expressions for the subleading finite $N$ contributions are
given in 
\cite[equ'n (2.4)]{Vafa:2004qa}, \cite[section 3]{Dijkgraaf:2005bp},
\cite{Rudd:1994ta,Baez:1994gk,Matsuo:2004nn,Kimura:2008gs}.  
We will not study the subleading
corrections in this paper.

As before, we simplify expression~(\ref{eq:chiral0}) for the
chiral component of the partition function expansion associated to
an irreducible representation $R$.
First, using identity~(\ref{eq:delta-vs-chi}),
we can replace the character with a delta function with a projector:
\begin{eqnarray} \label{eq:dimR-2-2p}
Z^+_R(A,p,N) & = &
N^{n_+ (2-2p)}\left( \frac{\dim r}{n_+!} \right)
\sum_{s_i, t_i  \cdots \in S_{n_+}}
\frac{
\delta\left( (\Omega_{n_+})^{2-2p} \left( \prod_{i=1}^p [s^+_i,t^+_i]
\right) P_r \right)
}{
\dim r } 
\nonumber \\
& & \hspace*{2in} \cdot
\exp\left( -  \frac{A}{\alpha'_{GT}} n_+ \right),
\\ 
\nonumber \\
Z^-_S(A,p,N) & = &
N^{n_- (2-2p)} \left( \frac{\dim s}{n_-!} \right)
\sum_{s^-_i, t^-_i \in S_{n_-}}
\frac{
\delta\left( (\Omega_{n_-})^{2-2p} \prod_{i=1}^p [s^-_i,t^-_i] P_{s^*}
 \right)
}{
\dim s }
\nonumber \\
& & \hspace*{2in} \cdot
\exp\left( -  \frac{A}{\alpha'_{GT}} n_- \right),
\end{eqnarray}
where $n_+$ is the number of boxes in the Young diagram for
$R$, $n_-$ is the number of boxes in the Young diagram for
$S$, and $P_r$, $P_{s^*}$ are the projectors~(\ref{eq:proj:defn}).

It will also be useful later to expand out the powers of
$\Omega_n$.  Following \cite[section 5.1]{Cordes:1994sd}
and using~(\ref{eq:omega-defn}), we find
\begin{eqnarray}
Z^+_R(A,p,N) & = &
N^{n(2-2p)} \sum_{s_i,t_i \in S_n} \sum_{L=0}^{\infty}
{\sum_{v_1, \cdots, v_L \in S_n}}^{\!\!\!\prime}
N^{\sum_j (K_{v_j} - n)} \, \frac{d(2-2p,L)}{n!}
\label{eq:full-chiral}
\\
& & \hspace*{1in} \cdot \,
\delta\left( v_1 \cdots v_L \left( \prod_{i=1}^p [s_i,t_i] \right) P_r \right)
\exp\left( -  \frac{A}{\alpha'_{GT}} n \right),
\nonumber
\end{eqnarray}
where
the primed sum means that the $v_i \neq 1$, and
\begin{equation}
(1+x)^m \: = \: \sum_{L=0}^{\infty} d(m,L) \, x^L.
\end{equation}
(As a consistency check, the reader may compare
\cite[equ'n (5.3)]{Cordes:1994sd}, which also sums over
irreducible representations.)
A similar expression, also involving an explicit projector,
arises in \cite[section 3]{Dijkgraaf:2005bp}.
We omit $Z_S^-(A,p,N)$ for simplicity, as the expression is very similar.

So far, we have given an explicit expression~(\ref{eq:full-chiral}) 
for $Z^+_R(A,p,N)$, the chiral
component of the partition function for a Nguyen-Tanizaki-\"Unsal universe
associated to a coupled representation $R\overline{S}$.
It will also be useful to write down the chiral component of the
full two-dimensional Yang-Mills partition function, which is given
by summing over contributions from the various irreducible
representations:
\begin{eqnarray}
Z^+(A,p,N) & = &
\sum_R Z^+_R(A,p,N), \\
& = &
\sum_{n=0}^{\infty}
N^{n(2-2p)} \sum_{s_i,t_i \in S_n} \sum_{L=0}^{\infty}
{\sum_{v_1, \cdots, v_L \in S_n}}^{\!\!\!\prime}
N^{\sum_j (K_{v_j} - n)} \, \frac{d(2-2p,L)}{n!}
\nonumber \\
& & \hspace*{1in} \cdot \,
\sum_r 
\delta\left( v_1 \cdots v_L \left( \prod_{i=1}^p [s_i,t_i] \right) P_r \right)
\\
& & \hspace*{1in} \cdot \,
\exp\left( -  \frac{A}{\alpha'_{GT}} n \right),
\nonumber \\
& = &
\sum_{n=0}^{\infty}
N^{n(2-2p)} \sum_{s_i,t_i \in S_n} \sum_{L=0}^{\infty}
{\sum_{v_1, \cdots, v_L \in S_n}}^{\!\!\!\prime}
N^{\sum_j (K_{v_j} - n)} \, \frac{d(2-2p,L)}{n!}
\nonumber \\
& & \hspace*{1in} \cdot \,
\delta\left( v_1 \cdots v_L \left( \prod_{i=1}^p [s_i,t_i] \right)  \right)
\label{eq:full:chiral:part}
\\
& & \hspace*{1in} \cdot \,
\exp\left( -  \frac{A}{\alpha'_{GT}} n \right).
\nonumber
\end{eqnarray}

Specifically, summing over the irreducible representations $R$ generates
a sum over degrees $n$, as well as a sum over the $S_n$ representations
$r$, removing the projector $P_r$ from the delta function.
(Both of these differences will be important in the subsequent
Nguyen-Tanizaki-\"Unsal analysis.)  The result is, in effect,
a weighted sum over two-dimensional Dijkgraaf-Witten theories for the
symmetric group $S_n$ for all values of $n$.

\subsection{The Dijkgraaf-Witten interpretation}
\label{sect:series:dw}

One interpretation of the terms above is in terms of the
correlation functions in a series of two-dimensional
Dijkgraaf-Witten theories\footnote{
Two-dimensional Dijkgraaf-Witten theories have been described in
many places, see for a few examples
\cite[appendix C.1]{Komargodski:2020mxz},
\cite[appendix C]{Ramgoolam:2022xfk}, 
\cite{Gardiner:2020vjp,mednyh,Snyder:2007ns,frobenius1,frschur,mulaseyu}.
} \cite{Dijkgraaf:1989pz}.
Briefly, this theory describes an orbifold of a point.
On a genus $p$ Riemann surface, the partition function of
(untwisted\footnote{
The `twist' of Dijkgraaf-Witten theory is a choice of discrete torsion
in the orbifold.  Only untwisted Dijkgraaf-Witten theories (without
discrete torsion) will be relevant for this paper.
}) Dijkgraaf-Witten theory with orbifold group $G$ is
\begin{equation}
Z_{DW, p} \: \propto \: \sum_{s_i, t_i \in G}
\delta\left( \prod_{i=1}^p [ s_i, t_i ] 
 \right),
\end{equation}
which the reader will already recognize is a component of the
expressions for the Gross-Taylor series expansion.
The operators of Dijkgraaf-Witten theory are twist fields,
which can be expressed as commuting linear combinations\footnote{
In the group algebra ${\mathbb C}[G]$.
}
of group elements (technically, elements of the center of the
group algebra).  A correlation function of twist fields can therefore
be expressed formally as a linear combination of a correlation
function of products of group 
elements $v_1, \cdots, v_L \in G$, which on a Riemann surface of genus $p$ 
is given by
\begin{equation}
\langle v_1 \cdots v_L \rangle_{DW,p} \: \propto \:
\sum_{s_i, t_i \in G}
\delta\left( v_1 \cdots v_L \left( \prod_{i=1}^p [ s_i,  t_i ]
 \right) \right).
\end{equation}
The chiral partition function $Z^+(A,p,N)$ given in
equation~(\ref{eq:full:chiral:part}) is clearly a linear combination
of such correlation functions, for $G = S_n$ (the symmetric group),
summed over $n$.

Now, any two-dimensional gauge theory in which a subgroup of the gauge
group acts trivially, has a global 1-form symmetry and hence
decomposes, into universes indexed by the irreducible representations
of the trivially-acting subgroup.
Indeed, this is at the heart of the Nguyen-Tanizaki-\"Unsal
decomposition of two-dimensional Yang-Mills theory.
For two-dimensional orbifolds, decomposition has been discussed in
for example
\cite{Hellerman:2006zs,Robbins:2020msp,Robbins:2021lry,Robbins:2021ibx,Robbins:2021xce,Sharpe:2021srf,Robbins:2022wlr,Sharpe:2022ene}.
Two-dimensional Dijkgraaf-Witten theory, in which the entire orbifold
group acts trivially, is simply a special case\footnote{
We can also see this from another perspective.  
Unitary two-dimensional topological field theories
also decompose, into a collection of invertible field theories, 
see e.g.~\cite{Durhuus:1993cq,Moore:2006dw,Huang:2021zvu,Komargodski:2020mxz}, 
and Dijkgraaf-Witten theory is also
a special case in that sense.  For our purposes, the orbifold perspective
is more relevant.
} of the orbifolds
considered in the references just cited, and the picture is
particularly simple.
In terms of the state space, for example, the projectors onto 
states in each universe are the projectors
$P_r$ (associated to irreducible representations $r$)
described in appendix~\ref{app:projs}.  In fact, when the entire group
acts trivially, it is a standard mathematics result that
there is a (noncanonical) one-to-one correspondence
between the possible twist fields and the projectors $P_r$,
as both form bases of the center of the group algebra ${\mathbb C}[G]$
for orbifold group $G$, see for example
\cite[section 6.3]{serrerep},
\cite{conlon,cheng,costache,karpilovsky,cr,collins,schur1,schur2,schur3}.
(In fact, those references describe a more general case, that of
twisted group algebras, which in an orbifold corresponds to adding
discrete torsion.  Here we only consider the untwisted case.)

A correlation function within a particular universe 
(corresponding to irreducible representation $r$) is obtained by
inserting a projection operator $P_r$, the same projector
appearing earlier:
\begin{equation}
\langle v_1 \cdots v_L \rangle_{DW,p,r} \: \propto \:
\sum_{s_i, t_i \in G}
\delta\left( v_1 \cdots v_L \left( \prod_{i=1}^p [ s_i ,  t_i ]
 \right) P_r \right).
\end{equation}
Comparing equation~(\ref{eq:full-chiral}), we see that the
chiral component $Z^+_R(A,p,N)$ is precisely a linear combination
of Dijkgraaf-Witten correlation functions restricted to universe $r$,
where $r$ is a representation of $S_n$ associated to the same
Young diagram as the representation $R$ of $SU(N)$.

In brief, we see that the Gross-Taylor expansion of a
Nguyen-Tanizaki-\"Unsal universe involves first restricting to
$n$ (equal to the number of boxes in the Young diagram for $R$),
then restricting to one universe in the decomposition of
Dijkgraaf-Witten theory for $S_n$.  In other words, the
Gross-Taylor expansion of a single Nguyen-Tanizaki-\"Unsal universe
involves a single universe of
the decomposition of two-dimensional untwisted Dijkgraaf-Witten theory.
The two decompositions are therefore closely and naturally linked.

In the next section, we shall see that the interpretation of the
components in terms of branched covers is more subtle.

\subsection{The branched cover interpretation}
\label{sect:series:cover}

We computed the series expansions of
a Nguyen-Tanizaki-\"Unsal universe associated to coupled
representation $R \overline{S}$ in equation~(\ref{eq:ntu-part:rs}),
and reviewed the series expansion of the full two-dimensional
Yang-Mills partition function~(\ref{eq:fullym:exp}), obtained by summing over
Nguyen-Tanizaki-\"Unsal universe partition functions.
Results for a Nguyen-Tanizaki-\"Unsal universe were
written in terms of the chiral partition function $Z^+_R(A,p,N)$,
and results for the full two-dimensional Yang-Mills partition function
were written in terms of the full chiral partition function
\begin{equation}
Z^+(A,p,N) \: = \: \sum_R Z^+_R(A,p,N).
\end{equation}

In this subsection we will review how to interpreted those chiral components
of partition functions
as a sum
over branched covers $\Sigma_W \rightarrow \Sigma_T$, following
\cite{Gross:1992tu,Gross:1993hu,Gross:1993yt,Cordes:1994sd,Moore:1994dk,Cordes:1994fc}.  
(This description is standard, but has not appeared recently in the literature,
and as we will be manipulating it extensively, we think it useful to review
in detail.)
Later, in
section~\ref{sect:disjoint}, we will reinterpret the terms above as a sum over
disjoint unions
$\hat{\Sigma}_W$ of stacky
copies of $\Sigma_T$.

The construction of the branched $n$-cover $\Sigma_W$
is described systematically
in
e.g.~\cite{fulton1,ezell}, \cite[section 5]{Cordes:1994fc},
which we briefly review here.  
The idea is that the elements $v_1, \cdots, v_L \in S_n$ define monodromies
about $L$ branch points in the base curve $\Sigma_T$.
The delta function is nonzero only when those monodromies are consistent
with the existence of
a smooth branched $n$-cover $\Sigma_W$
of a genus $p$ Riemann surface (specifically, 
$\Sigma_T$).

More systematically,
let $B$ be the branch locus on $\Sigma_T$ (the locations of
insertions $v_1, \cdots, v_L \in S_n$), and
define $X = \Sigma_T - B$,
then use the fact that there is a one-to-one correspondence between
conjugacy classes of subgroups of $\pi_1(X)$ and equivalence classes of
topological coverings of $X$ (see for example
\cite[theorem V.6.6, theorem V.10.2]{massey}),
which are glued according to the data of the homomorphism to
build the branched $n$-cover $\Sigma_W$.

To make this more clear, let us consider an example.
Specifically, consider $\Sigma_T = {\mathbb P}^1$ with two insertions
at positions denoted
$A$, $B$, and let $p \in {\mathbb P}^1$,
as illustrated in the figure below.
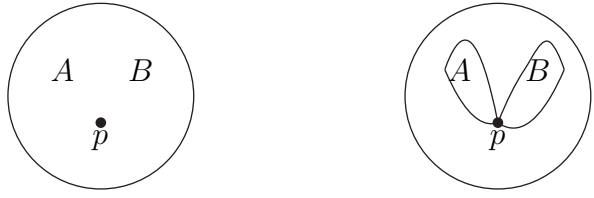
\begin{center}
\begin{picture}(70,70)(0,0)
\CArc(35,35)(35,0,360)
\Text(25,45)[r]{$A$}
\Text(55,45)[r]{$B$}
\Vertex(35,25){2}
\Text(35,22)[t]{$p$}
\end{picture}
\hspace*{1in}
\begin{picture}(70,70)(0,0)
\CArc(35,35)(35,0,360)
\Text(25,45)[r]{$A$}
\Text(55,45)[r]{$B$}
\Vertex(35,25){2}
\Text(35,22)[t]{$p$}
\Curve{(15,45)(17,50)(25,55)(35,25)}
\Curve{(15,45)(20,35)(35,25)}
\Curve{(35,25)(45,45)(55,55)(60,45)}
\Curve{(35,25)(55,35)(60,45)}
\end{picture}
\end{center}
Shown on the left is a schematic illustration of ${\mathbb P}^1$
with the two points $A$, $B$, and the basepoint $p$ for paths.
On the right is the same illustration with two nonintersecting
paths from $p$ marked.

Let the monodromies about the two points be
denoted $v_A$, $v_B$.
Suppose that $n=3$, so that $v_A, v_B, v_C \in S_3$, and take
\begin{equation}
v_A \: = \: (12)(3) \: = \: v_B.
\end{equation}
It is straightforward to check that the product
\begin{equation}
v_A v_B  \: = \: 1,
\end{equation}
so the delta function is nonzero.

In terms of branched 3-covers, the sheets of the cover in neighborhoods
of the two points take the form
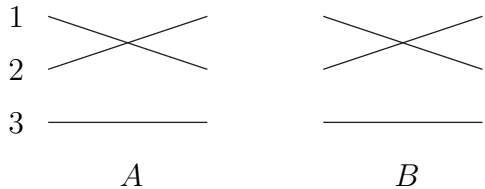
\begin{center}
\begin{picture}(60,60)(0,0)
\Line(0,60)(60,40)
\Line(0,40)(60,60)
\Line(0,20)(60,20)
\Text(27,0)[l]{$A$}
\Text(-15,60)[l]{$1$}
\Text(-15,40)[l]{$2$}
\Text(-15,20)[l]{$3$}
\end{picture}
\hspace*{0.5in}
\begin{picture}(60,60)(0,0)
\Line(0,60)(60,40)
\Line(0,40)(60,60)
\Line(0,20)(60,20)
\Text(27,0)[l]{$B$}
\end{picture}
\end{center}
Locally near $A$ and $B$, two of the sheets collide, but the third remains
disjoint.  The resulting branched 3-cover $\Sigma_W$ is a disjoint union
of one branched double cover of ${\mathbb P}^1$ (itself ${\mathbb P}^1$),
formed from the first two sheets,
and another ${\mathbb P}^1$ (the third sheet), so
\begin{equation}
\Sigma_W \: = \: {\mathbb P}^1 \coprod {\mathbb P}^1,
\end{equation}
and $\chi(\Sigma_W) = 4$.

Next, we will consider Euler characteristics, and review how in general,
the Euler characteristic of the branched cover $\Sigma_W$ always
matches that of the disjoint union $\hat{\Sigma}_W$, and both also match
the power of $N$ in the corresponding partition function term.

Consider the smooth branched cover $\Sigma_W$.
The powers of $N$ in the expansion above are
\begin{equation}
n (2-2p) \: + \: \sum_j (K_{v_j} - n),
\end{equation}
and using the Riemann-Hurwitz formula:
\begin{equation} \label{eq:riemann-hurwitz}
\chi(\Sigma_W) \: = \: n \chi(\Sigma_T) \: - \: \sum_{i=1}^L \left( n - K_{v_i}
\right),
\end{equation}
where $L$ is the number of branch points and $K_{v_i}$ is the number of
cycles in $v_i \in S_n$ corresponding to a branch point,
we see that
\begin{equation}
n (2-2p) \: + \: \sum_j (K_{v_j} - n) \: = \: \chi(\Sigma_W),
\end{equation}
and so the terms in $Z^+(0,p,N)$ are weighted by
\begin{equation}
N^{\chi(\Sigma_W)},
\end{equation}
as expected.

Rewriting in this language, we have that
\begin{eqnarray}
Z^+(A,p,N) & = &
\sum_{n=0}^{\infty} 
\sum_{s_i,t_i \in S_n} \sum_{L=0}^{\infty}
{\sum_{v_1, \cdots, v_L \in S_n}}^{\!\!\!\prime}
N^{\chi(\Sigma_W)} \, \frac{d(2-2p,L)}{n!}
\nonumber \\
& & \hspace*{0.75in} \cdot \,
\delta\left( v_1 \cdots v_L \left( \prod_{i=1}^p [s_i,t_i] \right)  \right)
\exp\left( -  \frac{A}{\alpha'_{GT}} n \right),
\label{eq:chiral-gt-sum}
\end{eqnarray}
where $\Sigma_W$ is a smooth $n$-fold cover of the genus $p$ base curve
$\Sigma_T$, branched over $L$ points.

In that language, the factor
\begin{equation}
 \frac{d(2-2p,L)}{n!}
\end{equation}
is interpreted in \cite{Cordes:1994sd,Cordes:1994fc} in terms of
the orbifold Euler characteristic of the (Hurwitz) moduli space of
maps $\Sigma_W \rightarrow \Sigma_T$.
We refer the reader to thsoe references for further details, which are beyond
the scope of this short overview.

Finally, the factor
\begin{equation}
\exp\left( -  \frac{A}{\alpha'_{GT}} n \right)
\end{equation}
is the weighting one expects in a sigma model describing maps of degree $n$.
Its presence merely serves to confirm the interpretation.

To summarize, we have reviewed how the partition function of two-dimensional
pure $SU(N)$ Yang-Mills on a Riemann surface $\Sigma_T$
can be rewritten in the form of a sum~(\ref{eq:chiral-gt-sum}) over
smooth branched covers $\Sigma_W \rightarrow \Sigma_T$,
which as noted in \cite{Gross:1992tu,Gross:1993hu,Gross:1993yt} is very
suggestive of an interpretation as the string field theory of a
sigma model with two-dimensional target space $\Sigma_T$.

Now, let us briefly compare to the corresponding chiral partition function of
an Nguyen-Tanizaki-\"Unsal universe~(\ref{eq:full-chiral}), explicitly
\begin{eqnarray}
Z^+_R(A,p,N) & = &
N^{n(2-2p)} \sum_{s_i,t_i \in S_n} \sum_{L=0}^{\infty}
{\sum_{v_1, \cdots, v_L \in S_n}}^{\!\!\!\prime}
N^{\sum_j (K_{v_j} - n)} \, \frac{d(2-2p,L)}{n!}
\\
& & \hspace*{1in} \cdot \,
\delta\left( v_1 \cdots v_L \left( \prod_{i=1}^p [s_i,t_i] \right) P_r \right)
\exp\left( -  \frac{A}{\alpha'_{GT}} n \right).
\nonumber
\end{eqnarray}

This is very similar to the expression above for the chiral contribution
to the full Yang-Mills partition function, with two differences:
\begin{itemize}
\item The full chiral partition function $Z^+(A,p,N)$ contains a sum over
values of $n$, whereas $Z^+_R(A,p,N)$ restricts to a single value
of $n$ (equal to the number of boxes in a Young diagram for the
representation $R$).
\item In $Z^+_R(A,p,N)$, the delta function contains a factor of a
projector $P_r$ (associated to a representation $r$ of $S_n$, associated
to the same Young diagram as $R$), which was not present in the full
chiral partition function $Z^+(A,p,N)$.
\end{itemize}

These two differences have the following effects.
\begin{itemize}
\item The restriction to a single $n$ means that $Z^+_R(A,p,N)$ only
receives contributions from maps of a single degree.  This is closely
analogous in a gauge theory to restricting to instantons of a single
degree.
\item Because of the projector $P_r$, there can be nonzero contributions from
monodromy insertions $v_i$ that are not allowed for a smooth branched
cover $\Sigma_W$.  For example, the exponent of $N$ can be odd,
which is not possible for the Euler characteristic of a smooth
oriented Riemann surface.
\end{itemize}

We will see these effects in concrete examples
in subsection~\ref{sect:exs}, and propose resolutions for the corresponding
physics puzzles in section~\ref{sect:interp}.

Before moving on, we note a few consequences of the
Riemann-Hurwitz theorem that will be relevant later:
\begin{itemize}
\item If $\chi(\Sigma_W) > \chi(\Sigma_T)$, then no holomorphic maps exist.
For example, there are no
holomorphic maps ${\mathbb P}^1 \rightarrow
T^2$, or more generally, from a lower-genus curve to a higher-genus curve.
To have maps, $\Sigma_W$ must have at least the same genus as $\Sigma_T$.
\item For fixed $\Sigma_W, \Sigma_T$ obeying the constraint above, maps
can exist of (nearly) any degree.  For example, in degree $n$,
for branchings in which only two sheets collide,
\begin{itemize}
\item maps ${\mathbb P}^1 \rightarrow {\mathbb P}^1$ exist and are branched
over $2(n-1)$ branch points,
\item maps $T^2 \rightarrow {\mathbb P}^1$ exist and are branched
over $2n$ branch points,
\item maps $\Sigma_W \rightarrow {\mathbb P}^1$ exist and are branched
over $2n - \chi(\Sigma_W)$ branch points,
\item unbranched maps $T^2 \rightarrow T^2$ exist for any $n$,
\item maps $\Sigma_W \rightarrow T^2$ exist and are branched over
$- \chi(\Sigma_W)$ points, for $\chi(\Sigma_W) < 0$.
\end{itemize}
That said, the orbifold Euler characteristic of the Hurwitz moduli space
will vanish in some cases\footnote{
For example, for $p=0$, $d(2,L) = 0$ for $L > 2$, hence the only branched covers
of ${\mathbb P}^1$ that contribute have no more than two branch points.
},
so not all such maps will be represented
explicitly in the Gross-Taylor expansion.
\end{itemize}

\subsection{Examples}
\label{sect:exs}

In this section we will walk through a number of examples, to make our
proposal more clear.

In each case, we will begin by reviewing the ordinary Gross-Taylor expansion,
for terms of fixed degree $n$ in the chiral partition functions
$Z^+(A,p,N)$ of theories in
the large $N$ limit, of the form
\begin{equation} \label{eq:source0}
N^{n(2-2p)} \sum_{s_i, t_i \in S_n}
\frac{1}{n!} \delta\left( (\Omega_n)^{2-2p} 
\prod_{i=1}^p s_i t_i s_i^{-1}
t_i^{-1}\right)
\exp\left( -  \frac{A}{\alpha'_{GT}} n \right),
\end{equation}
where
\begin{equation}
\Omega_n \: = \: \sum_{\sigma \in S_n} N^{K_{\sigma}-n} \sigma
\: = \: 1 \: + \: \sum_{\sigma \neq 1} \left( \frac{1}{N} \right)^{n-K_{\sigma}}
\sigma,
\end{equation}
(due to the fact that the identity in $S_n$ decomposes as $K_1 = n$ cycles).

Then, after discussing the interpretation of those terms as
branched covers,
we will describe the terms in the expansion of the
partition function of the corresponding
Nguyen-Tanizaki-\"Unsal universe.
The terms have
a similar form (at fixed degree) but also with
projectors $P_r$ inserted,
namely
\begin{equation}  \label{eq:source}
Z^+_R(A,p,N) \: = \: N^{n(2-2p)} \sum_{s_i, t_i \in S_n}
\frac{1}{n!} \delta\left( (\Omega_n)^{2-2p} 
\left( \prod_{i=1}^p s_i t_i s_i^{-1}
t_i^{-1} \right) P_r \right)
\exp\left( -  \frac{A}{\alpha'_{GT}} n \right).
\end{equation}

For simplicity, and to clean up notation, in the rest of the analysis
we will usually restrict to the zero-area limit.
(One should be slightly careful, as the exact expression for the
Yang-Mills partition function on surfaces of low genus
diverges in the zero-area limit and
must be regularized.  We will simply examine the interpretation of
individual terms, omitting areas merely for convenience, so the
convergence of the sum in that limit is not a concern.)

\subsubsection{$p=0, n=2$}
\label{sect:exs:peq0neq2}

First, consider the case that the Yang-Mills theory lives on
$\Sigma_T = {\mathbb P}^1$ (so that $p=0$), and restrict to maps of
degree $n=2$.  Since $S_2 = {\mathbb Z}_2$, we can write
\begin{eqnarray}
\Omega_{n=2} \: = \:
1 \: + \: \left( \frac{1}{N} \right) v,
\end{eqnarray}
for $v \in S_2$ the nontrivial element.

We begin by considering the pertinent part of the
ordinary Gross-Taylor expansion in this case.
Expanding~(\ref{eq:source0}) in the zero-area limit,
we find
\begin{eqnarray}
\frac{N^{2n}}{n!} \delta \left(  (\Omega_n)^{2} \right)
& = &
\frac{N^{2n}}{n!} \left( 1 \: + \: \left( \frac{1}{N} \right)^2 v^2 \right),
\\
& = & 
\frac{N^4}{2!} \: + \: \frac{ N^2 }{2!} \delta( v^2 ).
\label{eq:gt:peq0:neq2}
\end{eqnarray}

Following
\cite{Gross:1992tu,Gross:1993hu,Gross:1993yt,Cordes:1994sd,Moore:1994dk,Cordes:1994fc},
we interpret the first term as describing maps
${\mathbb P}^1 \coprod {\mathbb P}^1 \rightarrow {\mathbb P}^1$.
As a check,
note that $\chi( {\mathbb P}^1 \coprod {\mathbb P}^1) = 4$.
The first term is essentially describing a free $S_2 = {\mathbb Z}_2$
Dijkgraaf-Witten theory on $\Sigma_T = {\mathbb P}^1$, and as such,
from decomposition, it descomposes into a disjoint union of two
pieces, hence $\Sigma_W$ is a disjoint union.

We interpret
the second term as describing maps from a branched double cover
of ${\mathbb P}^1$, branched over two points (the locations of each $v$).
Such a double cover is precisely ${\mathbb P}^1$, and note
that the exponent of $N$ is correct for this case,
as $\chi({\mathbb P}^1) = 2$.
The factor of $1/2$ represents the orbifold Euler characteristic of the
Hurwitz moduli space of maps $\Sigma_W \rightarrow \Sigma_T$
\cite{Cordes:1994sd}.

In passing, note that the interpretation we have assigned to the two
Dijkgraaf-Witten partition functions above is ambiguous, simply because
$v^2 = 1$.  We could have equivalently written
\begin{equation}
\frac{N^{2n}}{n!} \delta \left(  (\Omega_n)^{2} \right)
\: = \:
\frac{N^4}{2!} \delta(v^2)  \: + \: \frac{ N^2 }{2!} \delta(1),
\end{equation}
and then tried to interpret the first term in terms of a branched
double cover of ${\mathbb P}^1$, and the second term in terms of a disjoint
union of two copies of ${\mathbb P}^1$, but this would not be consistent
with the exponentials of the $N$'s.
Our point, however, is that merely giving the Dijkgraaf-Witten partition
functions by themselves is ambiguous in this context.

Next, consider a Nguyen-Tanizaki-\"Unsal universe.
Expanding~(\ref{eq:source}) in the zero-area limit,
we find
\begin{eqnarray}
Z^+_R(0,p,N) & = &
\frac{N^{2n}}{n!} \delta \left(  (\Omega_n)^{2} P_r \right),
\\
& = &
\frac{N^{2n}}{n!} \delta \left(
(1) P_r \: + \: 2 \left( \frac{1}{N} \right) v P_r \: + \:
\left( \frac{1}{N} \right)^2 v^2 P_r \right),
\\
& = &
\frac{N^4}{2!} \delta( P_r) \: + \:
2 \frac{N^3}{2!} \delta(v P_r) \: + \:
\frac{N^2}{2!} \delta(v^2 P_r),
\end{eqnarray}
where in this case, $P_r = (1/2) (1 \pm v)$.
The first and last terms look essentially the same as in the ordinary
Gross-Taylor case, but we also have a new term, not present previously,
namely the term linear in $v$.  This term can survive here, whereas
previously it did not, because of the $P_r$.

Expanding out $P_r$, we have
\begin{eqnarray}
Z^+_R(0,p,N) & = &
\frac{N^4}{4} \: \pm \: \frac{N^3}{2} \: + \: \frac{N^2}{4}.
\end{eqnarray}
Adding two such contributions together, for each choice of projector
$P_r$, recovers the original Gross-Taylor result~(\ref{eq:gt:peq0:neq2}).

The first term can be interpreted as before, in terms of maps
${\mathbb P}^1 \coprod {\mathbb P}^1 \rightarrow {\mathbb P}^1$,
which is consistent with the fact that $\chi( {\mathbb P}^1 \coprod 
{\mathbb P}^2) = 2 \chi( {\mathbb P}^1) = 4$.

The second term, the new term, is more interesting.
The Euler characteristic
of any smooth closed Riemann surface is even, and here we need something
of Euler characteristic $3$.  It is constructed from a
Dijkgraaf-Witten correlation function with a single $v$, plus a projector.

Later in section~\ref{sect:interp} we will discuss these two issues -- the
construction of a theory that restricts to single instanton sectors,
and the interpretation of the extra terms.  Briefly, to address
the first issue, we will propose a new constraint on the 
Gross-Taylor sigma model, that it possesses a symmetry making such
a localization meaningful,
and for the second issue,
we will propose that the extra terms arise from stacky worldsheets
$\Sigma_W$.

\subsubsection{$p=0, n=3$}
\label{sect:exs:peq0neq3}

Next, we again consider the case of Yang-Mills theory on
$\Sigma_T = {\mathbb P}^1$, and restrict to maps of degree $n=3$.
Here, $|S_3| = 3! = 6$, so we write
\begin{eqnarray}
\Omega_{n=3} \: = \:
1 \: + \:  \sum_{v \neq 1}  \left( \frac{1}{N} \right)^{n-K_{v}} v.
\end{eqnarray}
For reference, the six elements of $S_3$ can be characterized as
\begin{equation}
(1)(2)(3), \: \: \:
(12)(3), (13)(2), (23)(1), \: \: \:
(123), (132).
\end{equation}
These six elements form three conjugacy classes, essentially labelled by
the orders of the cycles.

As before, we begin by considering the ordinary Gross-Taylor expansion
in this case (at fixed degree $n=3$).  Expanding~(\ref{eq:source0})
in the zero-area limit, we find
\begin{eqnarray}
\frac{N^{2n}}{n!} \delta \left(  (\Omega_n)^{2} \right)
& = &
\frac{N^{2n}}{n!} \delta \left( 1 \: + \: 
 \sum_{ij} 
\left( \frac{1}{N} \right)^{2n - K_{v_1} - K_{v_2}}
v_i v_j \right).
\end{eqnarray}
As before, each term can be interpreted as a 3-fold cover of
$\Sigma_T = {\mathbb P}^1$.

The first term can be interpreted as a disjoint union of three
copies
of ${\mathbb P}^1$.
As a consistency check, note that
\begin{equation}
\chi( {\mathbb P}^1 \coprod {\mathbb P}^1 \coprod {\mathbb P}^1 ) \: = \: 6,
\end{equation}
which matches the power of $N$ in that term.

Next, consider the second term.
There are two cases that contribute to the sum:
\begin{itemize}
\item If both $v_1$ and $v_2$ have order 2,
then 
$K_{v_1} = K_{v_2} = 2$,
so the term is of the form
\begin{equation}
\frac{N^6}{3!} \left( \frac{1}{N} \right)^{6-2-2} \delta(v_1 v_2)
\: = \: \frac{N^4}{3!} \delta(v_1 v_2),
\end{equation}
and $\Sigma_W = {\mathbb P}^1 \coprod {\mathbb P}^1$, with one copy of
${\mathbb P}^1$ a double cover of $\Sigma_W$, branched over two points,
and the second ${\mathbb P}^1$ a degree-one cover.  Note that
\begin{equation}
\chi( {\mathbb P}^1 \coprod {\mathbb P}^1 ) \: = \: 4,
\end{equation}
which matches the exponent of $N$.

\item The other case is that $v_1$ and $v_2$ have order 3.

In this case, from the Riemann-Hurwitz formula~(\ref{eq:riemann-hurwitz},
we see that $\Sigma_W$ has genus $0$, consistent with the fact that
this term is proportional to $N^2$.
Hence, this term describes degree-three
maps $\Sigma_W \rightarrow {\mathbb P}^1$
for $\Sigma_W = {\mathbb P}^1$.

\end{itemize}
No other cases can arise in this sum, given the delta function.

Next, we turn to the Nguyen-Tanizaki-\"Unsal decomposition,
meaning we incorporate the projector, and interpret the expression above
as the entire partition function, instead of just one term in a larger
function.
Here, from~(\ref{eq:source}), we have in the zero-area limit that
\begin{eqnarray} 
Z^+_R(0,p,N) & = &
\frac{N^{2n}}{n!} \delta \left(  (\Omega_n)^{2} P_r \right),
\\
& = &
\frac{N^{2n}}{n!} \delta \left( P_r
\: + \: 2 \sum_v \left(\frac{1}{N}\right)^{n-K_{v}} v P_r
 \: + \: 
 \sum_{ij} 
\left( \frac{1}{N} \right)^{2n - K_{v_1} - K_{v_2}}
v_i v_j \right),  \nonumber
\end{eqnarray}
for $n=3$ here.
The first and third terms can be interpreted as before, modulo the
addition of $P_r$.  $P_r$ itself depends upon the representation
$r$ of $S_n$, itself determined by the representation $R$ of $(S)U(N)$,
but our discussion below will apply to all cases.

It remains to discuss the middle term,
which does not appear in the original Gross-Taylor expansion.
Depending upon the order of $v$, there are two cases appearing in the
sum.
\begin{itemize}
\item First, consider the case that $K_v = 2$, for example if $v = (12)(3)$.
Here, the term takes the form
\begin{equation}
\frac{2}{3!} N^5 \delta \left( \sum_v v P_r \right).
\end{equation}

\item Next, consider the case that $K_v = 1$, for example if $v = (123)$.
Here, the term above takes the form
\begin{equation}
\frac{2}{3!} N^4  \delta \left( \sum_v v P_r \right).
\end{equation}

\end{itemize}
Both terms can be nonzero (because of the projector $P_r$).
The first would describe a worldsheet with odd Euler characteristic,
not possible for a smooth oriented closed Riemann surface.

\subsubsection{$p=1$}
\label{sect:exs:peq1}

Next, we turn to the case that $\Sigma_T = T^2$. and consider degree $n$
covering maps.  In~(\ref{eq:source0}), in the zero-area limit,
this corresponds to the terms
\begin{equation}
\frac{N^{0}}{n!} \sum_{s,t \in S_n} \delta\left( (\Omega_n)^0 
s t s^{-1} t^{-1} \right).
\end{equation}
In \cite{Gross:1992tu,Gross:1993hu,Gross:1993yt,Cordes:1994sd,Moore:1994dk,Cordes:1994fc},
these describe $n$-fold covers of $T^2$, with no branch points,
for which, from the Riemann-Hurwitz formula~(\ref{eq:riemann-hurwitz}),
the covering spaces are $\Sigma_W = T^2$, consistent with the power of $N$.

To be clear, these covering spaces $\Sigma_W$ are not necessarily
connected.  Whether the cover is one $T^2$ or several depends upon
$s$, $t$.  For example, if $s$ and $t$ each factorize suitably, the resulting
$n$-fold cover can be, for example, a disjoint union of a $k$-fold cover
and a $(n-k)$-fold cover.
Thus, this description encompasses both connected and disconnected covers.

Next, we turn to the Nguyen-Tanizaki-\"Unsal decomposition, meaning we
incorporate the projector.  Here, from~(\ref{eq:source}), 
in the zero-area limit, we have
\begin{eqnarray} 
Z^+_R(0,p,N) & = & 
1\frac{N^{0}}{n!} \sum_{s,t \in S_n} \delta\left( (\Omega_n)^0 
s t s^{-1} t^{-1} 
P_r \right).
\end{eqnarray}

First, consider the case $n=2$.  Since $S_2$ is abelian, there are no
extra terms arising from $P_r$, as the commutator $[s,t] = 1$ for all
$s, t \in S_2$.  The terms can be interpreted as follows:
\begin{itemize}
\item $s=t=1$:  Here, we get a disjoint union of two copies of $T^2$.
\item all other $s, t$:  Here, we get a single $T^2$, which
is an unbranched double cover of $T^2$.
\end{itemize}
For this case ($n=2$, $p=1$), and only this case, the Nguyen-Tanizaki-\"Unsal
universe appears to be a restriction to maps of a single degree,
without additional orbifold contributions.
For higher $n$, $S_n$ is nonabelian, and so there will be projector
contributions multiplying the commutator.

Now, let us turn to the Nguyen-Tanizaki-\"Unsal decomposition,
and add projectors.
In this case, from~(\ref{eq:source}), the partition function is
\begin{equation}
Z_R^+(0,p,N) \: = \:
\frac{N^{0}}{n!} \sum_{s,t \in S_n} \delta\left( (\Omega_n)^0 
s t s^{-1} t^{-1} P_r \right)
\end{equation}
for $r$ an irreducible representation of $S_n$.
This can admit additional contributions.  For example,
consider the case $n=3$.  Since $S_3$ is nonabelian, there are
additional terms involving the projector $P_r$, corresponding to
cases in which the commutator is different from the identity,
and instead has either order 2 or 3.

In section~\ref{sect:interp:exs:peq1} we will interpret the
terms of the expansion above, as a combination of a 
new symmetry in the Gross-Taylor sigma model
(to realize the restriction to specific degrees)
and by adding stacky worldsheets (to encompass the additional cases
not present previously).

\subsubsection{$p=2, n=2$}
\label{sect:exs:peq2neq2}

Next, we turn to the case of double covers of a genus-two Riemann
surface $\Sigma_T$.  As before, we begin with the
Gross-Taylor expansion at fixed degree $n=2$,
for which the relevant term in the
full partition function in the zero-area limit is
\begin{equation}
\frac{N^{-2n}}{n!} \sum_{s_i, t_i \in S_n} \delta\left( (\Omega_n)^{-2}
\prod_{i=1}^2 s_i t_i s_i^{-1} t_i^{-1}  \right)
\end{equation}

Now,
\begin{equation}
\Omega_2 \: = \: 1 + \left( \frac{1}{N} \right) v
\end{equation}
for $v$ the nontrivial element of $S_2 = {\mathbb Z}_2$,
and $\Omega_2^{-1}$ is the inverse element in the group algebra,
which one can quickly verify is
\begin{equation}
\Omega_2^{-1} \: = \: \frac{1}{1 - (1/N^2)} \left( 1 - \frac{v}{N} \right).
\end{equation}
Thus, expanding, the relevant terms are
\begin{eqnarray}
\lefteqn{
\frac{N^{-4}}{2!}  \sum_{s_i, t_i \in S_n}
 \delta\left( (\Omega_n)^{-2}
\prod_{i=1}^2 s_i t_i s_i^{-1} t_i^{-1}  \right)
} \nonumber \\
& = &
\frac{N^{-4}}{2!}  \sum_{s_i, t_i \in S_n} 
\left(  \frac{1}{1 - (1/N^2)} \right)^2 
\delta\left(  \left( 1 - \frac{v}{N} \right)^2
\prod_{i=1}^2 s_i t_i s_i^{-1} t_i^{-1}  \right),
\\
& = &   
\frac{N^{-4}}{2!}  \sum_{s_i, t_i \in S_n} 
\left(  \frac{1}{1 - (1/N^2)} \right)^2 
\delta\left( \left( 1 \: - \: \frac{2}{N} v \: + \: \frac{1}{N^2} v^2 \right)
\prod_{i=1}^2 s_i t_i s_i^{-1} t_i^{-1}  \right),
\end{eqnarray}

Let us interpret the three terms above systematically:
\begin{itemize}
\item First, consider the term
\begin{eqnarray}
\lefteqn{
\frac{N^{-4}}{2!}  \sum_{s_i, t_i \in S_n} 
\left(  \frac{1}{1 - (1/N^2)} \right)^2 
\delta \left( \prod_{i=1}^2 s_i t_i s_i^{-1} t_i^{-1}  \right)
} \nonumber \\
& = &
\sum_{k=0}^{\infty} \frac{N^{-4}}{2!} 
\left( \sum_{k=0}^{\infty} N^{-2k} \right)^2 
\sum_{s_i, t_i \in S_n} 
\delta \left( \prod_{i=1}^2 s_i t_i s_i^{-1} t_i^{-1}  \right)
\end{eqnarray}
To leading order in $N$, this term goes like $N^4$,
and describes an unbranched
double ($n=2$) cover of the genus-two Riemann surface
$\Sigma_T$.  From the Riemann-Hurwitz formula~(\ref{eq:riemann-hurwitz}),
that double cover $\Sigma_T$ must be a genus 3 surface, which is consistent
with the power of $N$.

\item Next, consider the term
\begin{equation}  \label{eq:p=2:n=2:reject}
- \frac{N^{-4}}{2!}  \sum_{s_i, t_i \in S_n} 
\left(  \frac{1}{1 - (1/N^2)} \right)^2 
\delta \left(  \frac{2}{N} v \prod_{i=1}^2 s_i t_i s_i^{-1} t_i^{-1}  \right).
\end{equation}

This term describes a branched double cover of $\Sigma_T$.
However, it does not make a nonzero contribution:
\begin{itemize}
\item Physically,
it is straightforward to check that the $\delta$ above vanishes.
Since $S_2$ is abelian,
\begin{equation}
\prod_{i=1}^2 s_i t_i s_i^{-1} t_i^{-1} \: = \: 1,
\end{equation}
and as $v \neq 1$, the argument of the delta function is not the identity.
\item Mathematically, from the Riemann-Hurwitz formula~(\ref{eq:riemann-hurwitz}),
\begin{equation}
\chi(\Sigma_W) \: = \: 2 \chi(\Sigma_T) - 1 \: = \: -5,
\end{equation}
which is odd, not consistent with the Euler characteristic of a
smooth curve.
\end{itemize}
When we add a projector, this term will reappear in the
Nguyen-Tanizaki-\"Unsal decomposition.

\item Finally, consider the term
\begin{equation}
\frac{N^{-4}}{2!}  \sum_{s_i, t_i \in S_n} 
\left(  \frac{1}{1 - (1/N^2)} \right)^2 
\delta \left(  \frac{1}{N^2} v^2 
\prod_{i=1}^2 s_i t_i s_i^{-1} t_i^{-1}  \right).
\end{equation}

This term describes a branched double cover of $\Sigma_T$,
branched over two points.
From the Riemann-Hurwitz formula~(\ref{eq:riemann-hurwitz}),
we see that $\Sigma_W$ is a genus-four Riemann surface,
which is consistent with the factor $N^{-6}$.

\end{itemize}

Now, we turn to the Nguyen-Tanizaki-\"Unsal universes.
We can add projectors and revisit these arguments to
interpret~(\ref{eq:source}) for $n=2$ in the zero-area limit, namely
\begin{eqnarray}
Z_R^+(0,p,N) & = & \frac{N^{-4}}{2!} \sum_{s_i,t_i \in S_2}
\delta\left( \left( \Omega_2 \right)^{-2} 
\left( \prod_{i=1}^2 s_i t_i s_i^{-1} t_i^{-1} \right) P_r \right),
\\
& = &
\sum_{k=0}^{\infty} \frac{N^{-4}}{2!} 
\left(  \frac{1}{1 - (1/N^2)} \right)^2 
\sum_{s_i, t_i \in S_n} 
\delta \left( \left( \prod_{i=1}^2 s_i t_i s_i^{-1} t_i^{-1}\right) P_r  \right)
 \\
& & \: - \:
\frac{N^{-4}}{2!}  \sum_{s_i, t_i \in S_n} 
\left(  \frac{1}{1 - (1/N^2)} \right)^2 
\delta \left(  \frac{2}{N} v 
\left( \prod_{i=1}^2 s_i t_i s_i^{-1} t_i^{-1} \right) P_r  \right)
\nonumber \\
& & \: + \:
\frac{N^{-4}}{2!}  \sum_{s_i, t_i \in S_n} 
\left(  \frac{1}{1 - (1/N^2)} \right)^2 
\delta \left(  \frac{1}{N^2} v^2 
\left( \prod_{i=1}^2 s_i t_i s_i^{-1} t_i^{-1} \right) P_r  \right).
\nonumber
\end{eqnarray}
We have already described how to interpret the first and third
terms as a disjoint union of stacky versions of $\Sigma_T$; it merely
remains to discuss the middle term.

Much as in the $p=0$ case, one effect of adding a projector is to add
a term not present previously, namely the middle term~(\ref{eq:p=2:n=2:reject})
that we
discarded above:
\begin{equation}
- \frac{N^{-4}}{2!}  \sum_{s_i, t_i \in S_n} 
\left(  \frac{1}{1 - (1/N^2)} \right)^2 
\delta \left(  \frac{2}{N} v 
\left( \prod_{i=1}^2 s_i t_i s_i^{-1} t_i^{-1} \right)
P_r \right).
\end{equation}
Previously, this term vanished, because the delta function could not
be nonzero.  Here, on the other hand, thanks to the insertion of $P_r$,
this term can be nonzero.

We will discuss our proposed interpretation of these terms in 
section~\ref{sect:interp:exs:peq2neq2}.
Briefly, we will propose that the Gross-Taylor sigma model has
a new symmetry that allows for a localization
on maps of a fixed degree, and also introduce additional (stacky) worldsheets
to describe the extra terms that appear in the presence of 
the projector $P_r$.

\section{Interpretation of the terms}
\label{sect:interp}

In this section we discuss the interpretation of the Gross-Taylor
series expansion
of the partition functions of the
distinct Nguyen-Tanizaki-\"Unsal
universes.  We now identify
the universes with\footnote{
In relating the Gross-Taylor combinatorics to individual
Nguyen-Tanizaki-\"Unsal universes, 
we are implicitly assuming that a coupled representation $R \overline{S}$
uniquely determines $R$ and $S$ separately, which is believed to hold in
the large $N$ limit, but not at finite $N$.
} coupled representation $R \overline{S}$, and
as outlined earlier,
factorize the partition function in the form
\begin{equation}
\left( \dim R \overline{S} \right)^{2-2p}
\exp\left( - g_{YM}^2 \frac{A}{2N} C_2(R \overline{S}) 
\right)
\: = \:
Z^+_R(A,p,N) \, Z^-_S(A,p,N)
\: + \: \mbox{subleading}.
\end{equation}
For simplicity, we focus on the chiral partition functions
$Z^+_R(A,p,N)$ for fixed representation $R$, 
given in equation~(\ref{eq:full-chiral}),
which we repeat below:
\begin{eqnarray}
Z^+_R(0,p,N) & = &
N^{n(2-2p)} \sum_{s_i,t_i \in S_n} \sum_{L=0}^{\infty}
{\sum_{v_1, \cdots, v_L \in S_n}}^{\!\!\!\prime}
N^{\sum_j (K_{v_j} - n)} \, \frac{d(2-2p,L)}{n!}
\nonumber \\
& & \hspace*{1in} \cdot \,
\delta\left( v_1 \cdots v_L \left( \prod_{i=1}^p [s_i,t_i] \right) P_r \right)
\exp\left( -  \frac{A}{\alpha'_{GT}} n \right).
\end{eqnarray}

This expression differs from the chiral partition function $Z^+(0,p,N)$
of the Gross-Taylor string in two important ways:
\begin{itemize}
\item First, there is no sum over degrees $n$, the expression above for
$Z_R^+(A,p,N)$ references one fixed degree (equal to the number of
boxes in the Young tableau for $R$).
Physically in the Gross-Taylor string, this corresponds to a restriction to
maps of fixed degree, which is analogous in a gauge theory to restricting
to sectors of fixed instanton number.  Restrictions on instantons to those
of instanton degree satisfying a divisibility criterion is common
in decomposition, but a restriction to a single instanton degree is
novel (and at least sometimes problematic).  
\item Second, for maps of any one degree $n$, the delta function includes
a projector $P_r$, not present in the original chiral
Gross-Taylor partition function $Z^+(A,p,N)$,
determined as in~(\ref{eq:proj:defn})
by an irreducible representation $r$ of $S_n$,
itself determined by $R$.  We
will see that
the projector $P_r$ in the expression above is going to result in new terms
not present in the original analysis, which can not be
interpreted in terms of ordinary covering maps $\Sigma_W \rightarrow
\Sigma_T$.  (Existence of extra contributions in individual universes
which cancel out when the universes are summed over is a typical feature
of decomposition, as we review in section~\ref{sect:interp:stack}.)

Such projectors have previously appeared in analyses of
finite $N$ contributions e.g.~\cite{Baez:1994gk,Matsuo:2004nn},
where they are interpreted as (nonperturbative)
open string contributions, but here they
arise in contributions in the large $N$ limit, where nonperturbative
contributions (in the string coupling, as opposed to $\alpha'$)
would not be expected.
\end{itemize}

In this section we will propose resolutions for both of these issues.
\begin{enumerate}
\item In subsection~\ref{sect:interp:restr}, we 
observe that a restriction to individual instanton sectors is consistent
if the theory admits a symmetry whose Noether current couples to the
instanton degrees, as either a 2-form or its Hodge dual.  
In other words, we propose that the Gross-Taylor sigma model must admit either
a global 1-form symmetry or a $(-1)$-form symmetry coupling to the
instanton degree.
We discuss a prototypical example
(two-dimensional pure Maxwell theory),
and also walk through a number of possible alternatives.

\item In subsection~\ref{sect:interp:stack}, we propose an
interpretion of the extra contributions
as arising from worldsheets that have orbifold points
(meaning, technically, certain kinds of smooth stacks),
in the special case that $\Sigma_T = {\mathbb P}^1$.
\end{enumerate}

In subsection~\ref{sect:interp:exs} we illustrate these ideas in examples.

\subsection{Restrictions on worldsheet instanton (map) degrees}
\label{sect:interp:restr}

Our computations earlier in this paper have suggested that,
if we have correctly tracked through the series expansion,
then
the Nguyen-Tanizaki-\"Unsal universes
are described by Gross-Taylor sigma models 
in
which maps are restricted to a single
degree (i.e.~fixed worldsheet instanton number).

From a field theory perspective, this seems particularly troubling.
Ordinarily we sum over all instantons.  Labelling field configurations
by instanton number is typically just an artifact of a semiclassical
expansion, and does not have an intrinsic meaning in quantum field theory.

Concretely, there is a standard old argument of Weinberg (see
e.g.~\cite[section 23.6]{Weinberg:1996kr}) linking cluster
decomposition to instanton sums in semiclassical expansions.  
The idea is as follows.  Imagine trying to restrict to instantons of a single
degree, then add a field configuration consisting of a closely-spaced
instanton-antiinstanton pair.  Now, move the centers of that pair far
apart.  If cluster decomposition holds, then asymptotically one has theories
with different numbers of instantons.
Any restriction on instantons can only arise via a violation of
cluster decomposition.

Now, there is a loophole in this argument: if the restriction
arises via a disjoint union of theories, each separately summing over
all instantons, with theta angles arranged
so that some instantons cancel out in the sum, one can arrange for a restriction
on instanton numbers, at the cost of violating cluster decomposition
in a very mild controllable fashion (as any disjoint union violates cluster
decomposition).  This 
is often exploited
in decomposition \cite{Hellerman:2006zs}.
For example,
theories with restrictions to instanton degree divisible by an integer
have been discussed in 
e.g.~\cite{Pantev:2005rh,Pantev:2005zs,Pantev:2005wj,Hellerman:2010fv}, 
but the resulting theories
are believed to be consistent only by virtue of decomposition, in the
sense that they are the result of
superimposing   
similar
theories with slightly different theta angles, as outlined above.
The different
theories (universes) each separately sum over all instanton degrees;
only in the sum does one see an apparent restriction on instanton degrees
(in the semiclassical expansion).

To restrict to a single
instanton degree, as we appear to see here,
and not just to instanton degrees satisfying a divisibility criterion,
would be considerably stronger than the examples above discuss.
Formally, the partition function of a single instanton sector
is given by integrating over values of
the theta angle, analogously to the ensemble\footnote{
It should be noted that
an ensemble is not the same as a decomposition, which becomes visible
in QFTs on spacetimes with multiple connected components.  (In the
former case, there is only one summand/integral over the ensemble,
whereas in the second, there are as many as connected components.)
} averaging discussed in
e.g.~\cite{Witten:2020bvl,Maloney:2020nni,Afkhami-Jeddi:2020ezh,Cotler:2020hgz,Schlenker:2022dyo}, 
essentially Fourier-transforming along
the theta angle to pick out the contribution from a single degree.
To interpret this directly as a 
a decomposition would require
an uncountable infinity of universes, 
parametrized by a nondynamical theta angle,
which we find unlikely.

We emphasize that a restriction to a single instanton sector will not
always be possible, even with a summation over countably infinitely
many universes, in an arbitrary quantum field theory.  
For example, consider ordinary
orbifolds.  As they are finite gauge theories, the instanton sectors
are precisely the twisted sectors, which are enumerated by
equivalence classes of bundles.  On $T^2$, for example, modular
invariance tightly constrains possible theories, and the only
individual instanton sector consistent with modular invariance
is the untwisted sector.
Restrictions to subsets of twisted
sectors do frequently arise in decomposition, but such restrictions
are always to modular-invariant subsets, never to a single
nontrivial twisted sector.

In this section,
we will describe several different potential proposals for possible resolutions
of this puzzle, but ultimately we will observe that there is (at least)
one  
set of circumstances where this would be consistent:  when the theory
has a symmetry whose Noether current couples to either the 
pullback of the K\"ahler form
or its dual
(at least perturbatively in the string coupling constant),
as the integral of that pullback is the worldsheet instanton number.
This would require the Gross-Taylor sigma model to admit either a global
1-form symmetry or a $(-1)$-form symmetry.
In the former case, the Gross-Taylor sigma model would decompose,
as we will discuss later.

\subsubsection{Proposal}

Before making our proposal, we shall first walk through several
possibilities.

Later in section~\ref{sect:disjoint} we will argue that the branched covers
$\Sigma_W$ can be replaced, for at least some purposes,
with disjoint unions, so one might suspect
that perhaps one can deform $\Sigma_W$ to a corresponding disjoint union.
Since the Gross-Taylor theory is a topological (string) theory, it should
be invariant under smooth deformations, suggesting that perhaps the
Gross-Taylor string on a branched cover is equivalent to the
Gross-Taylor string on a disjoint union, potentially shedding
light on the present question.  Unfortunately, we will see in
section~\ref{sect:disjoint:deform} 
that although such deformations exist, the disjoint
unions are not in the closure of the Hurwitz moduli space, one must instead
pass a finite distance through a larger moduli space, which invalidates this
potential argument.

Another option one might consider is that potentially the extra contributions
of section~\ref{sect:interp:stack}
might, conceivably,
cancel out some of the contributions from the regular curves, rendering the
sum of contributions trivial.  Unfortunately, the contributions from
section~\ref{sect:interp:stack} 
are weighted with different factors of $N$, so such a 
cancellation can not take place.

In appendix~\ref{app:ginsu} we will outline another proposal, which is
to interpret the single-instanton sectors as different
QFTs, obtained from localizing the original QFT onto the
desired instanton sectors by adding (analogues of) $BF$
terms.
This is
straightforward to describe in the bosonic case, but as we describe
in appendix~\ref{app:ginsu}, we run into a subtlety with the supersymmetrization
which appears to obstruct its application in cohomological field theories.

Having walked through several possibilities, we now turn to our proposal.

Briefly, we are proposing that the Gross-Taylor sigma model has a 
suitable symmetry which enables one to meaningfully select worldsheet
instanton sectors.
In other contexts,
for ordinary (invertible zero-form) symmetries, one would require that 
there be a conserved charge, so as to make the corresponding 
quantity meaningful in the
full quantum field theory, not just in some semiclassical expansion.

Here, we propose\footnote{
We would like to thank R.~Plesser for suggesting this direction.
} that
the Gross-Taylor sigma model admit
a symmetry whose Noether current is either the pullback of the 
K\"ahler form
or its Hodge dual (since the integral of that pullback is the
worldsheet instanton number), as these seem the most conservative options.
Given that the K\"ahler form is a
two-form, 
this suggests 
that the theory either has
\begin{itemize}
\item a global 1-form symmetry, or
\item a global $(-1)$-form symmetry,
\end{itemize}
(related to the worldsheet instanton degree).
If the (two-dimensional) theory has a 1-form symmetry, then
it decomposes.  That said, the Gross-Taylor
string theory is also expected to couple to worldsheet gravity,
so any such decomposition would not yield completely disjoint universes, 
but rather the
universes would still communicate via gravitational
interactions.  We discuss gravitational couplings and
decomposition in appendix~\ref{app:grav:decomp}.

We list two possibilities above, but these two possibilities are closely linked.
Recall that
in $d$ spacetime dimensions, if a theory has a global $(d-1)$-form
symmetry, it decomposes.  If one gauges that symmetry,
the resulting
theory is one of the universes of the decomposition \cite{Sharpe:2019ddn},
which
has a $(-1)$-form quantum symmetry, corresponding to
spacetime-filling defects.
If one then gauges
that quantum symmetry, one recovers the original theory (as expected
for a quantum symmetry).  Since the quantum symmetry is a $(-1)$-form symmetry,
gauging it corresponds to summing over spacetimes, and so one explicitly
recovers the original theory as sum over universes.  It is natural to speculate
that in the present circumstances, the Gross-Taylor string associated
to the whole two-dimensional Yang-Mills theory may have a global 1-form symmetry
and so decomposes (as $d-1 = 1$ here), in which case 
the string associated to one Nguyen-Tanizaki-"Unsal
universe of the Yang-Mills decomposition is obtained by gauging 
that 1-form symmetry, and so has a global $(-1)$-form symmetry.

This proposal may sound exotic, but in the next section we will describe
another two-dimensional theory where precisely this takes place.

Furthermore, in hindsight, existence of such a global symmetry in the
Gross-Taylor sigma model should be expected from standard yoga.
The restriction to a Nguyen-Tanizaki-\"unsal universe in the
decomposition of two-dimensional pure Yang-Mills should be equivalent
to gauging the corresponding one-form symmetry 
\cite{Sharpe:2019ddn}, and as is well-known,
target-space
gauge symmetries correspond to worldsheet global symmetries.
(For example, in a conventional string theory, the Noether current
for an ordinary worldsheet symmetry forms part of the vertex operator
for a target-space gauge field.)  Thus, at least in general terms,
in hindsight, it should not be surprising that the Gross-Taylor expansion
for a single Nguyen-Tanizaki-\"Unsal universe should possess a global
symmetry.
(That said, it is not clear how yoga alone would predict
coupling of that symmetry to worldsheet instanton degree, which is
our prediction here.)

Now, we are aware of two separate proposals for a worldsheet theory
of the Gross-Taylor string, namely the Cordes-Moore-Ramgoolam proposal
\cite{Cordes:1994sd,Moore:1994dk,Cordes:1994fc}, describing a sigma model
localizing on holomorphic maps,
and the Horava proposal \cite{Horava:1993aq,Horava:1995ic},
describing a sigma model localizing on harmonic maps,
which are both proofs-of-principle of the possible
existence of a Gross-Taylor sigma model.
We are not aware of a linearly-realized symmetry of the form proposed
above in either\footnote{
In the Horava proposal, it is tempting to consider an analogue of a 
$BU(1)$ symmetry acting 
as $d \phi \mapsto d \phi + \Lambda$, which naively is a symmetry of the
kinetic terms for constant metric, but we do not understand how
this would be consistent with nontrivial metrics or
curvature terms.
} of these
proposals.  
One possibility is that the symmetry is present, but nonobvious
(perhaps realized noninvertibly, much as in 
e.g.~\cite{Huang:2021zvu,Nguyen:2021yld,Nguyen:2021naa}).  
Another possibility is that there exists a third possible
worldsheet theory for the Gross-Taylor sigma model.
We leave this issue for future work.

We should add that although we believe the property above is a necessary
condition, it might not be sufficient.  Here we have in mind the example
of electric charge, which is conserved and defines a 
superselection rule (see e.g.~\cite{Wick:1970bd}), in which in the
superselection sectors, the total
charge is fixed.
Similarly, here, the Gross-Taylor expansion of the Nguyen-Tanizaki-\"Unsal universes
produces maps of one\footnote{
In examples in which
$\Sigma_W$ has multiple connected components, that fixed number arises as the
sum of the degrees of the maps from the components of $\Sigma_W$ to
$\Sigma_T$.
} fixed degree.  
In any event, to be consistent, more than a superselection rule is needed
here -- the Nguyen-Tanizaki-\"Unsal universes are separately well-defined
physical theories, and so too should the restrictions of the Gross-Taylor
sigma model to maps of total fixed degree be consistent.
For example, if the Gross-Taylor sigma model
has a 1-form symmetry (one of the options we list), then it also has a 
decomposition, which 
is certainly stronger than just superselection, and would yield
separately well-defined theories
(modulo its gravitational coupling,
see appendix~\ref{app:grav:decomp}).
We leave this question for future work.

In passing, we observe that two-dimensional cohomological field theories
(such as existing proposals for the Gross-Taylor string)
typically have semisimple operator algebras, and so decompose -- but
that is a decomposition only of the topological subsector, not 
necessarily of the
entire theory, and more to the point, it need not have any connection
to worldsheet instanton degrees.  As a result, it is unfortunately
not relevant for our purposes here.

\subsubsection{Prototype:  two-dimensional pure Maxwell theory}
\label{sect:interp:proto}

In this section we will study 
two-dimensional pure Maxwell theory, as a theory whose symmetries
are a prototype for those we propose for the Gross-Taylor string.
Specifically, it has a (one-form) symmetry with Noether current related
to the $U(1)$ bundle degree, also known as the
$U(1)$  monopole\footnote{We shall use the terms monopole number
and vortex number interchangeably in two-dimensional abelian theories to describe
the first Chern class of a $U(1)$ bundle.
We are avoiding referring to
`instanton numbers' in two-dimensional gauge theories in this section,
as in two dimensions this is sometimes used
to refer to the value of
\begin{equation}
\int_{\Sigma} {\rm Tr}\, F \wedge *F.
\end{equation}
Unlike four dimensions, in two dimensions this is not a topological
invariant, and indeed has a nonzero area dependence, but in the 
literature the term instanton number is often associated with such a term,
even in two-dimensional theories, see e.g.~\cite{Gross:1994mr}.
}
number (meaning, the first Chern class of the bundle).
Because it has a one-form symmetry,
it decomposes, and  
we shall see explicitly that the universes of the decomposition are
indexed by an integer which is `Poisson dual' to the 
$U(1)$  monopole
number,
reflecting the symmetries of the theory.
At the end of this section, we will also discuss an analogue of the
Witten effect in dyons in four dimensions, here interchanging
universes under rotations of the theta angle.

The pure Maxwell theory in any dimension has a global $BU(1)$ symmetry,
given explicitly by shifts $A \mapsto A + \Lambda$,
with Noether current $J^e = *F$,
associated to \cite[section 4.1]{Gaiotto:2014kfa}
an operator $U_{\alpha}(p) = \exp( i \alpha *F(p))$.
There is also a magnetic symmetry with current $J^m = F$,
associated with an operator
$U_{\beta}(\Sigma) = \exp(i \beta \int_{\Sigma}F)$
corresponding to a $(-1)$-form symmetry
\cite[foonote 10, section 4.1]{Gaiotto:2014kfa}.
(Noether currents of this form, or their Hodge duals,
are essentially our prediction for the Gross-Taylor sigma model.)

As has been argued elsewhere (see e.g.~\cite{Hellerman:2006zs,Sharpe:2022ene}),
a $d$-dimensional theory with a global $(d-1)$-form symmetry
should decompose.  Here, since two-dimensional pure Maxwell theory
has a $BU(1)$ symmetry, one should expect that it decomposes into
countably many universes, indexed by irreducible representations of
$U(1)$ (i.e.~integers).
Indeed,
in \cite{Cherman:2020cvw},
it was argued that two-dimensional pure Maxwell theory is equivalent
to a disjoint union of invertible theories, in much the same way
that \cite{Nguyen:2021yld,Nguyen:2021naa} later argued that two-dimensional pure
Yang-Mills theories decompose.

Given the fact that the Noether current for the global 1-form symmetry
is related to $F$, it is natural to expect some sort of correlation between
the universes (labelled by $U(1)$ representations) and 
and the $U(1)$
monopole
numbers (meaning, values of $\int F$),
and indeed, it was argued in e.g.~\cite{Paniak:2002fi,Paniak:2003gn,Caporaso:2006kk}
that they are Poisson dual to one another.
As this will be important for our analysis, and is slightly subtle,
we will review the relationship carefully.

Following section~\ref{sect:rev} and  \cite[section 2]{Vafa:2004qa},
\cite{Aganagic:2004js}
write the partition function for two-dimensional pure Maxwell
theory on a Riemann surface $\Sigma$ in the form
\begin{eqnarray}
Z(\Sigma) & = &
\sum_R \left( \dim R \right)^{\chi(\Sigma)}
\exp\left[ - g_{YM}^2 A C_2(R)  \right],
\\
& = &
\sum_{q \in {\mathbb Z}} \exp\left[ - g_{YM}^2 A q^2 
\right],
\label{eq:part:Maxwell:exact}
\end{eqnarray}
using the fact that for $G = U(1)$ \cite[chapter 7, table 7.1]{iachello}
\begin{equation}
C_2(R) \: = \: q^2.
\end{equation}
In the expression above, since all irreducible representations $R$ of
$U(1)$ are one-dimensional, we have replaced representations $R$
with charges $q$.  The universes of the decomposition are then
indexed by $q$, and the $q$th universe has partition function
\begin{equation} \label{eq:interp:maxwell:qth-part}
 \exp\left[ - g_{YM}^2 A q^2 
\right].
\end{equation}

Now, let us carefully study the relationship between
the $U(1)$ charge $q$ and $U(1)$ monopole numbers $\int F$.
First, note that the area dependence of
the exponent is linear in the area $A$,
but the kinetic
term for a gauge field of monopole number $q$ should\footnote{
We would like to thank R.~Szabo for making this observation,
and for pointing out the role of the Poisson resummation that will
appear momentarily.
} be inversely
proportional to the area, so the $U(1)$ charge $q$ cannot be
precisely the same as the monopole number.

Let us take a moment to check this carefully, as it is important
for our analysis,
For 
$U(1)$ monopole number $n$,
we claim that the Maxwell kinetic term
\begin{equation}
\frac{1}{g_{YM}^2} \int_{\Sigma} F^{\mu \nu} F_{\mu \nu} \: \propto \:
\frac{n^2}{g_{YM}^2 {\rm Area}},
\end{equation}
a completely different area dependence from that of
the partition function for the $q$th 
universe~(\ref{eq:interp:maxwell:qth-part}).
Explicitly, in the special case of $\Sigma = S^2$,
following \cite[section 3]{Gross:1994mr},
a classical gauge field configuration with $U(1)$ monopole number
$n$ is given by
\begin{equation}
A_{\theta} \: = \: 0, \: \: \: A_{\phi} \: = \: n \, 
\frac{1 - \cos \theta}{2},
\end{equation}
hence
\begin{equation}
F_{\theta \phi} \: = \: \frac{n}{2} \sin \theta,
\: \: \:
F^{\theta \phi} \: = \: \frac{1}{2 r^4 \sin \theta},
\end{equation}
using
\begin{equation}
ds^2 \: = \: r^2 \left( d \theta^2 \: + \: \sin^2 \theta \, d \phi^2 \right),
\end{equation}
so that
\begin{equation}
\sqrt{\det g} \: = \: r^2 \sin \theta,
\end{equation}
and
\begin{equation}
\int_{S^2} \sqrt{\det g} \, F^{\theta \phi} F_{\theta \phi} \: = \: 
\frac{\pi n^2}{r^2} \: \propto \: \frac{n^2}{\rm Area}.
\end{equation}
Thus, we see that the area dependence of the partition function for the
$q$th universe~(\ref{eq:interp:maxwell:qth-part}) is not consistent with
what one would expect if the universes were indexed by the $U(1)$
monopole number.  Instead, we shall argue that they are indexed by
a Poisson-dual number.

To understand the role of $U(1)$ monopoles and the two-dimensional
$\theta$ angle, we will 
outline the derivation of the exact expression for the partition function,
following the analysis of \cite{Blau:1993hj}, but including
a $\theta$ angle term.
We will see explicitly that the irreducible representation and the $U(1)$ monopole
number are related by a Poisson\footnote{
See also e.g.~\cite{Witten:1992xu,Gross:1994mr,Minahan:1993tp,Caselle:1993mq} 
for
related discussions of Poisson resummation in the context of two-dimensional
pure Yang-Mills theories,
and also e.g.~\cite{Fine:1990zz,Fine:1991ux} for more detailed computations.
} resummation.
Schematically, working on $\Sigma = S^2$ for simplicity,
for the two-dimensional pure Maxwell theory with 
action
\begin{equation}
S \: = \: \frac{1}{g_{YM}^2} \int_{\Sigma} 
F^{\mu \nu} F_{\mu \nu} \: + \:
\int_{\Sigma}  i \theta F ,
\end{equation}
the
partition function
can be expressed as a sum over contributions from $U(1)$ monopoles
(nontrivial $U(1)$ bundles) of charge $n$, in the form
\begin{eqnarray} \label{eq:part:Maxwell:int1}
Z(\Sigma_g) & = &
\sum_{n=-\infty}^{\infty} \left( \pi g_{YM}^2 A \right)^{-1/2}  \exp\left(
- \frac{n^2}{g_{YM}^2 A} \: + \: i \theta n \right),
\end{eqnarray}
(with overall factors $\left( \pi g_{YM}^2 A \right)^{-1/2}$
chosen in hindsight to make the result clean).
We can Poisson resum this expression as follows:
\begin{eqnarray}
Z(\Sigma_g) & = &
\int_{-\infty}^{\infty} d\lambda \sum_{n=-\infty}^{\infty} \delta(\lambda-n)
 \left( \pi g_{YM}^2 A \right)^{-1/2} 
\exp\left(
- \frac{\lambda^2}{g_{YM}^2 A} \: + \: i \theta \lambda \right),
\\
& = &
\int_{-\infty}^{\infty} d\lambda
\sum_{m=-\infty}^{\infty} \exp(2 \pi i m \lambda)
 \left( \pi g_{YM}^2 A \right)^{-1/2} 
\exp\left(
- \frac{\lambda^2}{g_{YM}^2 A} \: + \: i \theta \lambda \right),
\\
& = &
\int_{-\infty}^{\infty} d\lambda'
\sum_{m=-\infty}^{\infty}
 \left( \pi g_{YM}^2 A \right)^{-1/2}
\exp\left[ - \frac{ (\lambda')^2 }{ g_{YM}^2 A} 
\: - \: \frac{g_{YM}^2 A}{4} (\theta + 2 \pi m)^2 
\right],
\\
& = &
\sum_{m=-\infty}^{\infty}
\exp\left[
- \frac{g_{YM}^2 A}{4} (\theta + 2 \pi m)^2 
\right],
\label{eq:part:Maxwell:exact:theta}
\end{eqnarray}
where
\begin{equation}
\lambda' \: = \: \lambda \: - \: i \frac{g_{YM}^2 A}{2} \left( \theta + 2 \pi m
\right)^2.
\end{equation}
For simplicity we specialized to $\Sigma = S^2$; for other Riemann surfaces,
in principle one would need to take into account e.g.~moduli of flat connections
in the analysis.
However, the form of the result for the exact partition function,
as reviewed in section~\ref{sect:rev}, is universal.

Including the theta angle, we see that the partition function of the
$q=m$-th universe is
\begin{equation}
\exp\left[
- \frac{g_{YM}^2 A}{4} (\theta + 2 \pi m)^2 
\right].
\label{eq:part:Maxwell:exact:theta:univ}
\end{equation}
When $\theta = 0$, we see that this recovers the 
exact expression~(\ref{eq:part:Maxwell:exact}),
identifying the $U(1)$ charge $q$ with $m$, and absorbing factors of
$2$ and $\pi$ into $g_{YM}^2$.  For nonzero $\theta$, we can treat
the $\theta^2$ term as a contribution to an overall multiplicative
factor, leaving us just with two terms: one quadratic in $m$
(corresponding to the Casimir $C_2(R)$) and one linear in $m$ (corresponding to
the Casimir $C_1(R)$).  This form of the exact expression, taking
into account $\theta$ dependence, has also been discussed
in \cite[section 2]{Vafa:2004qa},
\cite{Aganagic:2004js},
albeit with a different normalization on the $\theta$ term.

In particular, 
comparing the area-dependence of the $m^2$ term to
the exact result for the partition function~(\ref{eq:part:Maxwell:exact:theta}),
we emphasize that
the irreducible $U(1)$ representation, the charge $q$, corresponds to
$m$, the Poisson-dual to the $U(1)$ monopole number, which was the
$n$ in equation~(\ref{eq:part:Maxwell:int1}), recovering the result in
\cite{Paniak:2002fi,Paniak:2003gn,Caporaso:2006kk}.
That said, although the universes of the decomposition are not indexed
by $U(1)$ monopole number, since the universes are all invertible
field theories with one-dimensional Fock spaces, the difference between
fixed $U(1)$ monopole number and fixed $U(1)$ charge is just a Fourier
transform, so it might still be meaningful to speak of universes associated
to fixed $U(1)$ monopole number.

The $\theta$ dependence of the partition 
function~(\ref{eq:part:Maxwell:exact:theta:univ}) suggests further
interesting physics, which we will return to shortly.
For the moment, we confirm the result above by
giving an alternative computation.
We can quickly outline a confirmation of this result\footnote{
We would like to thank C.~Closset for a discussion of this
computation.
} for the partition function as a function of
$m=q$, by semiclassically gauging the $BU(1)$ symmetry,
which in principle \cite{Sharpe:2019ddn} should select particular
universes.
Start with the pure Maxwell action
\begin{equation}
S \: = \: \frac{1}{g_{YM}^2} \int_{\Sigma} 
F^{\mu \nu} F_{\mu \nu} \: + \:
\int_{\Sigma}  i \theta F ,
\end{equation}
and gauge $BU(1)$.
This means we add a dynamical two-form tensor field potential
$B$, the gauge field for the gauged $BU(1)$,
whose gauge transformations couple to the gauge field
$A$ as follows:
\begin{eqnarray}
A & \mapsto & A - \Lambda,
\\ 
B & \mapsto & B + d \Lambda,
\end{eqnarray}
for $\Lambda$ any one-form.  The action of the pure Maxwell theory with the
gauged $BU(1)$ then takes the form\footnote{
A close analogue of this procedure was used in
four-dimensional pure Maxwell theory in
\cite[section 2.2]{Witten:1995gf}, \cite[section 2.4]{Gukov:2006jk}, 
to implement S-duality.  There, it was noted that just gauging $BU(1)$ left
a trivial theory, and so a dual gauge field was added, which coupled via
a topological term, a four-dimensional analogue of the $\int B$ term we
introduce above.  Here, by contrast, our goal is to
generate a trivial theory -- one of the universes of decomposition,
itself an invertible field theory in this case 
-- so no additional fields are needed.
}
\begin{equation}
S' \: = \:
 \frac{1}{g_{YM}^2} \int_{\Sigma} 
(F + B) \wedge *(F + B) \: + \:
\int_{\Sigma}  i \theta (F+B) .
\end{equation}

Now, in principle, to select one particular universe, we add
a theta angle term for the $BU(1)$ symmetry, parametrized by
irreducible representations of $U(1)$, namely ${\mathbb Z}$.
This is similar in principle to the procedure described
in \cite[section 8]{Sharpe:2019ddn}, where a $B {\mathbb Z}_k$
symmetry was gauged.  Explicitly, here,
we add a term proportional to $qB$, for $q \in {\mathbb Z}$ corresponding
to an irreducible representation of $U(1)$:
\begin{equation}
S'' \: = \:
 \frac{1}{g_{YM}^2} \int_{\Sigma} 
(F + B) \wedge *(F + B) \: + \:
\int_{\Sigma} \left( i \theta (F+B) + 2 \pi i q  B \right).
\end{equation}
We absorb $F$ into $B$ via the affine gauge transformation $A \mapsto A -
\Lambda$, to write
\begin{equation}
S'' \: = \:
 \frac{1}{g_{YM}^2} \int_{\Sigma} 
B \wedge *B \: + \:
\int_{\Sigma} i \left(  \theta  + 2 \pi q \right) B.
\end{equation}
Integrating out $B$ (and glossing over operator determinants,
see e.g.~\cite[section 2.2]{Witten:1995gf} for a more complete
analysis), we get
\begin{equation}
S'' \: = \:  \frac{1}{4} g_{YM}^2 A \left(
 \theta + 2 \pi q \right)^2
\end{equation}
matching the exponent of the partition function of
the theory~(\ref{eq:part:Maxwell:exact:theta:univ})
in any one fixed universe, determined by $q$ (up to factors of $2$, $\pi$),
which can be identified with the $m$ 
in~(\ref{eq:part:Maxwell:exact:theta:univ}).

In passing, we observe that the form of the exact
result for the partition function~(\ref{eq:part:Maxwell:exact:theta:univ}) 
for pure Maxwell theory
with a theta angle, suggests the existence of a two-dimensional
decomposition analogue of the four-dimensional Witten effect in
dyons \cite{Witten:1979ey}.  Recall that, under a rotation
$\theta \mapsto \theta + 2\pi$ of the four-dimensional theta angle,
dyon charges also rotate.  (The complete physical theory is invariant, as
this just interchanges existing dyon charges.)
Here, judging from the
exact partition function~(\ref{eq:part:Maxwell:exact:theta:univ}),
a rotation $\theta \mapsto \theta + 2\pi$ of the
two-dimensional theta angle is equivalent to shifting
$q \mapsto q + 1$, meaning that under a rotation of the theta
angle, the universe shifts.

We can also see the same result from the perspective of the
Hilbert spaces.  Recall that, at least for vanishing theta angle,
the Hilbert space of two-dimensional
pure Yang-Mills is given by the class functions on the Lie group
$G$, which has a basis of characters $\chi_R$ associated to irreducible
representations.
Because of the theta angle, a particle moving along a closed noncontractible
loop will pick up a phase, or more formally,
\begin{equation}
f(gz) \: = \: \lambda(z) f(g).
\end{equation}
(This is discussed in e.g.~\cite[section 2.4]{Sharpe:2014tca},
\cite{Tachikawa:2013hya} 
for discrete theta angles;
the present case is similar.  In essence, in two dimensions,
the $\theta$ angle acts as an electric field \cite{Coleman:1976uz}, 
hence it modifies
wavefunctions by phases.)
Here, for a fixed irreducible representation $R$ corresponding
to $U(1)$ charge $q$, and identifying $g \in U(1)$ with a phase $\alpha
= \exp(i \beta) \in {\mathbb C}^*$, $|\alpha|^2 = 1$, we can write
\begin{equation}
f(g) \: = \: \exp(i (\theta/2\pi) \beta) \exp( i \beta q)
\: = \:  \alpha^{q + \theta/(2\pi)}.
\end{equation}
Explicitly, rotating $\theta \mapsto \theta + 2 \pi$ is equivalent to
incrementing $q$, thereby shifting the universe, just as we saw
in partition functions.
(See also \cite{Nguyen:2022lie} for related remarks.)

So far we have discussed two-dimensional pure Maxwell theory as a possible
prototype for expected symmetries and properties of the Gross-Taylor string.
It should be noted in addition that, at least morally,
two-dimensional pure Yang-Mills theory is also similar, in the sense
that it has a global 1-form symmetry (realized 
noninvertibly \cite{Nguyen:2021yld,Nguyen:2021naa}), 
and the gauge instantons of the theory
are roughly Poisson dual to the representations (and hence universes)
\cite{Gross:1994mr,Minahan:1993tp,Caselle:1993mq}.

\subsection{The extra terms: stacky worldsheets}
\label{sect:interp:stack}

In the Gross-Taylor expansion of Nguyen-Tanizaki-\"Unsal universes
in section~\ref{sect:revisit-gt}, we came across two puzzles:
\begin{itemize}
\item terms that appear to correspond to contributions to sigma models
from maps of fixed degree,
\item terms that appear to arise from additional worldsheets,
often with powers of $N$ that could not be realized from smooth orientable
Riemann surfaces.
\end{itemize}
In the previous subsection, we proposed a solution for the first puzzle.
In this subsection, we will discuss the second puzzle,
and in the special case that
$\Sigma_T = S^2$, propose a possible interpretation of the extra terms.

Before going on, we should observe that such extra terms are common
in decomposition.  As a simple prototypical example, let us consider
the example of a two-dimensional $SU(2)$ gauge theory with
center-invariant matter.  This is equivalent to (decomposes into)
a pair of $SO(3)$ theories with different discrete theta angles,
schematically,
\begin{equation}
SU(2) \: = \: SO(3)_+ \, \coprod \, SO(3)_-.
\end{equation}
(See e.g.~\cite[section 2]{Sharpe:2022ene} and references therein.)
Each of the $SO(3)$ theories has nonperturbative sectors
($SO(3)$ bundles) not possessed by the $SU(2)$ theory.
However, because of their different weightings, those $SO(3)$-specific
bundles cancel out.  Essentially the same thing is happening
here:  we have extra worldsheets arising in the Gross-Taylor expansion of
a single Nguyen-Tanizaki-\"Unsal universe (analogous to
the non-$SU(2)$ $SO(3)$ bundles), 
whose contributions cancel out
when one sums over all universes.

A context that is more relevant for us is decomposition of sigma models.
We describe some analogous cases here:
\begin{itemize}
\item First, consider two-dimensional
orbifolds with trivially-acting subgroups.  A prototypical
example is $[X/D_4]$, where $D_4$ denotes the eight-element dihedral
group with center ${\mathbb Z}_2$.  If that center acts trivially, then
\cite[section 5.2]{Hellerman:2006zs}
\begin{equation}
[X/D_4] \: = \: [X/{\mathbb Z}_2 \times {\mathbb Z}_2] \, \coprod \,
[X/{\mathbb Z}_2 \times {\mathbb Z}_2]_{\rm d.t.},
\end{equation}
where each of the ${\mathbb Z}_2 \times {\mathbb Z}_2$ orbifolds
has twisted sectors not present in the $D_4$ orbifold
(analogous to the extra contributions present here).
Those additional sectors are weighted differently by discrete torsion,
and cancel out when the universes are summed together,
just as the extra contributions do here.
\item A related example is WZW models arising on boundaries of three-dimensional
Chern-Simons theories.  If one gauges a trivially-acting one-form
symmetry in the bulk three-dimensional theory, the boundary sees
an orbifold of a WZW model by a trivially-acting ordinary group
symmetry, see \cite{Pantev:2022pbf}.  Those WZW models are precisely
sigma models (whose targets are Lie groups, with background $H$ flux),
and so the story here is closely analogous.
\item Sigma models whose targets are gerbes,
which can be realized by two-dimensional gauged linear sigma models with
gauge groups in which a subgroup acts trivially \cite{Pantev:2005zs,Pantev:2005rh,Pantev:2005wj}.
Physically, these are equivalent to gauge theories with a restriction
on gauge instantons.
These also decompose into disjoint unions of ordinary sigma models on
underlying spaces (realized via ordinary gauged linear sigma models).
The constituent universes are described by gauge theories with no
restriction on gauge instantons -- so again, there are extra contributions
not present in the original theory -- which are weighted differently
by $B$ fields / theta angles.
\end{itemize}
As the Gross-Taylor string is believed to be a sigma model, these
examples are more directly relevant, albeit they all only refer to
cases involving finitely many universes.

Thus, existence of extra terms in the Gross-Taylor expansion
of individual Nguyen-Tanizaki-\"Unsal universes is not surprising.
However, 
their interpretation in the language of string field theories of
sigma models needs to be addressed, and we turn to this next.

Technically, the extra terms are the result of including a projection operator,
and such a projection operator has been previously discussed,
in the context of finite $N$ corrections, in
e.g.~\cite{Baez:1994gk}, \cite[section 3]{Matsuo:2004nn}.
Reference \cite{Baez:1994gk} describes the projection operator as
giving rise to a ``projection point,'' an analogue of the $\Omega$-points
(interpreted in terms of branched covers in
\cite{Cordes:1994sd}),
and  \cite[section 3]{Matsuo:2004nn} breaks the resulting contributions
up into a ``perturbative'' sector (with even powers of $N$) and
a ``residual'' sector (which contains terms with odd powers of $N$),
which they suggest should be attributed to open string worldsheets.
(See also e.g.~\cite[section 3.2]{Caporaso:2006kk}, which also
interprets odd powers of $N$ in $U(N)$ theories in terms of open
strings.)
As their focus is on finite $N$ corrections, which should incorporate
nonperturbative corrections, 
it is natural for them to assume open strings are involved in their analysis.

Here, we are seeing projection operators arising in the large $N$ limit,
not just in finite $N$ corrections, which distinguishes this
case from the analyses in e.g.~\cite{Baez:1994gk,Matsuo:2004nn}.
At least naively,
we would not expect nonperturbative\footnote{
Nonperturbative in the string dilaton, of course, as opposed to
$\alpha'_{GT}$.
} corrections to a large $N$ limit,
and hence, although it is still a potential interpretation,
 we would not expect open string
contributions in our case.  

We should quickly add that another places where
odd powers of $N$ arises is in
the Gross-Taylor expansion of $SO(N)$ and $Sp(N)$ gauge theories,
see e.g.~\cite{Naculich:1993ve,Naculich:1993uu,Naculich:1994kd,Ramgoolam:1993hh,Crescimanno:1996hx}, where it is said to reflect nonorientable Riemann
surfaces.

Briefly, one other interpretation of these terms is in terms of
singular branched covers, obtained as limits of smooth branched covers
in which the branch points collide.  We will describe this in examples,
and one might speculate that perhaps this is a reflection of some
subtle differences in contact terms arising in the Gross-Taylor
sigma model for the individual Nguyen-Tanizaki-\"Unsal universes.
We do not exclude this possibility.
In any event, in this section we will float a different proposal,
which will have another application later in section~\ref{sect:disjoint}.

In this subsection, 
we suggest another potential interpretation of
those extra terms arising in the 
Gross-Taylor expansion of the Nguyen-Tanizaki-\"Unsal universes
(and potentially also the ``projection points'' of \cite{Baez:1994gk}),
different from either open strings or nonorientable worldsheets,
in the special case $\Sigma_T = S^2$.
We propose that they may arise from sigma models with stacky
worldsheets, and will merely provide a `proof of principle' 
for such a description,
but, we emphasize, we are only making a proposal, a suggestion,
not a definitive statement.

Our proposal is that the extra contributions may rise from
worldsheets $\hat{\Sigma}_W$ that are disjoint unions of $n$ stacky
copies $S_i$ of $\Sigma_T$:
\begin{equation}
\hat{\Sigma}_W \: = \: \coprod_{i=1}^n S_i
\end{equation}
(the same $n$ of $S_n$, meaning that previously the worldsheet 
was an $n$-fold cover of $\Sigma_T$).
We will explicitly describe a construction of stacks
$S_i$ with matching combinatorial description, covering degree, and
Euler characteristic of $\hat{\Sigma}_W$ correctly matches
the power of $N$. 
(We will discuss orbifold Euler characteristics of Hurwitz moduli spaces
shortly.)
However, there will be multiple disjoint unions of this form which
satisfy those constraints, as we shall see.
Nevertheless, worldsheets of this form represent our current best guess
at an interpretation of these extra terms, at least in the case
$\Sigma_T = S^2$.

Stacks may sound exotic, but in many ways they are simple generalizations of
spaces.  They admit metrics, spinors, gauge fields, and so forth,
and can be dealt with using the usual tools of differential
geometry.  See for example \cite{metzler1,heinloth,behrendxu,noohi1,noohi2,noohi3},
\cite[lecture 3]{alper1} for some introductory material.
In any case, we will only require stacks\footnote{
Technically, we are working with smooth Deligne-Mumford stacks,
and the subset of those which are of the form of local orbifolds on
Riemann surfaces.  We will not use, for example,
gerbes on curves.
} that take the form
of Riemann surfaces with local orbifold points, 
which will simplify the discussion.  
We collect some relevant facts about stacks in 
appendix~\ref{app:basics-stacks}.

For these reasons, it is natural to conjecture that it may be possible
to define sigma models with stacky worldsheets,
not just stacky target spaces (as has been discussed in 
e.g.~\cite{Pantev:2005rh,Pantev:2005zs,Pantev:2005wj}).  
We shall assume that this is the case in the remainder
of this subsection, and turn to the construction of the stacky worldsheets
from the Gross-Taylor combinatorics.

Now, let us turn to the construction of our new worldsheets $\hat{\Sigma}_W$,
which are
disjoint unions of stacky copies $S_i$ of the 
Riemann surface $\Sigma_T$,
\begin{equation}  \label{eq:disjointunion}
\hat{\Sigma}_W \: = \: \coprod_{i=1}^n S_i,
\end{equation}
The stacks $S_i$ are constructed from the Gross-Taylor
combinatorics as follows.
Write each $v_a \in S_n$ as a product of cycles, of the form
$(1\cdots n_1) (n_2 \cdots n_3) \cdots$.  If the integer $i$ is an
element of a cycle of $v_a$ of length $k$,
put a ${\mathbb Z}_k$ orbifold point
on $\Sigma_T$, corresponding to $v_i$.

As a simple example, suppose $n=6$ and $v = (12)(345)(6) \in S_6$.
Then, of the six copies of $\Sigma_T$ appearing in the disjoint
union, two would have ${\mathbb Z}_2$ orbifold points, three
would have ${\mathbb Z}_3$ orbifold points, and one would have no
orbifold point at all.

The reader should note that even this prescription is not unique;
we could interchange which sheets $S_i$ receive which specific orbifold
points.  Such interchanges will preserve the Euler characteristic,
as will be clear momentarily.  
In section~\ref{sect:disjoint}, we will apply similar ideas to
replace branched cover interpretations, and
we will give a more systematic construction of such
disjoint unions from branched covers, which will eliminate such ambiguity.  

Now, let us compute the Euler characteristic of the disjoint
union~(\ref{eq:disjointunion}) above.  
The idea is to start with the Euler characteristic
of a disjoint union of $n$ copies of $\Sigma_T$, then subtract
the Euler characteristic of $nL$ disks, and add back the contribution
from disks containing single orbifold points.
In doing so, since a disk with a single
${\mathbb Z}_k$ orbifold point has Euler characteristic $1/k$ and appears
$k$ times, a collection of $k$ such disks will contribute $1$.
In the prescription above, the Euler characteristic contributed
by all of the disks determined by a single $v$ is therefore equal to the
number of cycles.

Assembling these pieces, we then see that
\begin{eqnarray}
\chi\left( \coprod_{i=1}^n S_i \right)
& = &
n \chi(\Sigma_T) \: - \: n L \: + \: \sum_{j=1}^L K_{v_j},
\\
& = & 
n (2-2p) \: + \: \sum_{j=1}^L \left( K_{v_j} - n \right),
\end{eqnarray}
which matches the exponent of $N$ in $Z_R^+(A,p,N)$.
The ambiguity mentioned above, namely redefining the $S_i$ by moving
orbifold points between different sheets, clearly
preserves $\chi( \coprod_i S_i )$.

It is important to note that this ansatz will only generate the
correct power of $N$ in the case of $\Sigma_T = S^2$, so that
$p=0$.  For $p > 0$, $p$ the genus of $\Sigma_T$,
there are additional group-theoretic factors of
the form
\begin{equation}
\prod_{i=1}^p \left[ s_i, t_i \right].
\end{equation}
Ordinarily these have to close up to the $v_i$ insertions, but,
in the presence of a projector $P_r$, they no longer need close,
and also do not come with any ameliorating factors of $N$.
In examples, we will see that they do not appear to have a natural
stacky interpretation, at least not following the ansatz above.
For this reason, we only apply our ansatz to the case $p=0$.

So far we have verified that the exponent of $N$ matches the
Euler characteristic of $\hat{\Sigma}_W$.  Ideally, to thoroughly
check this proposal, one would also
compute the Euler characteristics of the corresponding
Hurwitz moduli spaces, essentially to verify the
detailed prediction
\cite[equ'n (1.1)]{Cordes:1994sd}
\begin{equation}  \label{eq:cmr:Euler:predict}
Z^+(0,N,p) \: = \: \exp\left[
\sum_{h=0}^{\infty} N^{2-2h} \chi\left( \overline{H(h,p)} \right)
\right],
\end{equation}
where $H(h,p)$ denotes the Hurwtiz moduli space of maps from 
a connected worldsheet of genus $h$ to one of genus $p$.
Here, we are enlarging the number of possible worldsheets through the
addition of stacky points, and it is not entirely clear how the
different numbers of stacky points should be weighted, to extend the
expression above.  For example, formally, 
possible extensions of~(\ref{eq:cmr:Euler:predict}) include
the form
\begin{eqnarray}
Z^+(0,N,p) & =  & \exp\left[
\sum_{h=0}^{\infty}  N^{2-2h} \left(
\prod_{k=2}^{\infty} \left( \sum_{n_k=0}^{\infty} f_k(n_k) 
N^{(-1 + 1/k)n_k} \right) \right)
\chi\left( \overline{H(h,p,n_2,n_3,\cdots)} \right)
\right],  \nonumber
\end{eqnarray}
where $H(h,p,n_2,n_3,\cdots)$ denotes the Hurwitz moduli space  
with $n_k$ ${\mathbb Z}_k$ orbifold points for each $k$,
and
for unknown functions $f_k(n_k)$, defining relative multiplicities
of ${\mathbb Z}_k$ orbifold points,
(For example, $f_k(n) = 1$,
$f_k(n) = 1/n!$, and $f_k(n) = 1/(k! n!)$ all naively seem equally plausible.)
To understand this proposal at the level of orbifold Euler characteristics
of Hurwitz moduli spaces, we would need a proposal for those
functions.
Furthermore, as previously discussed, there are multiple possible
interpretations as stacky worldsheets for the extra terms.
We will discuss possibilities in examples, but,
for the reasons above and because
we are only attempting to provide a proof of principle, not a 
definitive answer,
we leave a detailed analysis of Hurwitz moduli spaces for future work.

For terms in which a smooth branched cover $\Sigma_W$ exists,
as in the Gross-Taylor interpretation, there exists a(t least one) disjoint
union $\coprod_i S_i$ of the form above, to which $\Sigma_W$
can be deformed (albeit not without leaving the Hurwitz moduli space).  
The details of this construction are
described in section~\ref{app:rh-revisited}.

Another aspect of those projectors is that
they weight the contributions by phases.  In the same spirit as the rest
of this section, we propose that those phases be interpreted as
resulting from $B$ fields on $\Sigma_W$, as we will elaborate
in examples.

Later in section~\ref{sect:disjoint}, we will utilize such disjoint
unions to give a different geometric picture of the Gross-Taylor
expansion of a single Nguyen-Tanizaki-\"Unsal universe.
Each component of the disjoint union, each stacky copy of
$\Sigma_T$, maps to $\Sigma_T$ with degree one, and we identify
each such contribution with an invertible field theory.
The projector $P_r$ then contributes weights which can be
interpreted in terms of flat $B$ fields on the various components of the
disjoint union replacing $\Sigma_W$.
We will propose to interpret those (stacky) copies of $\Sigma_T$,
replacing $\Sigma_W$, as a reflection of a decoposition of the
Gross-Taylor string to invertible field theories.

In passing, we find it interesting that stacky worldsheets appear
in this construction.  Certainly stacks have previously been
described as target spaces of strings, see in particular
\cite{Pantev:2005rh,Pantev:2005zs,Pantev:2005wj},
but there, the worldsheets were ordinary Riemann surfaces, whereas
here, the worldsheets are stacks.  On the other hand,
since maps from stacks factor through the underlying spaces, which is
an important part of the proposed dictionary, it is not entirely clear
to us how much weight should be ascribed to the difference between
a stacky copy $S$ of $\Sigma_T$ and $\Sigma_T$ itself.  We will leave
this for future work.

We will re-use this same proposal to a different end in
section~\ref{sect:disjoint},
where we give an alternative geometric interpretation of the terms
in the Gross-Taylor expansion, which better reflects the fact that the
Nguyen-Tanizaki-\"Unsal universes are invertible field theories.

\subsection{Examples}
\label{sect:interp:exs}

In this section we will walk through the same examples as in
section~\ref{sect:exs}, this time demonstrating how to describe the
extra contributions (arising from presence of projectors $P_r$)
in terms of stacky worldsheets.
As in section~\ref{sect:exs}, in examples
we will restrict to the zero-area limit, to simplify 
computations.

\subsubsection{$p=0, n=2$}
\label{sect:interp:exs:peq0neq2}

In this section we consider the special case $p=0$
(so that $\Sigma_T = {\mathbb P}^1$), and a representation $R$
described by a Young tableau with $n=2$ boxes.
From section~\ref{sect:exs:peq0neq2}, 
the Gross-Taylor expansion of the Nguyen-Tanizaki-\"Unsal
universe has the form
\begin{eqnarray}
Z^+_R(0,p,N) & = &
\frac{N^{2n}}{n!} \delta \left(  (\Omega_n)^{2} P_r \right),
\\
& = &
\frac{N^{2n}}{n!} \delta \left(
(1) P_r \: + \: 2 \left( \frac{1}{N} \right) v P_r \: + \:
\left( \frac{1}{N} \right)^2 v^2 P_r \right),
\\
& = &
\frac{N^4}{2!} \delta( P_r) \: + \:
2 \frac{N^3}{2!} \delta(v P_r) \: + \:
\frac{N^2}{2!} \delta(v^2 P_r),
\\
& = &  \label{eq:peq0:neq2:chiralpart}
\frac{N^4}{4} \: \pm \: \frac{N^3}{2} \: + \: \frac{N^2}{4}.
\end{eqnarray}

As discussed in section~\ref{sect:exs:peq0neq2}, the first and third 
terms above can be interpreted as contributions to the original
Gross-Taylor sigma model, from maps of fixed degree $n=2$.
We interpret these as reflecting the presence of a suitable
symmetry in the Gross-Taylor sigma model.

The middle term is novel to the expansion of a Nguyen-Tanizaki-\"Unsal
universe.  The exponent of $N$ is odd,
and so this term cannot correspond to a map from a smooth
curve $\Sigma_W$ to $\Sigma_T = {\mathbb P}^1$.
It is constructed from a
Dijkgraaf-Witten correlation function with a single $v$, plus a projector.

Now, how should this term be interpreted?

Suppose we try to interpret this term by expanding out the projector $P_r$,
picking out the $v$ term in $P_r$.  Then, this term is
\begin{equation}
\pm 2 \frac{N^3}{2!} \delta(v v).
\end{equation}
Following the usual prescription, this would naively appear
to be a branched double cover of ${\mathbb P}^1$, branched over two points,
which is another ${\mathbb P}^1$.  However, that has the wrong Euler
characteristic to match the power of $N$.

If we ignore the projector, this term appears to describe
a double cover of ${\mathbb P}^1$, branched over
a single point, which does not exist as a smooth manifold.

Instead, we propose to interpret this term as describing a stack.
Following the discussion in section~\ref{sect:interp:stack},
given the single factor of $v$, 
we propose to interpret this
in terms of a curve with a ${\mathbb Z}_2$ orbifold point, and in fact,
as $n=2$ copies of $\Sigma_T$, each with a ${\mathbb Z}_2$ orbifold
point, again restricted to maps of total degree $n=2$ as above.

To check this proposal, we check Euler characteristics.
Consider a stacky ${\mathbb P}^1$ with a single
${\mathbb Z}_2$ orbifold point, explicitly the weighted projective
stack ${\mathbb P}^1_{[1,2]}$.  Euler characteristics are additive,
and this is the union of a disk and a single ${\mathbb Z}_2$ orbifold,
so we can write
\begin{equation}
\chi( \mbox{stacky }{\mathbb P}^1) \: = \:
\chi({\rm disk}) \: + \: \chi(B {\mathbb Z}_2) \: = \: 1 \: + \: 1/2
\: = \: 3/2,
\end{equation}
using the fact that $\chi(B {\mathbb Z}_2) = 1/2$.
(See for example appendix~\ref{app:euler} for more details on Euler
characteristic computations in stacky curves.)

Now, this is just a single cover of $\Sigma_T = {\mathbb P}^1$,
but we can create
a double cover by taking two copies.  Note that the Euler characteristic
of two copies is given by
\begin{equation}
\chi( {\mathbb P}^1_{[1,2]} \coprod {\mathbb P}^1_{[1,2]}) \: = \:
(2)(3/2) \: = \: 3,
\end{equation}
which precisely matches the power of $N$ appearing in the middle term.
Furthermore, each stack separately has a projector to
${\mathbb P}^1$, so there certainly exists at least one map
${\mathbb P}^1_{[1,2]} \coprod {\mathbb P}^1_{[1,2]}
\rightarrow {\mathbb P}^1$.

The two choices of sign on the middle term of~(\ref{eq:peq0:neq2:chiralpart})
arise from the two possible
projectors $P_r$, for $r$ an irreducible representation of 
$S_2 = {\mathbb Z}_2$.  One might interpret these as
choices of $B$ fields on each of the
two copies of ${\mathbb P}^1$. 
For example, perhaps
in one universe, the two copies have the
same $B$ field (trivial), and in the other universe, one copy has trivial
$B$ field whereas the other has $\int B = -1$.
An alternate possible interpretation is that one picks the same $B$ field
on both elements of the disjoint union:  $+1$ for one $r$, $-1$ for the
other.
(To be clear, ordinarily in a sigma model, the $B$ field is defined
on the target $\Sigma_T$, not the worldsheet, whereas here we are fixing
a closed 2-form on the worldsheet.)

In principle, there is another stack that is a double cover of
${\mathbb P}^1$, with related combinatorics and matching Euler characteristics.
Specifically, consider the disjoint union of one ordinary
${\mathbb P}^1$ and one ${\mathbb P}^1$ with two ${\mathbb Z}_2$
orbifolds points.  
The ordinary ${\mathbb P}^1$ has Euler characteristic
two, and a ${\mathbb P}^1$ with two orbifold points has Euler characteristic
one, so the Euler characteristic of the disjoint union is three,
also matching the power of $N$.
(The existence of this second disjoint union
can be interpreted as an ambiguity in the disjoint union construction,
in which we have moved one of the ${\mathbb Z}_2$ orbifold points to
another sheet in the cover.)

Now, let us briefly comment on orbifold Euler characteristics of Hurwitz
moduli spaces.  From appendix~\ref{app:stacks:maps}, a map from a stacky
curve to $\Sigma_T$ is equivalent to an ordinary map from the underlying
curve to $\Sigma_T$, so it is natural to suspect that the orbifold
Euler characteristic of a Hurwitz moduli space of maps from a curve
with stacky points is the same as that of maps from a curve with 
the stacky points omitted.  (This might possibly ignore sublteties
in the compactification, however.)  In the present case, we use the
fact that the orbifold Euler characteristic of the Hurwitz moduli space
of unbranched maps ${\mathbb P}^1 \rightarrow {\mathbb P}^1$ of degree one is 
one.  We discussed two potential interpretations of the stacky curves
above:
\begin{itemize}
\item $S_1 \coprod S_1$, where $S_1$ denotes a ${\mathbb P}^1$ with a single
${\mathbb Z}_2$ orbifold point.  Assuming that one sums over all possible
orbifold point insertions with the same weighting, it is natural to 
speculate that the pertinent factor would be  
$1/2!$ (from expanding an exponential,
to get a disjoint union, in the form of~(\ref{eq:cmr:Euler:predict}).
Here, however, the extra terms have magnitude $1$.
\item ${\mathbb P}^1 \coprod S_2$, where $S_2$ denotes a
${\mathbb P}^1$ with two ${\mathbb Z}_2$ orbifold points inserted.
Here, there would not be a symmetry factor as appeared in the previous
example, so it is natural to speculate that the 
pertinent factor would be $1$, which does match the magnitude of
the extra terms.
\end{itemize}
We emphasize that this is not a definitive conclusion.

In passing, there is another interpretation of this term, which we will
not utilize here.
Specifically, there exist a singular branched double cover of
${\mathbb P}^1$.  This can be obtained as a limit of
a branched double cover of
${\mathbb P}^1$ which is branched over two points, by taking a limit that
the two points approach one another.  Since the resulting curve can
be constructed from two caps and a point, its Euler characteristic is $3$,
matching the exponent of $N$.

\subsubsection{$p=0, n=3$}
\label{sect:interp:exs:peq0neq3}

Next, we consider the case that $\Sigma_T = {\mathbb P}^1$
(so that $p=0$), and the case that the representation $R$ is described
by a Young tableau with $n=3$ boxes.
From section~\ref{sect:exs:peq0neq3},
the Gross-Taylor expansion of the Nguyen-Tanizaki-\"Unsal universe is
\begin{eqnarray} 
Z^+_R(0,p,N) & = &
\frac{N^{2n}}{n!} \delta \left(  (\Omega_n)^{2} P_r \right),
\\
& = &
\frac{N^{2n}}{n!} \delta \left( P_r
\: + \: 2 \sum_v \left(\frac{1}{N}\right)^{n-K_{v}} v P_r
 \: + \: 
 \sum_{ij} 
\left( \frac{1}{N} \right)^{2n - K_{v_1} - K_{v_2}}
v_i v_j P_r \right),  \nonumber
\end{eqnarray}

The first and third terms can be interpreted as in
section~\ref{sect:exs:peq0neq3}.  In particular, in neither case does
the $P_r$ make any difference, as only the identity element of the projector
contributes to the sum.  Both of these terms describe smooth covers
$\Sigma_W$ describing degree-three maps to the target $\Sigma_T = {\mathbb P}^1$.

The interpretation of the middle terms is more interesting, as here
the projector plays an important role.
Depending upon the order of $v$, there are two cases appearing in the
sum.
\begin{itemize}
\item First, consider the case that $K_v = 2$, for example if $v = (12)(3)$.
Here, the term takes the form
\begin{equation}
\frac{2}{3!} N^5 \delta \left( \sum_v v P_r \right).
\end{equation}
Following the prescription of section~\ref{sect:interp:stack},
this describes a disjoint union of a single ${\mathbb P}^1$
(mapped to itself by $v$) and two copies of
${\mathbb P}^1_{[1,2]}$, hence
$\Sigma_W = {\mathbb P}^1 \coprod {\mathbb P}^1_{[1,2]} \coprod
{\mathbb P}^1_{[1,2]}$, and since $\chi({\mathbb P}^1_{[1,2]}) = 3/2$,
as discussed previously, we see $\chi(\Sigma_W) = 2 + 3/2 + 3/2 = 5$,
also matching the exponent of $N$.

The disjoint union description is not unique; we can get another such 
description by moving orbifold points between sheets of the cover.
For example, another possible interpretation is as the disjoint
union ${\mathbb P}^1 \coprod {\mathbb P}^1 \coprod S$,
where $S$ is ${\mathbb P}^1$ with two 
${\mathbb Z}_2$ orbifold points.  This has Euler characteristic
$2+2+1 = 5$, again matching the exponent of $N$.

(Alternatively, we could interpret this as a singular curve, a 3-cover
branched over a single point, but we will not utilize that description
in this paper.)

\item Next, consider the case that $K_v = 1$, for example if $v = (123)$.
Here, the term above takes the form
\begin{equation}
\frac{2}{3!} N^4  \delta \left( \sum_v v P_r \right).
\end{equation}
Following the prescription of section~\ref{sect:interp:stack},
we interpret this as a disjoint union
$\Sigma_W = {\mathbb P}^1_{[1,3]} \coprod {\mathbb P}^1_{[1,3]} \coprod 
{\mathbb P}^1_{[1,3]}$, a disjoint union of three stacky copies of
$\Sigma_T = {\mathbb P}^1$, each with a single ${\mathbb Z}_3$ orbifold point.
A disk with one ${\mathbb Z}_3$ orbifold has Euler characteristic
$1/3$, so
$\chi( {\mathbb P}^1_{[1,3]}) = 4/3$,
hence $\Sigma_W$ has
Euler characteristic 4, matching the
exponent of $N$.

As before, the prescription of section~\ref{sect:interp:stack} is
not unique, and we can get other valid disjoint unions by moving orbifold
points between sheets.  Here, such possibilities are
\begin{itemize}
\item a disjoint union ${\mathbb P}^1 \coprod {\mathbb P}^1 \coprod S'$
where $S'$ is a 
${\mathbb P}^1$ with three ${\mathbb Z}_3$ orbifold points.  The
latter summand has Euler characteristic zero, so the Euler characteristic
of the disjoint union is $2+2+0 = 4$, matching the exponent of $N$,
\item a disjoint union with three summands:
one ordinary ${\mathbb P}^1$,
one ${\mathbb P}^1$ with one ${\mathbb Z}_3$ orbifold point,
and one ${\mathbb P}^1$ with two ${\mathbb Z}_3$ orbifold points.
The latter summand has Euler characteristic $2/3$, so the Euler
characteristic of the disjoint union is $2 + 4/3 + 2/3 = 4$, matching
the exponent of $N$.
\end{itemize}

(Alternatively, we could also interpret this in terms of a singular
$\Sigma_W$, a 3-cover branched over a single point, where all three sheets meet.
This curve can also be constructed from three caps and a point,
so its Euler characteristic is 4, which matches the exponent of
$N$.
As before, we will not utilize that interpretation here.)

\end{itemize}

Next, we outline a preliminary check of orbifold Euler characteristics
of Hurwitz moduli spaces, to compare amongst these possibilities.
As in the previous section, we use the fact that maps from curves
with orbifold points
factor uniquely through maps without orbifold points, and ignore
potential compactification subtleties.
\begin{itemize}
\item First, consider the case of extra terms with $v$ such that
$K_v = 2$.  Since there are three such $v \in S_3$, the magnitude of the
numerical factor appearing in the extra terms is
\begin{equation}
\frac{2}{3!} (3) \: = \: 1.
\end{equation}
We described two possibilities:
\begin{itemize}
\item ${\mathbb P}^1 \coprod S_1 \coprod S_1$, where $S_1$ denotes
a ${\mathbb P}^1$ with a single ${\mathbb Z}_2$ orbifold point.
Proceeding as in the previous example, naively we expect a $1/2!$
(as a symmetry factor, from expanding out an exponential).
\item ${\mathbb P}^1 \coprod {\mathbb P}^1 \coprod S_2$,
where $S_2$ denotes a ${\mathbb P}^1$ with two ${\mathbb Z}_2$ orbifold
points.  Proceeding as in the previous example, naively we expect
another $1/2!$, again as a symmetry factor, this time relating the
two copies of ${\mathbb P}^1$.
\end{itemize}
Neither separately has the right numerical factor, though we do observe
that the sum of these two possibilities does add
up correctly.
\item Next, consider the case of extra terms with $v$ such that
$K_v = 1$.  Since there are two such $v \in S_3$, the magnitude of the
numerical factor appearing in the extra terms is
\begin{equation}
\frac{2}{3!} (2) \: = \: \frac{2}{3}.
\end{equation}
We described three possibilities:
\begin{itemize}
\item $\tilde{S}_1 \coprod \tilde{S}_1 \coprod \tilde{S}_1$, where
$\tilde{S}_1$ denotes a ${\mathbb P}^1$ with one ${\mathbb Z}_3$ point.
Here, from the same reasoning as before, we naively expect a $1/3!$,
as a symmetry factor (from expanding an exponential).
\item ${\mathbb P}^1 \coprod {\mathbb P}^1 \coprod \tilde{S}_3$,
where $\tilde{S}_3$ denotes a ${\mathbb P}^1$ with three ${\mathbb Z}_3$
points.  Here, we expect a $1/2!$.
\item ${\mathbb P}^1 \coprod \tilde{S}_1 \coprod \tilde{S}_2$,
where $\tilde{S}_2$ denotes a ${\mathbb P}^1$ with two ${\mathbb Z}_3$
orbifold points.  Here, from the same reasoning, we naively
expect $1$, as there is no symmetry between the terms (unless the
presence of two orbifold points contributes a symmetry factor).
\end{itemize}
As before, none of these terms separately duplicates the whole expected
factor, though it is possible a sum might, at least if the terms
receive additional analogues of symmetry factors due to the presence of multiple
orbifold points.
\end{itemize}
As there are multiple possibilities,
we leave a detailed examination of orbifold Euler characteristics of
Hurwitz moduli spaces for other work.

\subsubsection{$p=1$}
\label{sect:interp:exs:peq1}

From section~\ref{sect:exs:peq1},
the Gross-Taylor expansion of the Nguyen-Tanizaki-\"Unsal universe is
\begin{eqnarray} 
Z^+_R(0,p,N) & = & 
1\frac{N^{0}}{n!} \sum_{s,t \in S_n} \delta\left( (\Omega_n)^0 
s t s^{-1} t^{-1} 
P_r \right),
\end{eqnarray}
for $r$ an irreducible representation of $S_n$.

If the commutator
\begin{equation}
[s,t] \: = \: s t s^{-1} t^{-1}
\end{equation}
equals the identity, then the projector $P_r$ is irrelevant, and one
recovers the same terms as in a Gross-Taylor expansion restricted to
degree $n$, namely, unbranched $n$-fold covers of $T^2$.

If the commutator is different from the identity, then unfortunately our 
ansatz does not apply in general.
For example, suppose that $n=3$, so that
$s, t \in S_3$.  Now, $S_3$ is a nonabelian group with three elements of
order 2 and two elements of order 3.
The commutator takes values in the commutator subgroup, consisting
of even permutations, denoted $A_3$, which has order 3 (and also contains
order-3 elements of $S_3$, as it happens).
If we try to interpret the result as a disjoint union
$S_1 \coprod S_2 \coprod S_3$ where the $S_3$ are copies of $T^2$
with orbifold points, we run into the issue that $\chi(S_i) \neq 0$
(unless $S_i = T_2$), and so if there are any stacky points present
at all, the resulting disjoint union has nonzero Euler characteristic,
which does not match the power of $N$.  

Examples of this form are the reason we restrict our ansatz for
the extra contributions to the special case $p=0$ ($\Sigma_T = S^2$).

Now, let us turn to the terms that can be interpreted merely
as the restriction to maps of a single degree.
Here, these are maps $T^2 \rightarrow T^2$.
Now, sigma models $T^2 \rightarrow T^2$ have, of course,
been extensively studied in the literature, as simple computable
examples of CFTs.  Using existing results, one can show that the
partition function of an A-twisted sigma model on $T^2$ with target
$T^2$ remains modular invariant even after restriction to maps of
a single degree.  Specifically, recall from
\cite[section 1]{Bershadsky:1993ta} that
the partition function on $T^2$ of the A model with target $T^2$ is
\begin{eqnarray}
{\rm Tr}\, (-)^F F_L F_R q^{L_0} \overline{q}^{\overline{L}_0}
& = &
\frac{t + \overline{t}}{4 \pi \tau_2}
\sum_{m,n,r,s} \exp\biggl[  
- \frac{t}{4 \tau_2 \rho_2} \left| (m + r \rho) 
- \overline{\tau}(n + s\rho) \right|^2
\\
& & \hspace*{1.5in}
\: - \:
\frac{\overline{t}}{4 \tau_2 \rho_2} \left|
(m + r \rho) - \tau(n + s \rho) \right|^2 \biggr]
\nonumber
\end{eqnarray}
where $\rho$ is the complex modulus of the target $T^2$, and the
(not necessarily invertible) matrix
\begin{equation}
R \: = \: \left[ \begin{array}{cc} r & m \\ s & n \end{array} \right]
\end{equation}
encodes windings, with the degree of the map $T^2 \rightarrow T^2$
given by $| \det R \, |$, as explained in \cite[section 1]{Bershadsky:1993ta}.
Now, it is straightforward to check that
a modular transformation of
\begin{equation}
F_1^{\rm top} \: = \: \int \frac{d^2 \tau}{\tau_2}
{\rm Tr}\, (-)^F F_L F_R q^{L_0} \overline{q}^{\overline{L}_0}
\end{equation}
by
\begin{equation}
\tau \: \mapsto \: \frac{ a \tau + b}{c \tau + d},
\: \: \:
\rho \mapsto \: \frac{ e \rho + f}{g \rho + h}
\end{equation}
for
\begin{equation}
{\cal A} \: = \: \left[ \begin{array}{cc} a & b \\ c & d \end{array} \right] \: \in \:
SL(2,{\mathbb Z}),
\: \: \:
{\cal B} \: = \: \left[ \begin{array}{cc} e & f \\ g & h \end{array} \right]  \: \in \:
SL(2,{\mathbb Z}),
\end{equation}
can be absorbed
into a redefinition of $R$ as
\begin{equation}
R \: \mapsto \: {\cal A}^{-1} R \, {\cal B}.
\end{equation}
However, since ${\cal A}$ and ${\cal B}$ both have unit determinant,
$| \det R \, |$ is invariant, and so restricting to maps of a single degree
does not break modular invariance.

\subsubsection{$p=2, n=2$}
\label{sect:interp:exs:peq2neq2}

From section~\ref{sect:exs:peq2neq2},
the Gross-Taylor expansion of the Nguyen-Tanizaki-\"Unsal universe is
\begin{eqnarray}
Z_R^+(0,p,N) & = & \frac{N^{-4}}{2!} \sum_{s_i,t_i \in S_2}
\delta\left( \left( \Omega_2 \right)^{-2} 
\left( \prod_{i=1}^2 s_i t_i s_i^{-1} t_i^{-1} \right) P_r \right),
\\
& = &
\sum_{k=0}^{\infty} \frac{N^{-4}}{2!} 
\left(  \frac{1}{1 - (1/N^2)} \right)^2 
\sum_{s_i, t_i \in S_n} 
\delta \left( \left( \prod_{i=1}^2 s_i t_i s_i^{-1} t_i^{-1}\right) P_r  \right)
 \\
& & \: - \:
\frac{N^{-4}}{2!}  \sum_{s_i, t_i \in S_n} 
\left(  \frac{1}{1 - (1/N^2)} \right)^2 
\delta \left(  \frac{2}{N} v 
\left( \prod_{i=1}^2 s_i t_i s_i^{-1} t_i^{-1} \right) P_r  \right)
\nonumber \\
& & \: + \:
\frac{N^{-4}}{2!}  \sum_{s_i, t_i \in S_n} 
\left(  \frac{1}{1 - (1/N^2)} \right)^2 
\delta \left(  \frac{1}{N^2} v^2 
\left( \prod_{i=1}^2 s_i t_i s_i^{-1} t_i^{-1} \right) P_r  \right).
\nonumber
\end{eqnarray}

Terms in the first and third lines we interpret as in
section~\ref{sect:exs:peq2neq2}, in terms of maps from smooth branched covers
$\Sigma_W$ of fixed degree $n=2$.
We propose that the restriction to fixed degree be implemented 
as described previously.

In the special case that the product of the two commutators,
we could interpret the middle term as describing
$\Sigma_W = C \coprod C$, where each stack
$C$ is $\Sigma_T$ with a single ${\mathbb Z}_2$ singularity.
The curve $C$ has $\chi(C) = -2 -1 + 1/2 = -5/2$,
so the disjoint union has $\chi(\Sigma_W = C \coprod C) = -5$,
matching the leading power of $N$.
(Alternatively, we could interpret $\Sigma_W$ as a singular curve,
a double cover of the genus-two curve $\Sigma_T$ branched over a single
point.  Since $\chi(\Sigma_T) = 2-2p = -2$, removing a disk yields
$\chi = -3$, and two copies with one disk added then have
$\chi = (2)(-3) + 1 = -5$, matching the exponent of $N$.)
In any event, for reasons previously described, we only
suggest this stacky interpretation in the case $p=0$,
so the present example is not applicable.

\section{Alternative geometric interpretation of the decomposition}
\label{sect:disjoint}

In this section we give an alternative geometric interpretation of the
terms in the Gross-Taylor expansion of a
Nguyen-Tanizaki-\"Unsal universe,
in the special case that $\Sigma_T = S^2$,
motivated by the fact that the Nguyen-Tanizaki-\"Unsal universes
are trivial quantum field theories, with a one-dimensional state space.  
Specifically, instead of interpreting the terms in terms of
sigma model
maps from branched covers of $\Sigma_T$, we instead interpret them in terms of
a counting problem, counting disjoint unions 
\begin{equation}
\coprod_{i} S_i
\end{equation}
of stacky copies $S_i$ of $\Sigma_T$.
The idea is that the Gross-Taylor expansion of a single universe is
really just reproducing a single numerical factor, 
which can be reproduced by a suitable counting problem,
hamely counting (stacky) copies of $\Sigma_T$.

In the case $\Sigma_T = S^2$,
we will provide a systematic procedure for reinterpreting Dijkgraaf-Witten
partition functions in terms of disjoint unions of
stacks instead of branched covers,
and will check that the result is consistent with powers of $N$,
in the sense that the Euler characteristic of the stack one obtains
matches the Euler characteristic of the branched cover.

We will not attempt to interpret
the numerical factors in terms of automorphisms of covering maps
(or orbifold Euler characteristics of Hurwitz moduli spaces),
as in this section we are not associating a nontrivial sigma model.
Instead, we interpret the
numerical factors merely as defining counterterms, and as such provide
no further consistency tests.

That said, in principle,
one could also imagine interpreting these terms in terms of a 
Gross-Taylor string, restricted to degree one maps.
As discussed in appendix~\ref{app:stacks:maps}, maps from a stacky curve
to a space factor through an underlying smooth curve.
In the present case, maps $S_i \rightarrow \Sigma_T$ factor through the projection to
the underlying ordinary curve $\Sigma_T$: 
\begin{equation}
S_i \: \stackrel{\pi}{\longrightarrow} \: \Sigma_T \:
\longrightarrow \: \Sigma_T,
\end{equation}
hence degree one maps $S_i \rightarrow \Sigma_T$ can be identified
with degree one maps $\Sigma_T \rightarrow \Sigma_T$.

This is another perspective on identifying each stacky curve $S_i$
with a copy of an invertible
field theory on $\Sigma_T$.
Put another way, roughly speaking,
we are constructing the Nguyen-Tanizaki-\"Unsal universes on
a two-dimensional space $\Sigma_T$
by summing over copies of degree one maps $\Sigma_T \rightarrow \Sigma_T$.

In any event, our perspective in this section is merely to provide a 
geometric counterpoint to the idea that the Nguyen-Tanizaki-\"Unsal
universes are invertible field theories, 
by rethinking the combinatorics as describing counting copies of 
$\Sigma_T$, rather than in terms of a path integral for a sigma model.

We emphasize that the existence of a decomposition of the Gross-Taylor
string is not in question -- it is automatic for any unitary two-dimensional
topological field theory (with semisimple local operator algebra).
What is more interesting is that there exists a structure in the combinatorics
used to justify the existence of the Gross-Taylor string, which appears to
reflect the presence invertible field theories.

We also note that the stacks we consider in this paper are all
smooth
Deligne-Mumford stacks, which are specifically
of the form of Riemann surfaces with
isolated local orbifold points.  More general Deligne-Mumford stacks 
describing e.g.~gerbe
structures will not appear in this paper.

\subsection{Disjoint unions instead of branched covers}
\label{app:rh-revisited}

As previously discussed, 
the chiral partition function of the Gross-Taylor expansion of a single
Nguyen-Tanizaki-\"Unsal universe has
the form~(\ref{eq:full-chiral}), namely
\begin{eqnarray}
Z^+_R(A,p,N) & = &
N^{n(2-2p)} \sum_{s_i,t_i \in S_n} \sum_{L=0}^{\infty}
{\sum_{v_1, \cdots, v_L \in S_n}}^{\!\!\!\prime}
N^{\sum_j (K_{v_j} - n)} \, \frac{d(2-2p,L)}{n!}
\\
& & \hspace*{1in} \cdot \,
\delta\left( v_1 \cdots v_L \left( \prod_{i=1}^p [s_i,t_i] \right) P_r \right)
\exp\left( -  \frac{A}{\alpha'_{GT}} n \right),
\nonumber
\end{eqnarray}
where $n$ is the number of boxes in the Young tableau for $R$.

Previously we have discussed how this can be interpreted in terms of
\begin{enumerate}
\item a sum over branched $n$-fold covers $\Sigma_W \rightarrow \Sigma_T$,
\item plus some extra contributions, arising from the presence
of the projector $P_r$.
\end{enumerate}
In the special case $\Sigma_T = S^2$, we have described how
the extra contributions can be interpreted as disjoint unions of stacky
copies of $\Sigma_T$.  
Here, for the case $\Sigma_T = S^2$,
we extend that alternative interpretation
to include terms previously interpreted as branched covers.
In other words, in this section we will describe how all of the terms
can be interpreted as a disjoint union of $n$ stacky copies of
$\Sigma_T$, with Euler characteristic matching each power of $N$.

To be clear, in this proposal in this section, 
we are not interpreting the terms physically
as a sigma model from that disjoint union; instead, we are setting up
a counting problem, which seems more nearly appropriate to the fact that
the Nguyen-Tanizaki-\"Unsal universes are trivial (invertible) field theories.

Since we are discarding the sigma model interpretation in this section,
we will not attempt to compare orders of automorphism groups, as
in this alternative interpretation, we are merely giving a geometric 
counterpoint to the description of invertible field theories.

For the remainder of this section, we will focus on understanding the
terms previously interpreted as branched covers in this language.

Previously, the elements $v_1, \cdots, v_L \in S_n$ defined monodromies
about $L$ branch points in the base curve $\Sigma_T$.
In the present interpretation, the orders of the cycles in each 
monodromy $v_i$ define orders of orbifold points on copies of the
Riemann surface $\Sigma_T$.

We can describe the construction systematically as follows.
Let $B$ be the branch locus on $\Sigma_T$, and $p$ some point
(to base loops).  We can associate either branched $n$-covers or
disjoint unions $\hat{\Sigma}_W$ of $n$ stacky copies $S_i$ of $\Sigma_T$
to homomorphisms $f: \pi_1(\Sigma_T - B, p) \rightarrow S_n$.
We have already discussed branched covers;
we construct the disjoint union of $n$ stacky copies of $\Sigma_T$
as follows.
Pick a set of nonintersecting paths emanating from the fixed point $p$,
each winding around a point in the branch locus $B$.
For each point $b$ in the branch locus $B$, let $v_b$ be the image
of the corresponding path under the homomorphism $f$.
We construct a set of stacky curves $S_i$, where each $S_i$ is a copy
of $\Sigma_T$ with orbifold structures at each $b \in B$.  If
$d(i,b)$ is the length of a cycle containing $i$ in the cycle decomposition
of $v_b$, then the orbifold structure on $S_i$ at $b \in B$ is
${\mathbb Z}_{d(i,b)}$.

Later in this section
we also describe a construction of
$\hat{\Sigma}_W$ directly from $\Sigma_W$.

To make this more clear, let us update our previous example.
Specifically, consider $\Sigma_T = {\mathbb P}^1$ with two insertions
at positions denoted
$A$, $B$, and let $p \in {\mathbb P}^1$,
as illustrated in the figure below.
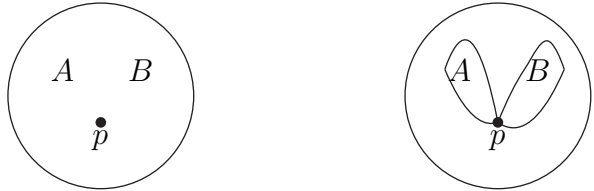
\begin{center}
\begin{picture}(70,70)(0,0)
\CArc(35,35)(35,0,360)
\Text(25,45)[r]{$A$}
\Text(55,45)[r]{$B$}
\Vertex(35,25){2}
\Text(35,22)[t]{$p$}
\end{picture}
\hspace*{1in}
\begin{picture}(70,70)(0,0)
\CArc(35,35)(35,0,360)
\Text(25,45)[r]{$A$}
\Text(55,45)[r]{$B$}
\Vertex(35,25){2}
\Text(35,22)[t]{$p$}
\Curve{(15,45)(17,50)(25,55)(35,25)}
\Curve{(15,45)(20,35)(35,25)}
\Curve{(35,25)(45,45)(55,55)(60,45)}
\Curve{(35,25)(55,35)(60,45)}
\end{picture}
\end{center}
Shown on the left is a schematic illustration of ${\mathbb P}^1$
with the two points $A$, $B$, and the basepoint $p$ for paths.
On the right is the same illustration with two nonintersecting
paths from $p$ marked.

Let the monodromies about the two points be
denoted $v_A$, $v_B$.
Suppose that $n=3$, so that $v_A, v_B, v_C \in S_3$, and take
\begin{equation}
v_A \: = \: (12)(3) \: = \: v_B.
\end{equation}
It is straightforward to check that the product
\begin{equation}
v_A v_B  \: = \: 1,
\end{equation}
so the delta function is nonzero.

We have already seen in section~\ref{sect:series:cover} how this translates into
branching data for branched 3-covers.  Briefly, locally near the points
$A$ and $B$, two of the sheets collide, but the third remains disjoint.
This results in a branched cover of the form
\begin{equation}
\Sigma_W \: = \: {\mathbb P}^1 \coprod {\mathbb P}^1,
\end{equation}
as already discussed,
and $\chi(\Sigma_W) = 4$.

In terms of disjoint unions of stacky curves,
the three stacky copies $S_{1,2,3}$ of ${\mathbb P}^1$ have local orbifolds as
illustrated in the diagram below:
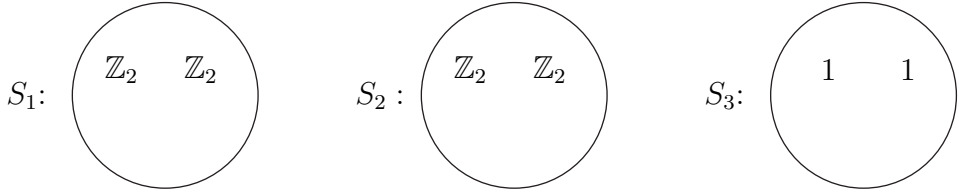
\begin{center}
\begin{picture}(70,70)(0,0)
\Text(-25,35)[l]{$S_1$:}
\CArc(35,35)(35,0,360)
\Text(25,45)[r]{${\mathbb Z}_2$}
\Text(55,45)[r]{${\mathbb Z}_2$}
\end{picture}
\hspace*{0.75in}
\begin{picture}(70,70)(0,0)
\Text(-25,35)[l]{$S_2:$}
\CArc(35,35)(35,0,360)
\Text(25,45)[r]{${\mathbb Z}_2$}
\Text(55,45)[r]{${\mathbb Z}_2$}
\end{picture}
\hspace*{0.75in}
\begin{picture}(70,70)(0,0)
\Text(-25,35)[l]{$S_3$:}
\CArc(35,35)(35,0,360)
\Text(25,45)[r]{$1$}
\Text(55,45)[r]{$1$}
\end{picture}
\end{center}
Using the methods in appendix~\ref{app:euler}, it is straightforward to check
\begin{equation}
\chi(S_1) \: = \: 1 \: = \: \chi(S_2), \: \: \:
\chi(S_3) \: = \: 2,
\end{equation}
hence
\begin{equation}
\chi\left( S_1 \coprod S_2 \coprod S_3 \right) \: = \: 4,
\end{equation}
matching the Euler characteristic of $\Sigma_W = {\mathbb P}^1 \times
{\mathbb P}^1$.

Next, we shall give a systematic description of how to replace
$n$-fold covers of $\Sigma_T$
with disjoint unions of $n$ copies of $\Sigma_T$ with orbifold points,
of matching Euler characteristic.

We will replace any $\sigma \in S_n$ by
a product of cyclic orbifolds defined by the permutation structure,
in the obvious way.  We list examples in the table below:
\begin{center}
\begin{tabular}{ccc}
$n$ & $\sigma \in S_n$ & Orbifold \\ \hline
$2$ & $1 = (1)(2)$ & $1$ \\
$2$ & $(12)$ & ${\mathbb Z}_2$ \\
$3$ & $(12)(3)$ & ${\mathbb Z}_2$ \\
$3$ & $(123)$ & ${\mathbb Z}_3$ \\
$4$ & $(12)(34)$ & ${\mathbb Z}_2 \times {\mathbb Z}_2$ \\
$4$ & $(123)(4)$ & ${\mathbb Z}_3$ \\
$4$ & $(1234)$ & ${\mathbb Z}_4$
\end{tabular}
\end{center}

The construction is as follows.  Given a smooth branched $n$-cover
$\pi: \Sigma_W \rightarrow \Sigma_T$, let $B \subset \Sigma_T$ denote
the branch locus.  Let $p \in \Sigma_T$, $p \not\in B$, and choose a set
of nonintersecting paths emanating from $p$, each winding around a point in
the branch locus $B$.  For each $b \in B$, let $v_b \in S_n$ denote the
monodromy about $b$.  Now, we construct a series of stacky curves
$S_i$, where each $S_i$ is associated to an element of $\pi^{-1}(p)$,
and is a copy of $\Sigma_T$, with orbifold structures at each
$b \in B \subset \Sigma_T$.  If $d(i,b)$ is the length of a cycle
containing $i$ in the cycle decomposition of $v_b$, then the orbifold
structure on $S_i$ at $b \in B$ is ${\mathbb Z}_{d(i,b)}$.

We then define $\hat{\Sigma}_W = \coprod_i S_i$.  Note that $\hat{\Sigma}_W$
itself depends upon a choice of paths; different paths will
change the $S_i$ and hence $\hat{\Sigma}_W$.
(As a result, since there is a braid
group action on the paths, there is a braid group action on the possible
choices of
$\hat{\Sigma}_W$.  We will see an example later in which the braid group
can change the disjoint union $\hat{\Sigma}_W$, for a fixed branched cover
$\Sigma_W$.)

We claim that
\begin{equation}
\chi(\Sigma_W) \: = \: \chi( \hat{\Sigma}_W ).
\end{equation}
We can see this as follows.
First, from results on orbifold Euler characteristics of
stacky curves in appendix~\ref{app:euler},
note that
\begin{equation}
\chi(S_i) \: = \: \chi(\Sigma_T) \: + \: \sum_{b \in B} \left[
\frac{1}{d(i,b)} - 1 \right].
\end{equation}
Then,
\begin{eqnarray}
\chi(\hat{\Sigma}_W) & = &
\chi\left( \coprod_i S_i \right) \: = \:
\sum_{i=1}^n \chi(S_i),
\\
& = &
n \chi(\Sigma_T) \: + \: \sum_{i=1}^n \sum_{b \in B} \left[
\frac{1}{d(i,b)} - 1 \right],
\\
& = &
n \chi(\Sigma_T) \: - \: n |B| \: + \:
\sum_{i=1}^n \sum_{b \in B} \frac{1}{d(i,b)}.
\end{eqnarray}
However,
\begin{eqnarray}
\sum_{i=1}^n \sum_{b \in B} \frac{1}{d(i,b)}
& = &
\sum_{b \in B} \left( \mbox{number of cycles in } v_b \right),
\\
& = &
\sum_{b \in B}\left( \mbox{number of ramification points over } b \right),
\\
& = & \mbox{degree of ramification divisor},
\end{eqnarray}
and so from the Riemann-Hurwitz formula,
\begin{equation} \label{eq:rh}
\chi(\Sigma_W) \: = \: n \chi(\Sigma_T) \: - \: \sum_{P \in \Sigma_W}
\left( e_P - 1 \right),
\end{equation}
where $\Sigma_W$ is a branched $n$-fold cover of $\Sigma_T$ and
$e_P$ is the ramification index at $P$,
we have that
the Euler characteristic of the disjoint union $\hat{\Sigma}_W$
matches that of the smooth branched cover $\Sigma_W$:
\begin{equation}
\chi(\hat{\Sigma}_W) \: = \: \chi(\Sigma_W).
\end{equation}

In passing, the construction above applies to $\Sigma_T$ of any genus.
We restrict to genus zero to accomodate interpretations of the additional
terms of section~\ref{sect:interp:stack}, 
which we only know how to interpret in terms of
disjoint unions of stacks in the special case of genus zero.

Below we list some examples.
\begin{itemize}
\item If $\Sigma_W$ is an unbranched $n$-sheeted cover of $\Sigma_T$,
we replace $\Sigma_W$ with a disjoint union of $n$ copies of
$\Sigma_T$.  From the Riemann-Hurwitz formula~(\ref{eq:rh}),
\begin{equation}
\chi(\Sigma_W) \: = \: \chi\left( \coprod_n \Sigma_T \right).
\end{equation}
\item Suppose $\Sigma_W$ is a branched double cover of $\Sigma_T$, branched
over $k$ points.  We replace $\Sigma_W$ by a disjoint union of two copies
of a stack $S$, where $S$ is $\Sigma_T$ with $k$ ${\mathbb Z}_2$ orbifold
points.

Let us check that these have the same Euler characteristic.
First, from Riemann-Hurwitz~(\ref{eq:rh}),
\begin{equation}
\chi(\Sigma_W) \: = \: 2 \chi(\Sigma_T) - k.
\end{equation}
To compute $\chi(S)$, remove $k$ disks from $\Sigma_T$ and replace each
with a disk containing a ${\mathbb Z}_2$ orbifold.  This modifies the
Euler characteristic as
\begin{equation}
\chi(S) \: = \: \chi(\Sigma_T) \: - \: k(1) \: + \: k (1/2),
\end{equation}
using the fact that the Euler characteristic of an ordinary disk is $1$,
and that of a disk containing a ${\mathbb Z}_m$ orbifold is $1/m$.
It follows immediately that
\begin{equation}
\chi(\Sigma_W) \: = \: \chi(S \coprod S) \: = \: 2 \chi(S).
\end{equation}
\item Let $\Sigma_W$ be a 3-fold cover of $\Sigma_T$, branched along
two points, with branching at each point described by elements of the
conjugacy class $(12)(3)$.  Then, $\Sigma_W$ is a disjoint union of
one copy of
$\Sigma_T$ and one branched double-cover of $\Sigma_T$, branched over two
points, hence
\begin{equation}
\chi(\Sigma_W) \: = \: \chi(\Sigma_T) \ + \: 2 \chi(\Sigma_T) - 2.
\end{equation}
The corresponding stack is a disjoint union $\Sigma_T \coprod S \coprod S$,
where $S$ is $\Sigma_T$ with a pair of ${\mathbb Z}_2$ orbifold points.
As a result, from the analysis above,
$\chi(S) = \chi(\Sigma_T) -1$, hence
\begin{equation}
\chi( \Sigma_T \coprod S \coprod S) \: = \:
3 \chi(\Sigma_T) - 2,
\end{equation}
matching $\chi(\Sigma_W)$.
\item Suppose there are two branch points.  If the monodromy about one is
$v \in S_n$, then the monodromy about the other is $v^{-1}$.  Assume $v$
has $m$ cycles, in which the $i$th cycle has $k_i$ elements.  (The same will
be true of $v^{-1}$.)  Define $S_i$ to be a ${\mathbb P}^1$ with two
${\mathbb Z}_{k_i}$ orbifolds.  Then,
\begin{equation}
\hat{\Sigma}_W \: = \: \coprod_{j = 1}^m \left( 
\coprod_{k_j} S_j \right).
\end{equation}
Now, let us compare Euler characteristics.  For a single curve,
\begin{equation}
\chi(S_i) \: = \: \chi({\mathbb P}^1) - 2 + 2/k_i,
\end{equation}
hence
\begin{eqnarray}
\chi\left( \hat{\Sigma}_W \right)  & = &
\sum_{j=1}^m k_j \chi(S_j),
\\
& = & \sum_{j=1}^m k_j \left( \chi({\mathbb P}^1) - 2 + 2/k_j \right),
\\
& = & n \chi({\mathbb P}^1) - 2n + 2m,
\end{eqnarray}
using the fact that
\begin{equation}
\sum_{j=1}^m k_j \: = \: n.
\end{equation}
By comparison,
\begin{equation}
\chi(\Sigma_W) \: = \: n \chi({\mathbb P}^1) - \sum_{i=1}^2 (n-m),
\end{equation}
which matches.
\item Let $\Sigma_W$ be a 3-fold cover of $\Sigma_T$, branched along
three points, with branching at each point described by elements of the
conjugacy class $(123)$.  Then,
\begin{equation}
\chi(\Sigma_W) \: = \: 3 \chi(\Sigma_T) - (3)(2).
\end{equation}
The corresponding stack is a disjoint union $S \coprod S \coprod S$,
where each $S$ is a copy of $\Sigma_T$ with three ${\mathbb Z}_3$
orbifold points.  It is straightforward to compute
\begin{equation}
\chi(S) \: = \: \chi(\Sigma_T) - 3 + (3)(1/3) \: = \: \chi(\Sigma_T)
- 2,
\end{equation}
hence
\begin{equation}
\chi(S \coprod S \coprod S) \: = \: 3 \chi(\Sigma_T) - 6,
\end{equation}
matching $\Sigma_W$.
\item Let $\Sigma_W$ be a 3-fold cover of $\Sigma_T$, branched along
three points, with branching at the three points described by
$(12)(3)$, $(13)(2)$, and $(23)(1)$.  (In other words, at each branch point,
two of the three sheets intersect, but a different pair at each one.)
In this case,
\begin{equation}
\chi(\Sigma_W) \: = \: 3 \chi(\Sigma_T) - 3(3-2) \: = \:
3 \chi(\Sigma_T) - 3.
\end{equation}
We can replace $\Sigma_W$ by a disjoint union of three copies of the
stack $S$, where $S$ is $\Sigma_T$ with two ${\mathbb Z}_2$ orbifold points.
It is straightforward to compute
\begin{equation}
\chi(S) \: = \: \chi(\Sigma_T) - 2 + 2(1/2) \: = \: \chi(\Sigma_T) - 1
\end{equation}
(by omitting two disks and replacing them with disks with
${\mathbb Z}_2$ orbifolds), hence
\begin{equation}
\chi(S \coprod S \coprod S) \: = \: 3 \chi(\Sigma_T) - 3,
\end{equation}
matching $\chi(\Sigma_W)$.
\item Finally, we consider an example which will illustrate the path-dependence
of the prescription, and demonstrate that in general one will get different
disjoint unions from different choices of path.

Begin with an $n=4$ cover of $\Sigma_T = {\mathbb P}^1$, branched over five
points.  As before, fix a base point and pick a set of paths around the five
branch points $b_{1-5}$ with monodromies
\begin{eqnarray}
v_1 & = & (12)(34),
\\
v_2 & = & (13)(24),
\\
v_3 & = & (14)(23),
\\
v_4 & = & (13),
\\
v_5 & = & (13).
\end{eqnarray}
It is straightforward to check that the product of the $v_i$ is $1$,
so this is well-defined on ${\mathbb P}^1$, and from the Riemann-Hurwitz
formula, the genus of the branched cover is $1$.  As a disjoint union of
stacks, our prescription gives a disjoint union of four $S_i$, 
each a stacky version of ${\mathbb P}^1$, given as follows:
\begin{itemize}
\item $S_1$ has five ${\mathbb Z}_2$ orbifold points, one at each of the original
branch points.
\item $S_2$ has three ${\mathbb Z}_2$ orbifold points, at the
branch points, $b_{1-3}$.
\item $S_3$ has five ${\mathbb Z}_2$ orbifold points, one at each of the
original branch points.
\item $S_4$ has three  ${\mathbb Z}_2$ orbifold points, at the
branch points, $b_{1-3}$.
\end{itemize}
Now, consider a braid group action that maps the monodromies above to
\begin{eqnarray}
v_1' & = & v_1 \: = \: (12)(34), 
\\
v_2' & = & v_2 \: = \: (13)(24), 
\\
v_3' & = & v_4^{-1} v_3 v_4^{-1} \: = \: (12)(34),
\\
v_4' & = & v_4^{-1} v_3^{-1} v_4 v_3 v_4 \: = \: (24),
\\
v_5' & = & v_5 \: = \: (13).
\end{eqnarray}
The branched cover is the same, but the disjoint union of stacks differs.
Here, we have a disjoint union of four stacky copies $S'_i$ of ${\mathbb P}^1$,
each with four ${\mathbb Z}_2$ orbifold points:
\begin{itemize}
\item $S'_1$ has ${\mathbb Z}_2$ orbifold points at $b_{1-3}, b_5$.
\item $S'_2$ has ${\mathbb Z}_2$ orbifold points at $b_{1-4}$.
\item $S'_3$ has ${\mathbb Z}_2$ orbifold points at $b_{1-3}, b_5$.
\item $S'_4$ has ${\mathbb Z}_2$ orbifold points at $b_{1-4}$.
\end{itemize}
Thus, we see here explicitly that the path dependence means one can obtain
different disjoint unions of stacks from the same branched cover.
\end{itemize}

\subsection{Deforming branched covers to disjoint unions}
\label{sect:disjoint:deform}

Intuitively, one might expect these disjoint unions
$\hat{\Sigma}_W$ constructed from branched covers
to be limits of deformations of the branched covers $\Sigma_W$.
However, matters are more complex.
Let us walk through this carefully.

If we fix the complex structure of the base $\Sigma_T$ and the location of the
branch points, then there are only finitely many covers for fixed
$n$.  Varying the complex structure on $\Sigma_T$ and the location of
the branch points gives the Hurwitz moduli space.  If that Hurwitz
moduli space is connected, then the disjoint union $\hat{\Sigma}_W$ cannot
be a limit of points on that moduli space, because of Zariski's main
theorem, which says that if a proper family of varieties (here, curves)
has a connected general fiber, then every fiber must be connected.
So the disjoint unions $\hat{\Sigma}_W$ automatically describe points in
a different component of the Hurwitz moduli space.

Similarly, there are no metric deformations (Ricci-flat or otherwise)
relating the branched cover $\Sigma_W$ to the disjoint union $\hat{\Sigma}_W$.
A simple example involves unbranched covers of $T^2$, which are also themselves
$T^2$.  A metric deformation would deform one $T^2$ (the cover) to a disjoint
union of multiple $T^2$'s; however, as a Calabi-Yau, the space of
Ricci-flat metrics on $T^2$ is well-known, and
no such deformation through Ricci-flat
metrics exists.

One could consider more general metric deformations, not necessarily
through Ricci-flat metrics, but the
same problem arises.
All metric deformations induce conformal deformations, or put another
way, each conformal class contains a metric of constant scalar curvature,
and so Zariski's connectedness argument above applies again.

That said, there are (at least) two options for larger moduli spaces
which can link the covers to disjoint unions.
(Neither of these is a metric deformation, however.)

One option is to view the covers and disjoint unions
as points in a larger moduli space, distinct from the
Hurwitz moduli space.  To that end, we can view the Hurwitz moduli space
as parametrizing tuples $(C,B,\rho,p)$ where $C$ is a curve,
$B$ the branch locus, $p \in C-B$, and $\rho: \pi_1(C-B,p) 
\rightarrow S_n$ is a homomorphism, and then to get a larger moduli space,
embed $S_n$ into some continuous group such as $GL(n,{\mathbb C})$.
Let $\lambda$ denote the composition of $\rho$ with the embedding
of $S_n$ into a larger group.  If one fixes the conjugacy classes of the
images of the loops around the points of $B$ (but as conjugacy classes
in $GL(n,{\mathbb C})$ instead of $S_n$), then this space is connected
and all of the Hurwitz moduli spaces with local monodromies in these
conjugacy classes will embed as closed subvarieties (but will be disjoint
inside it).

A second option exists, which changes the genus of the cover
at intermediate points between the Hurwitz moduli space and the disjoint
union.
For example,
given an $n$-sheeted etale cover
of an elliptic curve $E$, there is a moduli space $M$ of ramified
$n$-sheeted covers of deformations of $E$ such that general points of $M$
represent smooth genus $n$ ramified covers of a deformation of $E$.  You
can also arrange for $M$ to contain two disjoint special closed
subvarieties $N_{c}$ and $N_{d}$ parametrizing singular ramified $n$-sheeted
covers of deformations of $E$ (of arithmetic genus $n$). Furthermore,
starting from
the universal family of curves over $N_{c}$,
the
family of normalizations of the fibers is a family of smooth
and unramified connected covers of deformations of $E$, i.e.~of smooth
curves of genus one. Similarly, passing to the normalizations of
the fibers of the universal family over $N_{d}$ gives a family of
$n$ disconnected copies of deformations of $E$.
In any event, this is not a metric deformation.

\subsection{Examples}

\subsubsection{$p=0, n=2$}

First, consider the case that the Yang-Mills theory lives on
$\Sigma_T = {\mathbb P}^1$ (so that $p=0$), and restrict to maps of
degree $n=2$, as previously discussed in 
sections~\ref{sect:exs:peq0neq2}, \ref{sect:interp:exs:peq0neq2}.
The Gross-Taylor expansion of the Nguyen-Tanizaki-\"Unsal universe
is
\begin{eqnarray}  \label{eq:peq0:neq2:int1}
Z^+_R(0,p,N) & = &
\frac{N^4}{4} \: \pm \: \frac{N^3}{2} \: + \: \frac{N^2}{4}.
\end{eqnarray}
Adding two such contributions together, for each choice of projector
$P_r$, recovers the original Gross-Taylor result~(\ref{eq:gt:peq0:neq2}).

The first term can be interpreted as before, in terms of maps
${\mathbb P}^1 \coprod {\mathbb P}^1 \rightarrow {\mathbb P}^1$,
which is consistent with the fact that $\chi( {\mathbb P}^1 \coprod 
{\mathbb P}^2) = 2 \chi( {\mathbb P}^1) = 4$.

The middle term we interpret as previously in 
section~\ref{sect:interp:exs:peq0neq2},
as a disjoint union of two copies of a stacky ${\mathbb P}^1$,
each copy with a single ${\mathbb Z}_2$ orbifold point.

Finally, we turn to the third term in~(\ref{eq:peq0:neq2:int1}).
Previously, we interpreted this term as
describing maps from a branched double cover $\Sigma_W$
of ${\mathbb P}^1$, branched over two points (the locations of each $v$).
(In fact, $\Sigma_W$ is itself another ${\mathbb P}^1$.)
Here, we interpret the last term as describing
a stacky $\Sigma_W$, given by a disjoint union of two stacky
${\mathbb P}^1$, each with two ${\mathbb Z}_2$ orbifold points.

As a consistency check, let us compute Euler characteristics.
A stacky ${\mathbb P}^1$ with two ${\mathbb Z}_2$ orbifold points
can be described as a cylinder with two ${\mathbb Z}_2$ orbifolds.
Since the Euler characteristic is additive, we can write it as
\begin{equation}
\chi({\rm cylinder}) \: + \: 2 \chi(B {\mathbb Z}_2) \: = \:
0 \: + \: 2(1/2) \: = \: 1.
\end{equation}
So, any one stacky ${\mathbb P}^1$ with two ${\mathbb Z}_2$ orbifold
points has Euler characteristic 1, half of the Euler characteristic
of a ${\mathbb P}^1$.  A disjoint union of two such stacky
${\mathbb P}^1$'s has the same Euler characteristic as a single
ordinary ${\mathbb P}^1$, and so, this has the same Euler
characteristic as the branched double cover utilized in
\cite{Gross:1992tu,Gross:1993hu,Gross:1993yt,Cordes:1994sd,Moore:1994dk,Cordes:1994fc}.
(See appendix~\ref{app:euler} for more information on Euler characteristics
of the stacky curves appearing in this paper.)

To summarize, we have interpreted the terms in the Gross-Taylor
expansion of the Nguyen-Tanizaki-\"Unsal universe~(\ref{eq:source})
as disjoint unions $\prod_i S_i$
of stacky copies $S_i$ of $\Sigma_T$, such that each component
$S_i$ of the disjoint union maps (at degree one) to $\Sigma_T$,
with the property that the Euler characteristic of the
disjoint union $\prod_i S_i$ matches the power of $N$.
Each of the maps $S_i \rightarrow \Sigma_T$ factors through
the base $\Sigma_T$,
\begin{equation}
S_i \: \stackrel{\pi}{\longrightarrow} \: \Sigma_T \:
\longrightarrow \: \Sigma_T,
\end{equation}
and we identify them combinatorially with copies of
invertible field theories on $\Sigma_T$.

\subsubsection{$p=0, n=3$}

Next, we again consider the case of Yang-Mills theory on
$\Sigma_T = {\mathbb P}^1$, and restrict to maps of degree $n=3$,
as discussed previously in sections~\ref{sect:exs:peq0neq3},
\ref{sect:interp:exs:peq0neq3}.
Here, $|S_3| = 3! = 6$, so we write
\begin{eqnarray}
\Omega_{n=3} \: = \:
1 \: + \:  \sum_{v \neq 1}  \left( \frac{1}{N} \right)^{n-K_{v}} v.
\end{eqnarray}
For reference, the six elements of $S_3$ can be characterized as
\begin{equation}
(1)(2)(3), \: \: \:
(12)(3), (13)(2), (23)(1), \: \: \:
(123), (132).
\end{equation}
These six elements form three conjugacy classes, essentially labelled by
the orders of the cycles.

From~(\ref{eq:source}), we have
\begin{eqnarray} 
Z^+_R(0,p,N) & = &
\frac{N^{2n}}{n!} \delta \left(  (\Omega_n)^{2} P_r \right),
\\
& = &
\frac{N^{2n}}{n!} \delta \left( P_r
\: + \: 2 \sum_v \left(\frac{1}{N}\right)^{n-K_{v}} v P_r
 \: + \: 
 \sum_{ij} 
\left( \frac{1}{N} \right)^{2n - K_{v_1} - K_{v_2}}
v_i v_j \right),  \nonumber
\end{eqnarray}
for $n=3$ here.

Previously, we interpreted the first term as a disjoint union of three
copies of ${\mathbb P}^1$, which also holds in the present setting.

Next, we discuss the middle term.
We interpret this in the same fashion described earlier in
section~\ref{sect:interp:exs:peq0neq3}, in terms of stacky copies of
$\Sigma_T = {\mathbb P}^1$.  Specifically,
\begin{itemize}
\item for $K_v = 2$, we interpret $\Sigma_W = 
{\mathbb P}^1 \coprod {\mathbb P}^1_{[1,2]} \coprod 
{\mathbb P}^1_{[1,2]}$, in other words three copies of
${\mathbb P}^1$, two each with one ${\mathbb Z}_2$ orbifold point.

\item for $K_v=1$, we interpret $\Sigma_W$ as a disjoint union of
three copies of ${\mathbb P}^1_{[1,3]}$, meaning three copies of
${\mathbb P}^1$ each with a single ${\mathbb Z}_3$ orbifold point.
\end{itemize}
This is consistent with the displayed combinatorics and also has the
same Euler characteristic as the exponent of $N$.

Now, we turn to the last term.
There are two cases that contribute to the sum:
\begin{itemize}
\item If both $v_1$ and $v_2$ have order 2,
then 
$K_{v_1} = K_{v_2} = 2$,
so the term is of the form
\begin{equation}
\frac{N^6}{3!} \left( \frac{1}{N} \right)^{6-2-2} \delta(v_1 v_2)
\: = \: \frac{N^4}{3!} \delta(v_1 v_2),
\end{equation}
Previously we interpreted $\Sigma_W = {\mathbb P}^1 \coprod {\mathbb P}^1$, with one copy of
${\mathbb P}^1$ a double cover of $\Sigma_W$, branched over two points,
and the second ${\mathbb P}^1$ a degree-one cover.  
Here, we interpret $\Sigma_W = {\mathbb P}^1 \coprod S \coprod S$, where each $S$
is a ${\mathbb P}^1$ with two ${\mathbb Z}_2$ orbifold points, as will
be relevant later.
It is straightforward to see that $\chi(S) = (2)(1/2) = 1$,
so $\chi(\Sigma_W) = 2 + 1 + 1 = 4$, matching that of the interpretation above.

\item The other case is that $v_1$ and $v_2$ have order 3.

Previously, we interpreted $\Sigma_W = {\mathbb P}^1$,
a three-fold cover of ${\mathbb P}^1$ branched over three points.
Here, we interpret $\Sigma_W = S' \coprod S' \coprod S'$, where each $S'$ is a
stacky ${\mathbb P}^1$ with two ${\mathbb Z}_3$ orbifold points.
It is straightforward to see that $\chi(S') = (2)(1/3) = 2/3$,
hence $\chi(\Sigma_W) = (3)(2/3) = 2$, matching that of the
interpretation above. 
\end{itemize}

\subsubsection{$p=1$}

Our proposal is only intended for the case $p=0$,
because, we do not always have a stacky interpretation
of all of the extra terms of
section~\ref{sect:interp:stack}.

Nevertheless, it may help the reader to understand the construction
to examine some examples of ordinary terms (interpretable as branched
covers) at higher genus, so in this and the next section,
we discuss some examples.

Consider the case that $\Sigma_T = T^2$. and consider degree $n$
covering maps, as previously discussed in sections~\ref{sect:exs:peq1},
\ref{sect:interp:exs:peq1}. 
Here, from~(\ref{eq:source}), we have
\begin{eqnarray} 
Z^+_R(0,p,N) & = & 
1\frac{N^{0}}{n!} \sum_{s,t \in S_n} \delta\left( (\Omega_n)^0 
s t s^{-1} t^{-1} 
P_r \right).
\end{eqnarray}

First, consider the case $n=2$.  Since $S_2$ is abelian, there are no
extra terms arising from $P_r$, as the commutator $[s,t] = 1$ for all
$s, t \in S_2$. 
There are four terms, corresponding to two cases:
\begin{itemize}
\item $s=t=1$:  Previously, we interpreted this as
a disjoint union of two copies of $T^2$.
\item $s \neq 1$ or $t \neq 1$:  Previously, we interpreted this as
a single $T^2$, which
is an unbranched double cover of $T^2$.
\end{itemize}

Each of these cases 
is interpreted here as a disjoint union of $2$ copies of $\Sigma_T = T^2$, without any orbifold points.  The reader should
note that this $\Sigma_W$ has the same Euler characteristic as the
$\Sigma_W$ of the Gross-Taylor expansion, so powers of $N$ match.

\subsubsection{$p=2, n=2$}

Again, our construction is only intended to apply to the case
$p=0$, but it may help the reader to see how terms interpretable
as branched covers can be reinterpreted as disjoint unions of
stacks, at higher genus.  (We do not have an interpretation of the
extra terms arising from the presence of a projector $P_r$, but our
description does apply to the ordinary terms arising in the 
branched cover description.)

Consider the case of double covers of a genus-two Riemann
surface $\Sigma_T$, as previously discussed in 
sections~\ref{sect:exs:peq2neq2}, \ref{sect:interp:exs:peq2neq2}.  
Here, from equation~(\ref{eq:source}), the Gross-Taylor expansion
of the Nguyen-Tanizaki-\"Unsal universe is
\begin{eqnarray}
Z_R^+(0,p,N) & = & \frac{N^{-4}}{2!} \sum_{s_i,t_i \in S_2}
\delta\left( \left( \Omega_2 \right)^{-2} 
\left( \prod_{i=1}^2 s_i t_i s_i^{-1} t_i^{-1} \right) P_r \right),
\\
& = &
\sum_{k=0}^{\infty} \frac{N^{-4}}{2!} 
\left(  \frac{1}{1 - (1/N^2)} \right)^2 
\sum_{s_i, t_i \in S_n} 
\delta \left( \left( \prod_{i=1}^2 s_i t_i s_i^{-1} t_i^{-1}\right) P_r  \right)
 \\
& & \: - \:
\frac{N^{-4}}{2!}  \sum_{s_i, t_i \in S_n} 
\left(  \frac{1}{1 - (1/N^2)} \right)^2 
\delta \left(  \frac{2}{N} v 
\left( \prod_{i=1}^2 s_i t_i s_i^{-1} t_i^{-1} \right) P_r  \right)
\nonumber \\
& & \: + \:
\frac{N^{-4}}{2!}  \sum_{s_i, t_i \in S_n} 
\left(  \frac{1}{1 - (1/N^2)} \right)^2 
\delta \left(  \frac{1}{N^2} v^2 
\left( \prod_{i=1}^2 s_i t_i s_i^{-1} t_i^{-1} \right) P_r  \right).
\nonumber
\end{eqnarray}

The first term above was intepreted previously as giving
$\Sigma_W$ as a unbranched double cover of the genus-two
Riemann surface $\Sigma_T$, which means $\Sigma_W$ has genus 3.
Here, we can interpret this term in terms of
$\Sigma_W = S \coprod S$, where $S$ is an ordinary Riemann surface of
genus $2$, and hence $\chi(S) = 2 - (2)(2) = -2$.  As a result,
$\chi(\Sigma_W) = -4$, matching that of the genus 3 surface in the
interpretation above.

We interpret the middle term as in section~\ref{sect:interp:exs:peq2neq2},
as describing
$\Sigma_W = C \coprod C$, where each stack
$C$ is $\Sigma_T$ with a single ${\mathbb Z}_2$ singularity.

Finally, consider the term
\begin{equation}
\frac{N^{-4}}{2!}  \sum_{s_i, t_i \in S_n} 
\left(  \frac{1}{1 - (1/N^2)} \right)^2 
\delta \left(  \frac{1}{N^2} v^2 
\prod_{i=1}^2 s_i t_i s_i^{-1} t_i^{-1}  \right).
\end{equation}
Previously, we took this term to describe $\Sigma_W$ as a 
branched double cover of $\Sigma_T$,
branched over two points, which from Riemann-Hurwitz implies that
$\Sigma_W$ is a genus-four Riemann surface.
Here, we take $\Sigma_W = S' \coprod S'$, where
$S'$ is a stacky genus-two surface with two ${\mathbb Z}_2$ orbifold points.
It is straightforward to check that
$\chi(S') = -2 - 2 + (2)(1/2) = -3$ (cutting out two ordinary disks and
replacing each with a ${\mathbb Z}_2$ orbifold),
so $\chi(\Sigma_W) = -6$, the same as the Euler characteristic in the
first interpretation, as a genus-four Riemann surface.

\section{Conclusions}

In this paper we have discussed the Gross-Taylor expansion of
universes of the decomposition of two-dimensional pure Yang-Mills.

One open problem is to find a (large $N$) interpretation of the
extra terms arising from the presence of the projector $P_r$,
at genus $p > 0$.  We have suggested one possible interpretation,
in terms of stacky worldsheets, that at least has some desired
properties, but until we can understand all cases, we consider
that proposal to be merely tentative.

We also find it intriguing that in our proposal, stacky worldsheets
arise.  Now, in string theory we are very used to working with
orbifolds of target spaces, and in computations in target space orbifolds,
one sometimes considers orbifold structures on worldsheets, but we
are not aware of previous work in which the worldsheet theories were
defined on two-dimensional worldsheet stacks (though see
\cite{Horava:1990ee} for an analogous special case in three-dimensional
Chern-Simons theories).  We leave this as a consideration
for the future.

We have focused on two-dimensional pure Yang-Mills with gauge
group $SU(N)$ for large $N$; however, the Gross-Taylor expansion has
also been computed for $SO(N)$, $Sp(N)$ groups,
see e.g.~\cite{Naculich:1993ve,Naculich:1993uu,Naculich:1994kd,Ramgoolam:1993hh,Crescimanno:1996hx}.
We expect similar results will hold there, but leave the details for future
work.

We have also focused on the large $N$ limit.
It would be interesting to understand finite $N$ corrections to the
Gross-Taylor expansion of the Nguyen-Tanizaki-\"Unsal universes.
Finite $N$ corrections to the original Gross-Taylor expansion
are discussed in e.g.~\cite[section 3]{Dijkgraaf:2005bp},
\cite{Rudd:1994ta,Baez:1994gk,Matsuo:2004nn,Vafa:2004qa}.
We leave that for future work.

In passing, we should also mention that two-dimensional pure Yang-Mills partition
functions have been related to black holes and higher-dimensional topological
strings in
e.g.~\cite{Ooguri:2004zv,Vafa:2004qa,Aganagic:2004js,Dijkgraaf:2005bp,Aganagic:2006je,Aganagic:2012si}.  
It is tempting to ask whether
decomposition of two-dimensional pure Yang-Mills
\cite{Nguyen:2021yld,Nguyen:2021naa}
may be applied to those results, perhaps for example to
the fragmentation processes described in
\cite{Maldacena:1998uz}.
We leave this for future work.

\section{Acknowledgements}

We would like to thank S.~Ramgoolam for initial collaboration on 
the material in section~\ref{sect:revisit-gt},
and A.~Cherman, C.~Closset, O.~Ganor, J.~Heckman,
S.~Hellerman, P.~Horava, T.~Jacobson, S.~Katz,
I.~Melnikov, A.~Perez-Lona, R.~Plesser, E.~Poppitz, S.~Ramgoolam, 
R.~Szabo, Y.~Tanizaki, W.~Taylor,
and M.~\"Unsal for useful discussions.
T.P.~was partially supported by NSF/BSF grant DMS-2200914, 
NSF FRG grant DMS 2244978, and by Simons
HMS Collaboration grant  347070.
E.S.~was partially supported by NSF grant PHY-2014086.

\appendix

\section{Some identities}
\label{app:identities}

To make this paper self-contained, we include here a handful of pertinent
character identities.

\subsection{Orthogonality relations}

For $\Gamma$ a finite group,  \cite[section 2]{serrerep},
\cite[chapter V, section 31.1]{cr},
\cite[section 7.3]{karpilovsky}, \cite[chapter 2.1]{collins},
\begin{equation} \label{eq:orthog1}
\frac{1}{|\Gamma|} \sum_{g \in \Gamma} \chi_r(ag) \chi_s(g^{-1}b) \: = \:
\frac{\delta_{r,s}}{\dim r} \chi_r(ab),
\end{equation}
\begin{equation} \label{eq:orthog2}
\frac{1}{|\Gamma|} \sum_{g \in \Gamma} \chi_r(agbg^{-1}) \: = \:
\frac{1}{\dim r} \chi_r(a) \chi_r(b),
\end{equation}
\begin{eqnarray} 
\sum_r 
\chi_r(g) \chi_r(h^{-1})
& = & \left\{ \begin{array}{cl}
0 & g, h \mbox{ not conjugate}, \\
\frac{ |G|}{|[g]|}  & g, h \mbox{ conjugate},
\end{array} \right.
\label{eq:char-master2}
\end{eqnarray}
where the sum is over all (honest, non-projective)
irreducible representations of $\Gamma$.
(Analogues exist for projective representations, but such will not be
used in this paper, and so are omitted for brevity.)

For example, if $b$ is in the center of the group algebra, then
\begin{eqnarray}
\chi_r(a) \chi_r(b)
& = & \frac{\dim r}{|\Gamma|} \sum_{g \in \Gamma} \chi_r(a g b g^{-1}),
\\
& = & \frac{\dim r}{|\Gamma|} \sum_{g \in \Gamma} \chi_r(a b) 
\: = \: (\dim r) \chi_r(ab).
\label{eq:char-central-product}
\end{eqnarray}
Similarly, the results above imply \cite[equ'n (B.10)]{Ramgoolam:2022xfk}
\begin{equation}  \label{eq:delta}
\sum_r \frac{ (\dim r) }{|\Gamma|} \chi_r(g) \: = \:  \delta(g).
\end{equation}

\subsection{Delta functions and projectors}
\label{app:projs}

In this section, we shall demonstrate that
\begin{equation}
\delta(g P_r) \: = \: \frac{ \dim r}{|\Gamma|} \chi_r(g),
\end{equation}
where $P_r$ is the projector \cite[equ'n (2.17)]{Ramgoolam:2022xfk}
\begin{equation}  \label{eq:proj:defn}
P_r \: = \: \frac{\dim r}{|\Gamma|} \sum_{g \in \Gamma} \chi_r(g^{-1}) \tau_g.
\end{equation}

To that end, first note that
\begin{eqnarray}
\chi_s(h P_r) & =  &
\frac{\dim r}{|\Gamma|} \sum_g \chi_r(g^{-1}) \chi_s(hg),
\\
& = &
\frac{\dim r}{|\Gamma|} \frac{|\Gamma|}{\dim r} \,  \delta_{r,s} \, \chi_r(h),
\\
& = &
\delta_{r,s} \, \chi_r(h),
\end{eqnarray}
where we used the identity \cite[equ'n (B.6)]{Ramgoolam:2022xfk}
\begin{equation}
\frac{1}{|\Gamma|} \sum_{g \in \Gamma} \chi_r(ag) \chi_s(g^{-1}b)
\: = \: \frac{\delta_{r,s}}{\dim r} \, \chi_r(ab)
\end{equation}
for untwisted finite group representations.

Next, using the identity~(\ref{eq:delta}),
we expand
\begin{eqnarray}
\delta(g P_r) & = &
\sum_s \frac{\dim s}{|\Gamma|} \chi_s( g P_r ),
\\
& = & \sum_s \frac{\dim s}{|\Gamma|} \delta_{r,s} \chi_r(g),
\\
& = &
\frac{\dim r}{|\Gamma|} \chi_r(g).  \label{eq:delta-vs-chi}
\end{eqnarray}

As a consistency check, note that summing both sides of the identity
above over irreducible representations $r$ yields $\delta(g)$,
on the left because of completeness of the projectors, and on the right
from~(\ref{eq:delta}).

\section{Some basics of stacks}
\label{app:basics-stacks}

In this appendix we collect a few facts about stacks that will be important
in the main discussion.  Introductions to topological and smooth
stacks include \cite{metzler1,heinloth,behrendxu,noohi1,noohi2,noohi3},
\cite[lecture 3]{alper1}.  Other details are given in
sections~\ref{sect:interp:stack} and \ref{sect:disjoint}.
For example, as discussed there, for many purposes, smooth stacks
can be treated as if they were smooth manifolds, in that they have
metrics, spinors, and all the other structures needed to define
quantum field theories.  (See also
\cite{Pantev:2005rh,Pantev:2005zs,Pantev:2005wj}, 
where stacks were discussed as targets of sigma models,
instead of worldsheets.)

\subsection{Euler characteristics of stacky curves}
\label{app:euler}

In this paper we will compute and utilize orbifold Euler characteristics of
stacky curves, which we will briefly outline in this appendix.

First, our stacky curves will all be Deligne-Mumford stacks,
with generic stabilizer $1$ (hence, not gerbes), and isolated points of
nontrivial stabilizer, which in this paper will always be local\footnote{
In passing, as varieties, ${\mathbb C}/{\mathbb Z}_n = {\mathbb C}$;
however, as stacks, $[{\mathbb C}/{\mathbb Z}_n] \neq {\mathbb C}$.
}
${\mathbb Z}_n$ orbifolds.

Next, we will use additivity and the fact that
the orbifold Euler characteristic of a disk with a single ${\mathbb Z}_n$
orbifold in its interior is $1/n$.

Since the Euler characteristic is additive, the Euler characteristic
of any curve with a finite number of orbifold points can be computed by
adding the Euler characteristic of the curve with disks about the orbifold
points excised, to the Euler characteristics of the disks.

For example, consider a genus $g$ curve $\Sigma$ with $k$ ${\mathbb Z}_n$
orbifold points.  Now, the Euler characteristic of a smooth genus
$g$ curve is $2-2g$, and if we excise $k$ disks,
\begin{equation}
\chi\left(\mbox{genus $g$ curve minus $k$ disks}\right) \: = \: 2 - 2g - k.
\end{equation}
We can then add back the Euler characteristics of each of the
disks with orbifold points, which in this case means adding
$(k)(1/n)$, to get the Euler characteristic of the stacky curve
$\Sigma$:
\begin{equation}
\chi(\Sigma) \: = \: 2 - 2g - k + k(1/n).
\end{equation}
For example, an $S^2$ with four ${\mathbb Z}_2$ orbifold points has
$\chi = 0$, same as an ordinary torus $T^2$.

\subsection{Maps from stacky curves}
\label{app:stacks:maps}

In the text, we sometimes discuss maps from stacky curves to
ordinary curves.
There is a one-to-one correspondence between
maps $S \rightarrow X$, for $S$ a stacky curve,
and $M \rightarrow X$, where $M$ is the moduli space of $S$,
and $X$ is a space.
This correspondence revolves around the canonical
projection map $\pi: S \rightarrow X$,
as follows:
\begin{itemize}
\item First, consider a map $f: S \rightarrow X$.
Since $X$ is a space, $f$ must send any nontrivial stabilizers on $S$
to the trivial stabilizers on $X$, hence $f$ factors through $\pi$:
\begin{equation}
f \: = \: g \circ \pi
\end{equation}
for $g: M \rightarrow X$ some map.
\item Conversely, given a map $g: M \rightarrow X$, we can compose with $\pi$
to get $f = g \circ \pi: S \rightarrow X$.
\end{itemize}

\section{Gravitational coupling and decomposition}
\label{app:grav:decomp}

We have argued that the Gross-Taylor string sigma model should admit
either a $(-1)$-form or 1-form global symmetry, the latter of which
would imply a decomposition.  Since the Gross-Taylor string is also
coupled to worldsheet gravity,
in this appendix we outline basics of
decomposition
in the presence\footnote{
Decomposition is a property of theories in $d$ spacetime dimensions with
a global $(d-1)$-form symmetry; however, as is now well-known,
it is believed that quantum gravity in dimensions
greater than two cannot have ungauged global symmetries.
We restrict to two dimensions here.
} of gravitational couplings.

Briefly, given a decomposing two-dimensional theory,
gravitational couplings will result in the universes communicating 
gravitationally, but we do not expect any non-gravitational interactions
between the universes.
We will study this in examples.

Consider first
abelian $BF$ theory in two dimensions at level $k$ (not yet coupled to
worldsheet gravity).
As a unitary two-dimensional topological field theory with a semisimple
local operator algebra, it decomposes
\cite{Durhuus:1993cq,Moore:2006dw,Huang:2021zvu,Komargodski:2020mxz},
in this case to a disjoint union
of $k$ distinct invertible field theories.  The projection operators
are linear combinations of the local operators
\begin{equation}
{\cal O}_m \: = \: : \exp(i m B ) :,
\end{equation}
which have clock-shift commutation relations with the Wilson lines.

Now, consider coupling\footnote{
It will not be relevant here, but nonabelian $BF$ theory, for gauge
group $PSL(2,{\mathbb R})$, is itself a model of two-dimensional gravity,
see e.g.~\cite[section 6.2.8]{Birmingham:1991ty},
\cite{Fukuyama:1985gg,Isler:1989hq,Chamseddine:1989yz}.
}
$BF$ theory to worldsheet gravity.
The $BF$ theory itself does not involve the worldsheet metric at all,
hence the `coupling' is trivial:  at least naively,
this is just a disjoint union of an
ordinary $BF$ theory and a pure worldsheet gravity theory.
The $BF$ theory itself still decomposes, but there is only one
worldsheet gravity sector.
One might describe this as a limiting case of a mostly-disjoint union,
with universes that interact with one another only gravitationally,
and which become a true disjoint union in the limit that the
gravitational sector is unbound.

In the same spirit, two-dimensional Dijkgraaf-Witten theory also decomposes,
as was discussed earlier in section~\ref{sect:series:dw}.
One could similarly `couple' Dijkgraaf-Witten theory to worldsheet gravity,
though again, one would expect that the coupling is trivial,
yielding a disjoint union of a Dijkgraaf-Witten theory (which decomposes)
and one copy of worldsheet
gravity.

Now, to further confuse matters, the Gross-Taylor sigma model is 
described in current proposals \cite{Cordes:1994sd,Moore:1994dk,Cordes:1994fc,Horava:1993aq,Horava:1995ic}
as a cohomological field theory, a topologically-twisted
supersymmetric theory with
a
topological subsector.  In such a theory, although the topological subsector
may formally decompose
\cite{Durhuus:1993cq,Moore:2006dw,Huang:2021zvu,Komargodski:2020mxz}, 
the entire QFT does not necessarily decompose.

As a prototype for such details, consider the A model with target
${\mathbb P}^1$, coupled to topological gravity.  This theory has
been considered in e.g.~\cite{Witten:1989ig,Eguchi:1994in}.
The puncture operators and gravitational descendants obey recursion relations
(see e.g.~\cite[equ'ns (5.4)-(5.5)]{Verlinde:1990ku}).
For example, in the A model with target ${\mathbb P}^1$
coupled to gravity, the puncture operator $P$
and the operator $Q$ generating ordinary A model correlation functions obey
\cite[equ'n (2.26)]{Witten:1989ig}, \cite{Eguchi:1994in}
\begin{equation}
\langle \sigma_n(\Phi) XY \rangle \: = \:
n \langle \sigma_{n-1}(\Phi) P \rangle \langle Q XY \rangle \: + \:
n \langle \sigma_{n-1}(\Phi) Q \rangle \langle P XY \rangle
\end{equation}
for $\Phi$ either $P$ or $Q$, and where $\sigma_n$ denotes
gravitational descendants.
For example,
\begin{eqnarray}
\langle \sigma_1(P) P P \rangle & = &
\langle P P \rangle \langle Q P P \rangle \: + \:
\langle P Q \rangle \langle P P P \rangle,
\\
\langle \sigma_1(P) P Q \rangle
 & = & \langle P P \rangle \langle Q P Q \rangle
\: + \: 
\langle P Q \rangle \langle P P Q \rangle.
\end{eqnarray}

If we restrict to the topological subsector of the
A model with target ${\mathbb P}^1$, so as to get a decomposition\footnote{
But only of the topological subsector, not the entire theory.
},
then the projecction operators
are of the form
\begin{equation}
\Pi_{\pm} \: = \: \frac{1}{2\sqrt{q}} \left( Q \pm \sqrt{q} \right),
\end{equation}
where $Q^2 = q$.
From the expressions above,
\begin{eqnarray}
\langle \sigma_n \Pi_{\pm} \Pi_{\pm} \rangle
& = &
2 n \langle \sigma_{n-1}(\Phi) P \rangle \langle Q \Pi_{\pm} \Pi_{\pm} \rangle
\: + \:
2n \langle \sigma_{n-1}(\Phi) Q \rangle \langle P \Pi_{\pm} \Pi_{\pm} \rangle,
\\
& = &
\langle \sigma_n(\Phi) \Pi_{\pm} \rangle,
\end{eqnarray}
trivially.
Formally, this behavior of the topological subsector is consistent with
earlier observations about $BF$ theory coupled to worldsheet gravity:
on on the face of it, if a decomposition arises in the theory without
gravity, then after coupling to gravity, one has a partial decomposition in
which the different universes can
(only) interact gravitationally.

\section{Potential alternative interpretations}
\label{app:ginsu}

As possible alternative interpretations of the localization onto
sectors of distinct instanton number in section~\ref{sect:interp:restr},
we outline here two proposals for constructions of distinct quantum field
theories which localize onto specific instanton sectors.

\subsection{Single instanton restriction as a limit}

In the spirit of \cite{Tanizaki:2019rbk},
consider\footnote{
We would like to thank Y.~Tanizaki and M.~\"Unsal for a discussion of
such instanton restrictions in their model.
} a sigma model with a restriction to instantons of
degree divisible by $k \in {\mathbb Z}$.
This restriction can be accomplished in a local action as follows.
Begin with a standard nonlinear sigma model action
$S_0$, and add two new fields (a circle-valued scalar $\tilde{\varphi}$
and a $U(1)$ gauge field $A$), and terms
\begin{equation} \label{eq:constr1}
\int_{\Sigma} \tilde{\varphi}\left( \phi^* \omega - k F \right)
\end{equation}
where $\omega$ is the K\"ahler form on the target space,
$\phi$ is the map into the target space,
and $F = dA$ is the curvature of a worldsheet $U(1)$ gauge field.

Integrating out $\tilde{\phi}$ gives the constraint
\begin{equation}
\phi^* \omega = k F
\end{equation}
so that $\phi^* \omega$ is constrained to be $k$ times an integer.
(This accomplishes the restriction on instanton degrees.)

Integrating out $A$ gives the constraint
\begin{equation}
k d \tilde{\varphi} \: = \: 0,
\end{equation}
forcing $\tilde{\varphi}$ to be a constant taking values (on connected
components of $\Sigma$) in $k$th roots of unity.
If we proceed down this road, the path integral is written as a sum
over (constant) values of $\tilde{\varphi}$.
Assuming for simplicity and without loss of generality
that $\Sigma$ is connected, the path integral
can be written in the form
\begin{eqnarray}
Z & = & \sum_{\tilde{\varphi}} \int [D \phi]
\exp(-S_0) \exp\left( - \int_{\Sigma} \phi^* (\tilde{\varphi} \omega) \right),
\\
& = &
\sum_{n=0}^{k-1} \int [D\phi] \exp(-S_0) \exp\left(
- \frac{n}{k} \int_{\Sigma} \phi^* \omega \right),
\label{eq:decomp-to-k-pieces}
\end{eqnarray}
which matches the path integral for a sum of universes
(indexed by $\tilde{\varphi}, n$) with variable $B$ fields
(given on universe $n$ by $(n/k) \omega$).
As is typical in examples of this form, the sum over universes enforces
the restriction on instanton degree:  instantons of the wrong degree
cancel out of the sum, leaving only instantons of degree divisible by $k$.

So far, we have generated a restriction to instantons satisfying a divisibility
criterion, which (as expected from decomposition) 
can be described via a sum over
universes.  We want something stronger -- a restriction to instantons of
a single possible degree.
On the face of it,
there are two natural ways one might try to get such a restriction
on instantons.  One way is to take $k \rightarrow 0$; however, this
limit is ill-behaved \cite{yuyapriv}.

Instead, one could try to interpret\footnote{
This conclusion was previously reached by Y.~Tanizaki and
M.~\"Unsal \cite{yuyapriv}.
} the limit $k \rightarrow \infty$
as defining a restriction to a single instanton sector.
Formally, this would correspond to a sum over infinitely many universes.

Fourier analysis suggests a related interpretation.
Write a partition function as
\begin{equation}
Z(\theta) \: = \: \sum_{n \in {\mathbb Z}} Z(n) \exp(i n \theta).
\end{equation}
Then,
\begin{equation}
Z(n) \: = \: \int_0^{2\pi} Z(\theta) \exp(-i \theta n) \frac{d\theta}{2\pi}.
\end{equation}
Formally, this would naively suggest an interpretation as a sum over
uncountably many universes, indexed by $\theta$ angles, though as that
would also appear to imply an uncountably infinite number of dimension-zero
operators, which we cannot reconcile with other results at this time,
we will not pursue such an interpretation here..

In passing, we observe that similar integrals have arisen in
ensemble averaging, see
e.g.~\cite{Witten:2020bvl,Maloney:2020nni,Afkhami-Jeddi:2020ezh,Cotler:2020hgz,Schlenker:2022dyo}.
It should be noted that
an ensemble is not the same as a decomposition, which becomes visible
in QFTs on spacetimes with multiple connected components.  (In the
former case, there is only one summand/integral over the ensemble,
whereas in the second, there are as many as connected components.)

\subsection{Direct single instanton restriction}

In this subsection we outline a proposal for a direct restriction to
single instantons.  It
is well-defined in bosonic theories,
but does not quite apply to cohomological field theories,
as we shall discuss.

In brief, this alternative proposal is to understand the
restriction to single instanton degrees via a construction of
countably infinitely many new quantum field theories, separately consistent,
which localize onto sectors of the original theory.  The idea is to
promote the theta angle to an axion, albeit with a kinetic term.

As a prototype, consider two-dimensional pure Maxwell theory.
This theory has classical action
\begin{equation}
S \: = \: \int_{\Sigma} F \wedge *F.
\end{equation}
Since it is a $U(1)$ gauge theory, it admits nontrivial $U(1)$ bundles,
which are classified by their first Chern classes, which
on a connected two-dimensional surface $\Sigma$ are elements of
$H^2(\Sigma,{\mathbb Z}) \cong {\mathbb Z}$.

Now, corresponding to the values $n$ of the first Chern class,
we consider a countable family of theories with local\footnote{
A nonlocal alternative would be to add a term of the form
\begin{equation}
\int_{\Sigma} B \left[ \left( \int_{\Sigma} F \right) - n \right],
\end{equation}
for $B$ a Lagrange multiplier.
This would clearly select out contributions from $U(1)$ instanton number $n$.
However, as a nonlocal theory, it is unclear to us to what extent it can be
renormalized in general, and hence it is unclear to what extent it would
exist as a quantum theory.  Instead, we shall work with a local action,
in which we use a Lagrange multiplier to force $F$ itself (not its integral)
to match a harmonic representative of the desired cohomology class.
As every cohomology class has a unique harmonic representative,
this should accomplish the same goal, while retaining locality.
} actions
\begin{equation}
S_n \: = \: \int_{\Sigma}\left(
F \wedge *F \: + \: B \left( F - n \, {\rm vol} \right) \right),
\end{equation}
where $n$ vol denotes a harmonic representative of
$[n] \in H^2(\Sigma, {\mathbb Z})$,
and $B$ is a circle-valued field.  (The theory is well-defined under
$B \mapsto B + 2 \pi$ because $\exp(2\pi i n) = 1$.)

In passing, this is similar in spirit to the
Duistermaat-Heckman localization as described in \cite{Witten:1992xu},
where one has a moment map $\mu \propto F$ \cite[equ'n (1.9)]{Witten:1992xu},
so that $\mu=0$ is a critical locus of the Yang-Mills action.

As $B$ is a dynamical field acting as a Lagrange multiplier,
each theory $S_n$ effectively localizes onto
gauge field configurations of first Chern class $n$.  Furthermore, since the
$BF$ terms are relevant deformations, we expect that they dominate in the IR.
(Put another way, $B$ is an axion without kinetic terms, a dynamical
theta angle, hence its zero mode integral recovers an average of the
form discussed in \cite{Maloney:2020nni}.)

For $n=0$, the theory is effectively then, in the IR, ordinary level 1 $BF$
theory.  For $n \neq 0$, pick a representative gauge field $\tilde{A}$
such that $\tilde{F} = n {\rm vol}$, and write the action as
\begin{equation}
S_n \: = \: \int_{\Sigma}\left(
F \wedge *F \: + \: B \left( F - \tilde{F} \right) \right),
\end{equation}
By defining a new gauge field to be the difference $A - \tilde{A}$,
we see that the $B(F - \tilde{F})$ theory effectively reduces to another
$BF$ theory at level 1, in the sense that it has the same operators and
OPEs (though in principle the partition function would be slightly different).

The theories defined by the $S_n$ form a countably infinite set of
theories which localize onto single instanton sectors of the original
(pure Maxwell) theory.  It is natural to speculate that these pieces
are Poisson-dual to the universes of the decomposition,
following the analysis in section~\ref{sect:interp:proto}.

This construction is analogous to a Fourier series.
To that end, recall that the central identity there
is Poisson resummation, which can be expressed as
\begin{equation}
\sum_n \exp(inx) \: = \: 2 \pi \sum_n \delta(x - 2 \pi n).
\end{equation}
Here, an analogous expression in path integrals would be
\begin{equation}
\sum_n \exp\left( i \int B n {\rm vol} \right) \: = \: 
\sum_n \delta\left[ B - 2 \pi n \right],
\end{equation}
Applying this analogue naively, one finds that
\begin{eqnarray}
\sum_n \int [D B] \exp\left( i \int  B \left( F - n {\rm vol} \right)
 \right) 
& = &
\int [DB] \exp\left( i \int B F \right) \delta[B],
\\
& = & 1.
\end{eqnarray}
In general, we would not expect such a naive computation to necessarily
survive quantum corrections, but it does suggest that
in special cases, it may be possible to recover the original theory through
some sort of sum over the pieces $S_n$.
In the next section, we will discuss how this may happen in this
particular example, two-dimensional pure Maxwell theory.

To be relevant for us, we would need a supersymmetrized version of this
procedure.  We shall next outline how that could be accomplished,
and the puzzles that result.

Recall for example from \cite[section 2]{Witten:1993yc} that in
two-dimensional (2,2) supersymmetric gauge theories, the
theta angle can be encoded in a twisted superpotential.
That references defines $\Sigma$ to be a twisted chiral superfield
given as superderivatives of the vector superfield, with components
\cite[equ'n (2.16)]{Witten:1993yc}
\begin{equation}
\Sigma \: = \: \sigma - i \sqrt{2} \theta^+ \overline{\lambda}_+
- i \sqrt{2} \overline{\theta}^- \lambda_- + \sqrt{2} \theta^+
\overline{\theta}^-(D - i F_{01}) + \cdots,
\end{equation}
from which one computes \cite[equ'n (2.26)]{Witten:1993yc}
\begin{eqnarray}
\int d \theta^+ d \overline{\theta}^- \Sigma |_{ \theta^- = \overline{\theta}^+
= 0} & = & \sqrt{2} \left( D - i F_{01} \right),
\\
\int d \theta^- d \overline{\theta}^+ \overline{\Sigma} |_{
\theta^+ = \overline{\theta}^- = 0}
& = &
\sqrt{2}\left( D + i F_{01} \right),
\end{eqnarray}
hence if one defines
\begin{equation}
t \: = \: i r + \frac{\theta}{2\pi},
\end{equation}
where $r$ is the FI parameter and $\theta$ the theta angle, then
\cite[equ'n (2.27)]{Witten:1993yc}
\begin{equation}
\frac{it}{2 \sqrt{2}} \int d \theta^+ d \overline{\theta}^- \Sigma |_{ \theta^- = \overline{\theta}^+
= 0}
\: - \: 
\frac{ i \overline{t}}{2 \sqrt{2}} \int d \theta^- d \overline{\theta}^+ \overline{\Sigma} |_{
\theta^+ = \overline{\theta}^- = 0}
\: = \:
- r D + \frac{\theta}{2\pi} F_{01}.
\end{equation}

With this in mind, we can define a supersymmetric axion in two dimensions
by promoting $t$ to a twisted chiral superfield $T$, with components
\begin{equation}
T \: = \: t - i \sqrt{2} \theta^+ \overline{\gamma}_+ 
-i \sqrt{2} \overline{\theta}^- \gamma_- + \sqrt{2}
\theta^+ \overline{\theta}^- F + \cdots,
\end{equation}
and then consider a twisted superpotential
\begin{eqnarray}
\lefteqn{
\frac{i}{2 \sqrt{2}} \int d \theta^+ d \overline{\theta}^-
(T \Sigma)  |_{ \theta^- = \overline{\theta}^+
= 0}
\: - \: 
\frac{i}{2 \sqrt{2}} \int d \theta^- d \overline{\theta}^+
( \overline{T} \overline{\Sigma} ) |_{ \theta^+ = \overline{\theta}^- = 0}
} \\
& = & - r D + \frac{\theta}{2\pi} F_{01} \: + \:
\frac{i}{2} \left[ \sigma F - \overline{\sigma} \overline{F}
- \sqrt{2} \overline{\gamma}_+ \lambda_- + \sqrt{2} \gamma_- \overline{\lambda}_+
+ \sqrt{2} \gamma_+ \overline{\lambda}_- -
\sqrt{2} \overline{\gamma}_- \lambda_+ \right]
\nonumber
\end{eqnarray}
(in the conventions of \cite{Witten:1993yc}).

In passing, this is closely analogous to axion couplings in four
dimensions, which are given by superpotential terms
${\cal A} W_{\alpha} W^{\alpha}$, for ${\cal A}$ the superfield
containing the axion, see for example \cite[equ'n (7)]{Banks:2002sd}.

It remains to supersymmetrize the coupling of the axion to 
a nontrivial cohomology class.  We will encounter a fatal flaw when we
try to do so.  For completeness, we outline the analysis in the remainder
of this appendix.

Let $\omega_0$ denote a harmonic representative of the
desired cohomology class.
Since $\omega_0$ is a differential form on the worldsheet, by itself
it is a constant in superspace.  A twisted superpotential term
$T \omega_0$ will yield a bosonic term $F \omega_0$, instead of the
desired $t \omega_0$.

To get the desired bosonic $t \omega_0$ term, we can instead think of
$\omega_0$ as the auxiliary field component of a superfield whose
other components vanish.  In the present case, in terms of
twisted chiral superfields, that means we imagine a superfield
\begin{equation}
\Omega_0 \: = \: -i 2 \sqrt{2} \theta^+ \overline{\theta}^- (\omega_0)_{01}.
\end{equation}
Given this superfield, we can add the twisted superpotential term
$T \Omega_0$, which generates the desired bosonic term $t \omega_0$.

For example, \cite[section 8.1]{Sharpe:2019ddn} gives a
twisted chiral multiplet for a dynamical two-form field, which arises
in the $\theta^+ \overline{\theta}^-$ component of the multiplet.
If we reduce to vevs, so that the two-form field is nondynamical and there
are no other components, this twisted chiral multiplet reduces
to the $\Omega_0$ above.

Unfortunately, this method has the disadvantage that it breaks
supersymmetry, as for example we have given a vev to the F term.
More explicitly, the twisted superpotential term $T \Omega_0$ gives
the desired bosonic term $t \omega_0$ but with no other fermionic partners,
which will not be closed under supersymmetry.

However, for purposes of understanding a cohomological field theory,
our requirements are weaker.  All we really need is for the BRST symmetry
to be preserved, which in the untwisted theory is only half of
the (2,2) supersymmetry transformations.
Furthermore, from \cite[equ'n (3.49)]{Morrison:1994fr},
the scalar component of $\Sigma$ is BRST-closed under the A-twist
discussed in that reference.  Given the same twist here, we see that a
Lagrangian term
\begin{equation}
t (\omega_0)_{01},
\end{equation}
though it would be closed under only half of supersymmetry,
would be BRST-closed
under a
(suitably chosen) A-twist, and so could be consistently added to a
cohomological field theory.

We emphasize that the key feature of a cohomological field theory is the
existence of a nilpotent scalar charge -- the BRST operator.
Existence of a BRST symmetry is what enables a topological field
theory to be studied semiclassically.  (The same symmetry is at the heart
of gauge-fixing in Yang-Mills theories, and is the reason for the shared
name.)  We ordinarily obtain cohomological field theories by topologically
twisting a supersymmetric theory -- but the key outcome of that process
is a BRST symmetry, which follows from only half of the supersymmetry.
Thus, breaking half of the supersymmetry but retaining the BRST symmetry
is sufficient for our purposes.

The reader should note that although the twisted superpotential
term $T \Omega_0$ will be BRST closed, its hermitian conjugate would not.
(For example, in the A-twisted GLSM of \cite{Morrison:1994fr},
$\sigma$ is BRST-closed but $\overline{\sigma}$ is not.)
Hence, one can only add a $T \Omega_0$ term, and not its
hermitian conjugate.

To summarize, our proposal to implement the Lagrange multiplier restriction
in an A-twisted GLSM is to add the terms
\begin{eqnarray}  \label{eq:twistsup:terms:fin}
\lefteqn{
\frac{i}{2 \sqrt{2}} \int d \theta^+ d \overline{\theta}^-
\left( T (\Sigma - \Omega_0) \right)  |_{ \theta^- = \overline{\theta}^+
= 0}
\: - \: 
\frac{i}{2 \sqrt{2}} \int d \theta^- d \overline{\theta}^+
( \overline{T} \overline{\Sigma} ) |_{ \theta^+ = \overline{\theta}^- = 0}
} \\
& = &
- r D + \frac{\theta}{2\pi} F_{01} \: + \:
\frac{i}{2} \left[ \sigma F - \overline{\sigma} \overline{F}
- \sqrt{2} \overline{\gamma}_+ \lambda_- + \sqrt{2} \gamma_- \overline{\lambda}_+
+ \sqrt{2} \gamma_+ \overline{\lambda}_- -
\sqrt{2} \overline{\gamma}_- \lambda_+ \right]
\nonumber \\
& & \: - \: t (\omega_0)_{01},
\\
& = &  
r\left( -D - i (\omega_0)_{01} \right)
+ \frac{\theta}{2\pi} \left( F_{01} - (\omega_0)_{01} \right)
\nonumber \\
& & \: + \:
\frac{i}{2} \left[ \sigma F - \overline{\sigma} \overline{F}
- \sqrt{2} \overline{\gamma}_+ \lambda_- + \sqrt{2} \gamma_- \overline{\lambda}_+
+ \sqrt{2} \gamma_+ \overline{\lambda}_- -
\sqrt{2} \overline{\gamma}_- \lambda_+ \right].
\end{eqnarray}
The resulting theory will still be BRST closed under an A-twist of the
form of \cite{Morrison:1994fr}.
The reader should note that the dynamical circle-valued
field $\theta$ acts as the
desired Lagrange multiplier above, forcing $F_{01} = \omega_0$.
The remaining terms are the result of supersymmetrization.

As a consistency test, in a GLSM without (untwisted) superpotential,
in the notation of \cite{Witten:1993yc},
the A-twist of \cite{Morrison:1994fr} twists the supersymmetry parameters
$\epsilon_+$, $\overline{\epsilon}_-$ to scalars,
hence using the supersymmetry transformations
\cite[equ'n (2.12)]{Witten:1993yc}, we have BRST transformations including
\begin{equation}
\delta \sigma \: = \: 0 \: = \: \delta t,
\end{equation}
and we see that $\overline{\lambda}_-$,
$\overline{\gamma}_-$, $\lambda_+$, and $\gamma_+$ are twisted to scalars.
The fermion bilinears in~(\ref{eq:twistsup:terms:fin}) are all
either scalars or 2-forms, so we see explicitly that
promoting the theta angle to an axion
is compatible with the topological twist.

Unfortunately, at this point we now encounter a basic problem we have
not been able to solve, arising from the
\begin{equation}
r\left( -D - i \omega_0 \right)
\end{equation}
terms.  Since the superfield $T$ is dynamical, the field $r$ is a
Lagrange multiplier, forcing
\begin{equation}
D \: = \: - i (\omega_0)_{01}.
\end{equation}
However, $D$ is real (by construction), and $i (\omega_0)_{01}$ is
pure imaginary, so this has no solutions.
For this reason, we do not utilize this supersymmetrized constraint
framework to try to understand the restriction on map degrees
arising in the Gross-Taylor expansion of
Nguyen-Tanizaki-\"Unsal universes.

In passing, promoting the FI parameter to a dynamical field often
arises in constructions of $H$ flux in GLSMs,
see for example
\cite{Israel:2023itj,Israel:2016xfu,Caldeira:2018ynv,Melnikov:2012nm,Quigley:2012gq,Quigley:2011pv,Adams:2012sh,Adams:2009tt,Adams:2009zg,Adams:2009av,Adams:2006kb,Hori:2002cd}.
For example, GLSMs with (2,2) supersymmetry, a single $U(1)$,
and a gauged FI parameter with a nonzero kinetic term are discussed in
\cite[section 3.1.1]{Caldeira:2018ynv}.
The result of gauging the FI parameter in the ${\mathbb P}^n$ model,
is described in \cite{Caldeira:2018ynv}
as a ${\mathbb P}^n$ fibered over a cylinder
parametrized by the FI parameter, giving altogether what is described
there as a trumpet geometry, with nonzero $H$ flux (roughly, the wedge
product of the Fubini-Study form on the projective space and a one-form
along the cylinder).  (Since the geometry is realized by a mix
of ordinary chiral and twisted chiral multiplets, the geometry is
necessarily an example of a generalized geometry in the
sense of \cite{Rocek:1991ze,Hitchin:2010qz}.)
This theory is also described as having a nontrivial
IR limit.
It also has a kinetic term for the axion.  The case of no kinetic
term is described in  \cite[equ'n (3.23)]{Caldeira:2018ynv}
as the limit $b_{\alpha}
\rightarrow \infty$, which the authors describe as the
``no squashing'' limit.

In the case of the Gross-Taylor sigma model, similar ideas would
apply, as well as the same fatal flaw.  For completeness, we outline
the details here.

At heart, the Cordes-Moore-Ramgoolam picture of the Gross-Taylor
sigma model is
a (supersymmetric,
topologically-twisted) sigma model, so the $F$ of two-dimensional Maxwell
theory is replaced by $\phi^* \omega$,
for $\phi: \Sigma_W \rightarrow \Sigma_T$ the map between worldsheets,
and $\omega$ the K\"ahler form on the target $\Sigma_T$.
The analogue of the procedure above would be
accomplished by
adding a periodic scalar $\varphi$ (without kinetic term, i.e.~a Lagrange
multiplier)
and local\footnote{
As discussed earlier, one could imagine adding a nonlocal term to the
Lagrangian, of the form
\begin{equation}
\varphi \int \left(  \phi^* \omega - \omega_0 \right).
\end{equation}
The nonlocal certain would certainly force the cohomology classes
to match, not just the representatives, but would be nonlocal,
hence renormalizability of the resulting theory is unclear.
On the other hand, since $\omega_0$ is harmonic and every cohomology
class has a unique harmonic representative, the local proposal above
should accomplish the same goal.  
}
bosonic terms in the Lagrangian
\begin{equation}
\varphi \left( \phi^* \omega - \omega_0 \right),
\end{equation}
where $\omega_0$ is a fixed harmonic two-form on the worldsheet $\Sigma_W$,
and $\omega$ is the K\"ahler form on the target.
The two-form $\omega_0$ is a fixed harmonic two-form on the worldsheet 
$\Sigma_W$, which should also capture the area of the worldsheet.
A natural candidate is the worldsheet K\"ahler
form, as on any K\"ahler manifold, the K\"ahler form is harmonic.
(This ultimately follows from the fact that $[L,\Delta_d] = 0$ for
$L$ the Lefschetz operator \cite[section 0.7]{gh},
\cite{so-harmonic}.)

As the Gross-Taylor theory is a topologically-twisted supersymmetric
theory,
one would need a supersymmetric version of the terms above.
We do not expect this to result in a well-defined theory, for the
same reasons as above, but for completeness we outline a few details.
Following \cite[equ'n (23)]{Rocek:1991ze},
\cite[section 4.2]{Adams:2023imc},
and in close analogy with our analysis of topologically twisted
gauge theories in the previous subsection,
we can write the supersymmetric extension of the terms above as
a twisted chiral superpotential
\begin{equation}
\int d \theta^+ d \overline{\theta}^- \, \tilde{\Phi} \left( \omega_{i \overline{\jmath}} 
D_- \Phi^{i}
\overline{D}_+ \overline{\Phi}^{\overline{\jmath}} \: - \: \Omega_0 \right)
\: + \:
\int d \theta^- d \overline{\theta}^+ \overline{\tilde{\Phi}} \left(
\omega_{i \overline{\jmath}} 
D_+ \Phi^i
\overline{D}_- \overline{\Phi}^{\overline{\jmath}} \right)
\end{equation}
where $\tilde{\Phi}$ is a twisted chiral multiplet whose
scalar component includes $\theta$, the Lagrange multiplier theta angle,
the scalar component of $\Phi^{\mu}$ is $\phi^{\mu}$,
and $\Omega_0 \propto \theta^+ \overline{\theta}^- \omega_0$ was
defined in the previous section..
In passing, note that
\begin{equation}
\omega_{i \overline{\jmath}} 
D_+ \Phi^{i}
\overline{D}_- \Phi^{\jmath}
\end{equation}
will be a twisted chiral multiplet precisely when $\omega$ is a closed
(1,1) form.
Just as in GLSMs,
the $\tilde{\Phi} \Omega_0$ term in the twisted superpotential
breaks supersymmetry, but is compatible with an A-twist, in the sense
that the twisted theory still possesses the BRST symmetry.
However, the same fatal flaw will arise here as arose
in the previous supersymmetric example, so we do not advocate this
approach as a means of understanding the Gross-Taylor expansion of
Nguyen-Tanizaki-\"Unsal universes.

\end{document}